\documentclass{aa}  

\usepackage{graphicx}
\usepackage{amssymb}  
\usepackage{amsmath}  
\usepackage{verbatim} 
\usepackage{textcomp} 
\usepackage{fp}       
\usepackage{pgf}      
\usepackage{xcolor}   
\usepackage{txfonts}

\def\kms{~km~s$^{-1}$}
\def\ha{{H$\alpha$}}
\def\hb{{H$\beta$}}

\def\oiii{[\ion{O}{III}]}
\def\nii{[\ion{N}{II}]}

\def\hi{\ion{H}{I}}
\def\hii{\ion{H}{II}}
\def\h2{\element[][][][2]{H}}
\def\lprimeco{$\rm L^{\prime}_{CO}$}
\def\alphaco{$\rm \alpha_{CO}$}
\def\kinemetry{\textit{Kinemetry}}
\def\micron{$\mu$}
\def\co2-1{CO J:2-1}
\def\coo3-2{CO J:3-2}
\def\msk-2{0.27~Jy/beam.km/s}        
\def\mask2{0.35~Jy/beam.km/s}        
\def\xx1y1-2{[67,62]}                
\def\x2y22{[147,142]}                
\def\galaxy{NGC~1566}                
\def\off_2{12.8}                     
\def\offset_2{+8.4}                  
\def\systvel{1485}                   
\def\streamvel{50~km~s$^{-1}$}       
\def\distangal{10} 
\def\maxoutflow{157}                 
\FPeval\systemic{round(\systvel -\off_2 :1 )} 
\FPeval\syst{round(\systvel -\offset_2 :1 )}  
\FPeval\restangent{round( tan((1/3600) *(pi/180))   :12)}          
\FPeval\scaleimg{round( (\restangent) *(\distangal) * (10^6) :0)}  
\FPeval\radbulge{round( (\scaleimg) * 3.5 :0)}                     
\begin{document}

   \title{Outflows in the inner kiloparsec of NGC~1566 as revealed by molecular (ALMA) and ionized gas (Gemini-GMOS/IFU) kinematics }

   \author{R. Slater\inst{1,2}
          \and
          C. Finlez\inst{1}
          \and
          N. M. Nagar\inst{1}
          \and
          A. Schnorr-M\"uller\inst{3,4}
          \and
          T. Storchi-Bergmann\inst{5}  
          \and
          D. Lena\inst{6,7}
          \and
          V. Ramakrishnan\inst{1}
          \and
          C. G. Mundell\inst{8}
          \and
          R. A. Riffel\inst{9}
          \and
          B. Peterson\inst{10,11}
          \and
          A. Robinson\inst{12}
          \and
          G. Orellana\inst{13}
          }

   \institute{Departamento de Astronom\'ia, Universidad de Concepci\'on, Casilla 160-C, Concepci\'on, Chile\\
             \email{royslater@astro-udec.cl,nagar@astro-udec.cl} 
         \and
             Departamento de F\'isica, Universidad de Concepci\'on, Casilla 160-C, Concepci\'on, Chile 
         \and
             Max-Planck-Institut f\"ur extraterrestrische Physik, Giessenbachstr. 1, D-85741, Garching, Germany 
         \and
             CAPES Foundation, Ministry of Education of Brazil, 70040-020, Bras\'ilia, Brazil 
         \and
             Instituto de F\'isica, Universidade Federal do Rio Grande do Sul, 91501-970, Porto Alegre, RS, Brasil 
          \and
             SRON, Netherlands Institute for Space Research, Sorbonnelaan 2, NL-3584 CA Utrecht, the Netherlands 
          \and
             Department of Astrophysics/IMAPP, Radboud University, Nijmegen, PO Box 9010, NL-6500 GL Nijmegen, the Netherlands 
          \and
             Department of Physics, University of Bath, Claverton Down, Bath, BA2 7AY, UK 
          \and
             Departamento de F\'isica/CCNE, Universidade Federal de Santa Maria, 97105-900, Santa Maria, RS, Brazil 
          \and
             Department of Astronomy, The Ohio State University, 140 W 18th Avenue, Columbus, OH 43210, USA 
          \and
             Center for Cosmology and AstroParticle Physics, The Ohio State University, 191 West Woodruff Avenue, Columbus, OH 43210, USA 
          \and
             School of Physics and Astronomy, Rochester Institute of Technology, 85 Lomb Memorial Dr., Rochester, NY 14623, USA   
          \and
             Instituto de F\'isica y Astronom\'ia, Universidad de Valpara\'iso, Avda. Gran Bretaña 1111, Valpara\'iso, Chile 
}
   \date{Received xx xx, xx; accepted March 29, 2018}


  \abstract
   {Tracing nuclear inflows and outflows in AGNs, determining the mass of gas involved in these, and their impact on the host galaxy and nuclear black hole, requires 3-D imaging studies of both the ionized and molecular gas.}
   {We aim to map the distribution and 
    kinematics of molecular and ionized gas in a sample of active galaxies, to quantify the nuclear inflows and outflows. Here, we analyze the nuclear kinematics of \galaxy\ via ALMA observations of the \co2-1\ emission at 24~pc spatial and $\sim$2.6\kms\ spectral resolution, and Gemini-GMOS/IFU observations of ionized gas emission lines and stellar absorption lines at similar spatial resolution, and 123\kms\ of intrinsic spectral resolution.}
   {The morphology and kinematics of stellar, molecular (CO) and ionized (\nii) emission lines are compared to the expectations from rotation, outflows, and streaming inflows.}
   {While both ionized and molecular gas show rotation signatures, there are significant non-circular motions in the innermost 200~pc and along spiral arms in the central kpc (CO). 
   The nucleus shows a double-peaked CO profile (Full Width at Zero Intensity of 200\kms), and prominent ($\sim$80\kms) blue and redshifted lobes are found along the 
    minor axis in the inner arcseconds.
    Perturbations by the large-scale bar can qualitatively explain all features in the observed velocity field. We thus favour the presence of a molecular outflow in the disk with true velocities of $\sim$180\kms\ in the nucleus and decelerating to 0 by $\sim$72~pc. 
   The implied molecular outflow rate is $\rm 5.6~[M_{\odot}yr^{-1}]$, with this gas  accumulating in the nuclear 
    2\arcsec\ arms.
   The ionized gas kinematics support an interpretation of a similar, but more spherical, outflow in the inner 100~pc, with no signs of deceleration.  There is some evidence of streaming inflows of $\sim$50\kms\ along specific spiral arms, and the estimated molecular mass inflow rate, $\rm \sim0.1~[M_{\odot}yr^{-1}]$, is significantly larger than the SMBH accretion rate ($\dot{m}=4.8\times10^{-5}~[M_{\odot}yr^{-1}]$).
}
   {}
   \keywords{Galaxies: nuclei -- Galaxies: active -- 
             Galaxies: kinematics and dynamics -- Galaxies: Seyfert
               }
\titlerunning{Outflows in the inner kpc of NGC~1566}
\authorrunning{Roy Slater C.}
   \maketitle

\section{Introduction}
 \label{introsect}
 Supermassive black holes (SMBH) are thought to be ubiquitous in galaxies with bulges and may be key to the formation and evolution of galaxies \citep{korandho2013}. The correlation between the host bulge and central black hole mass \citep{ferrandmerr2000,gebhardt2000,tremaine2002,ferrandford2005,gultekin2009,korandho2013} has been argued to imply a direct causal relationship between the accretion of material by the black hole, the host galaxy star formation and AGN-driven feedback, but direct observational evidence of the mechanisms responsible has remained elusive \citep{heckandbest2014}. Over the past decade, there have been a growing number of facilities providing 3-D spectroscopic imaging observations which have been supporting the study of gaseous and stellar kinematics in active and inactive galaxies at radio \citep[e.g.][]{morganti2009,nesvadba2010}, infrared wavelengths \citep[e.g.][]{storchi2010,riffel2013,diniz2015} and optical \citep[e.g.][]{dumas2007,storchi2007,dicaire2008,westoby2012,schnorr2014a,lena2015,roche2016}. The combination of enhanced sensitivity at unprecedented spatial and spectral resolution provided by ALMA has opened a new window on molecular gas dynamics to study the central kiloparsec of local galaxies where the dynamical and AGN-activity timescales become comparable, and nuclear fueling, AGN feedback and host galaxy quenching can be probed directly. 

\galaxy, a nearly face-on barred spiral galaxy (morphological type SAB) 
is the dominant \citep{vauco73} and brightest member of the 
Dorado group \citep{bajaja95,aguero2004,kilborn2005}, and one of 
the nearest and brightest Seyfert galaxies. Despite having 
many features of a Seyfert 1, several studies 
\citep{alloin85,bottema92,ehle96,kawamuro2013} 
have indicated this nature as uncertain.
\galaxy\ has an intermediate-strength bar \citep[projected radius 33\arcsec\ 
or $\sim$1.5~kpc and P.A. $\sim$0\degr;][]{aguero2004}, and two strongly 
contrasted spiral arms.
Both the assumption of trailing spiral arms and the more marked dust obscuration on the NW side 
(dust in the disk obscuring light from the bulge) seen in Hubble Space Telescope (HST) imaging \citep{malkan98}, 
point to the NW side as the near side and the SE as the far side of the disk.

Despite its proximity, the distance of \galaxy\ is controversial. 
Several studies using the Tully Fisher Relation (TFR) have claimed distances between 18 Mpc 
(EDD\footnote{http://edd.ifa.hawaii.edu/dfirst.php}) to around 6 Mpc \citep{sorce2014,tully2013}.
The \hi\ spectra used in these studies, though of high signal to noise, clearly show a double-peaked
structure, which could lead to significant underestimations of the rotation velocity and thus the 
TFR based distance. Consequently, in this work, we use the mean distance of \distangal\ Mpc from NED\footnote{The NASA/IPAC Extragalactic Database (NED)
is operated by the Jet Propulsion Laboratory, California Institute of Technology,
under contract with the National Aeronautics and Space Administration.}, 
in agreement with the distance used by \citet[][hereafter C14]{combes2014}. 
At this adopted distance, the linear scale in our images is \scaleimg\ pc/arcsec.   
\begin{figure*}
\centering         
  \includegraphics[bb=0 0 640 265,width=0.95\textwidth,clip]{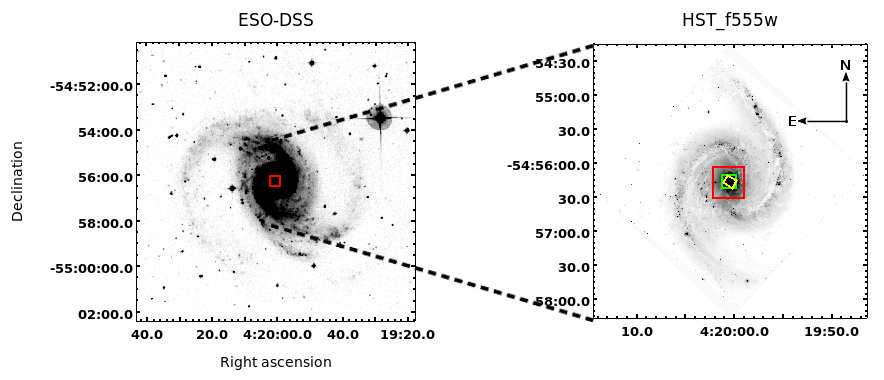}
  \caption{Wide field images of \galaxy. The left panel (ESO-DSS image from the UK Schmidt Telescope) shows the full galaxy, while the right panel (ID:13364, PI:Calzetti. HST image taken with the F555W filter) shows the inner morphology, and highlights the inner spirals arms and the $\sim$1\arcmin\ bar in PA $\sim$0. In both panels the fields of view of the datasets used in this work are shown in yellow (GMOS/IFU), green (inner ALMA FOV: 12\arcsec\ $\times$ 12\arcsec) and red (full ALMA FOV: 27\arcsec\ $\times$ 27\arcsec) squares.
}
  \label{largefieldfig}
\end{figure*}

The systemic velocity of \galaxy\ is 1504 \kms\ from \hi\ observations (NED), but there is wide
range in the optical-spectroscopy based recession velocity values found by different authors. 
C14 found a systemic velocity of 1516\kms\ for \coo3-2; offset $\sim$12\kms\ 
from the \hi- derived value.
We (see below) 
find that the galaxy's integrated \co2-1\ profile is centered on a systemic velocity of $\systvel$\kms.
As we discuss in this work, the nuclear CO profiles are highly perturbed and non-axisymmetric about the
nucleus. Thus the molecular gas derived systemic velocities do not necessarily trace the true systemic velocity of the nucleus of \galaxy. 

The position angle of the major axis of \galaxy\ is $\sim$45\degr\  
\citep[HyperLEDA;][]{LEDAIII}\footnote{http://leda.univ-lyon1.fr/},
and the inclination of the disk was found to be i=35\degr\ \citep[C14,][]{aguero2004}.
In this work we use this major axis position angle and a galaxy disk inclination of i=33\degr, 
as derived from our CO data (Sect.~\ref{obsersect}).

\hi\ studies of the local group 
of \galaxy\ (a sub-part of the Dorado group) show that \galaxy\ is
interacting with its smaller companions \citep{kilborn2005},
and this finding is reinforced 
with the strong correlation found between galaxies with prominent barred 
structures and companions in the Dorado group \citep{kendall2011}.
Given its strong and symmetric spiral arms, its active nucleus (AGN), and its 
proximity, \galaxy\ has been the
subject of great interest within the community, and has extensive
studies of its spiral arm formation \citep{korchagin2000,ma2001,erwin2004,kendall2011}, its gas kinematics \citep{pence90,bottema92,bajaja95,aguero2004,dicaire2008,mezcua2015} and the 
feeding and feedback of the SMBH in its center \citep{elvis89,schm-kinn96,combes2014,smajic2015,davies2016,dasilva2017}.

An early kinematic study of \galaxy\ in \hi\ and \ha\ \citep{pence90}
showed that the most significant \ha\ kinematic feature (after subtraction
of regular rotation) was a spiral arc 
located 26\arcsec\ from the nucleus towards the main spiral arm on the southeast (far) 
side of the galaxy. This spiral arc has a redshifted velocity of 60\kms, i.e.
gas moving away from the nucleus under the assumption of motion in the disk of the
galaxy. Under this assumption, \citet{pence90} estimated outflow velocities,
most plausibly driven by the AGN, of 130\kms\  in the plane of the galaxy, i.e., an equatorial outflow.
\citet{schm-kinn96} also supported the presence of an outflow when  
analyzing the morphology of the nuclear \oiii\ emission; they observed a total extension of 
$\sim$0\farcs7, mainly to the SE, which they interpreted as the base of a conical NLR originating in the nucleus and oriented perpendicular to the plane of the disk, i.e. a polar outflow. 
An extension in the nuclear \oiii\ emission to the SW
was also found by \citet{dasilva2017} in integral field unit (IFU) imaging. They interpreted the morphology
and blue-shifted kinematics of the \oiii\ line as being consistent with an outflow of $\sim$500--800 \kms\ 
driven by the AGN perpendicular to the plane of the disk. 
\citet{aguero2004} found a \hii\ deficiency in the 
inner regions of \galaxy\ \citep[see also][]{pence90}, reinforcing the evidence of 
outflows to the SE, and posited that the blueshifted knot found $\sim$8\arcsec\ from 
the nucleus on the far side of the disk signaled the presence of inflows along 
the galaxy minor axis. Using optical integral field spectroscopy, \citet{davies2016} 
found a high ($\sim$100--200 \kms) dispersion in the \ha\ line over a region $\sim$200pc to the SW of the nucleus,  which they
interpret as most likely due to an outflowing gas illuminated by the radiation field of the AGN. They find that the latter
is sufficiently high to drive outflows in this galaxy.
In X-rays, \citet{elvis89} found extended X-ray emission centered at 
a position $\sim$10\arcsec\ from the nucleus along PA=308\degr\ and at 30\arcsec\ from
the nucleus on the (roughly) opposite side (PA=130\degr). 
\citet{pence90} compared their posited outflow model with these extended X-ray emission 
regions and found that they share the same center. 
Radio imaging with \textit{Australia Telescope Compact Array} (ATCA) at 3.5 cm \citep[1.3 $\times$ 0.75 arcsec synthesized beam;][]{morganti99}
detected the nucleus in continuum, with a potential extension in PA $\sim$10\degr, and a weak 
radio blob 3 arcsec to the N (PA $\sim$10\degr).
The nucleus is detected by the \textit{Parkes Tidbinbilla Interferometer} (PTI) at 13 cm \citep[5 mJy; ][]{roy94}, i.e. it hosts a compact radio source. 
The previous reports of outflows in \galaxy\ are consistent with a picture of a nuclear outflow driven by the AGN
in which the compact base detected
in blue-shifted \oiii\ is primarily from a polar ionization cone tilted towards the observer and close to face-on and
a more extended (out to 1 kpc) equatorial outflow component detected  in a \ha\ arc in the disk of the galaxy. The reason that only
the blueshifted inner (<1\arcsec) ionization cone  has been detected towards the far side of the galaxy disk most likely lies in the dusty features 
seen on the opposite side (NW) of the nucleus (see Fig.~8 of  \citet{dasilva2017} and our structure map in Fig.~\ref{momentsgmosfigs}).

C14 have presented \coo3-2\ observations of \galaxy\ using 
ALMA in Cycle 0: their relatively low spectral resolution
($\sim$10.2\kms\ per channel) and relatively sparse $uv$ coverage limited the 
interpretation of the molecular
gas kinematics in the nuclear region. Their kinematic study of the CO 
emission showed a relatively regular rotational velocity field with 
redshifted streaming motions on the far side of the minor axis, and 
blueshifted streaming motions on the near side, both centered on, and within
a few arcsec of the nucleus. 
However, they argued that the small velocity amplitudes 
(total width $\lesssim$100 \kms) of these perturbations, and the fact that they were only seen in the central 1\arcsec, 
makes an outflow scenario improbable, and instead attributed the non-circular rotational
motions to other factors, e.g. streaming and bars. 
More recently \citet{smajic2015} extended the same study by adding 
SINFONI observations of near-IR molecular emission lines, and came to
similar conclusions, i.e. reinforcing the idea that the velocity perturbations
are more easily explained by streaming motions along the minor axis as a consequence 
of the central bar, rather than outflows.  
\begin{figure*}
\centering            
  \includegraphics[bb=0 50 560 290,width=0.9\textwidth,clip]{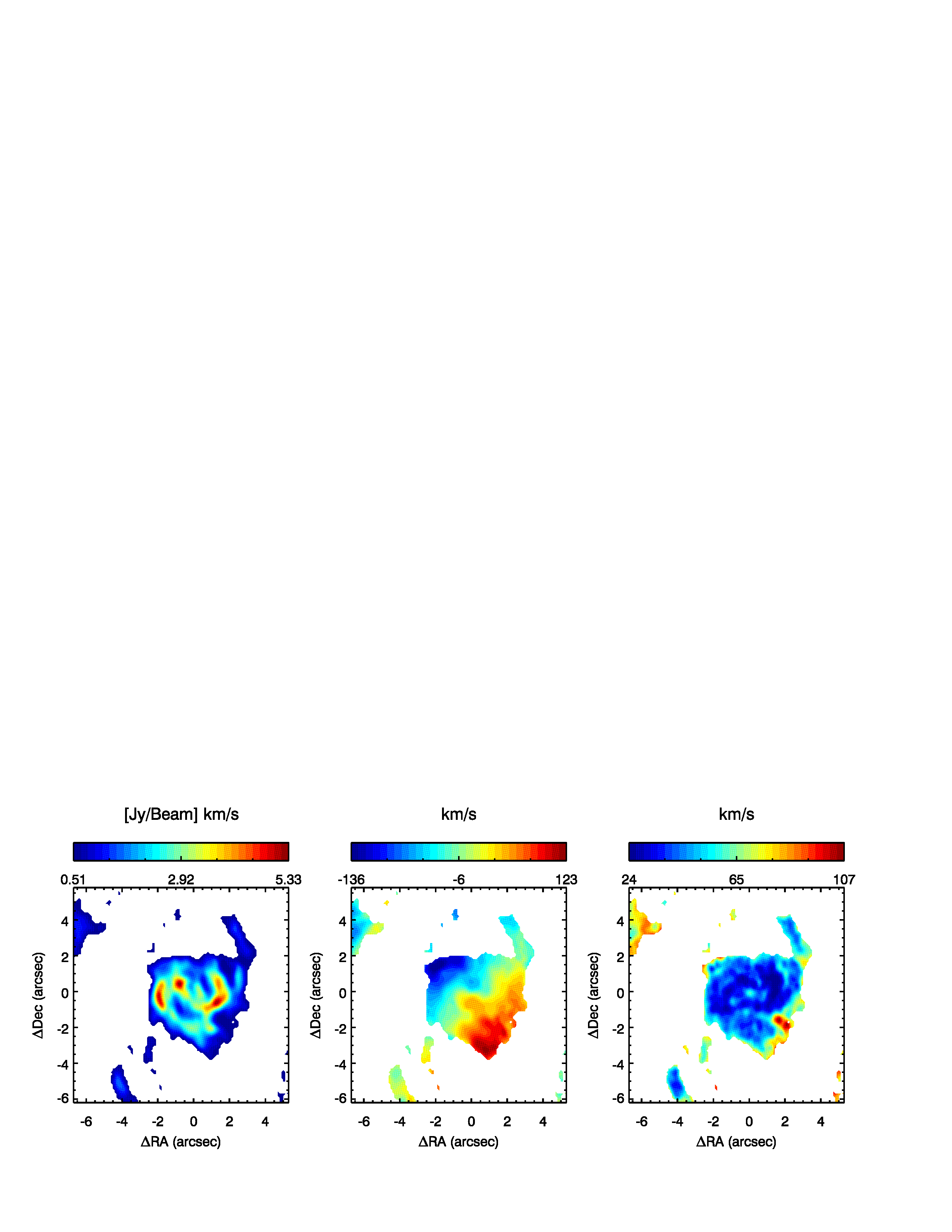}
   \includegraphics[bb=0 50 560 290,width=0.9\textwidth,clip]{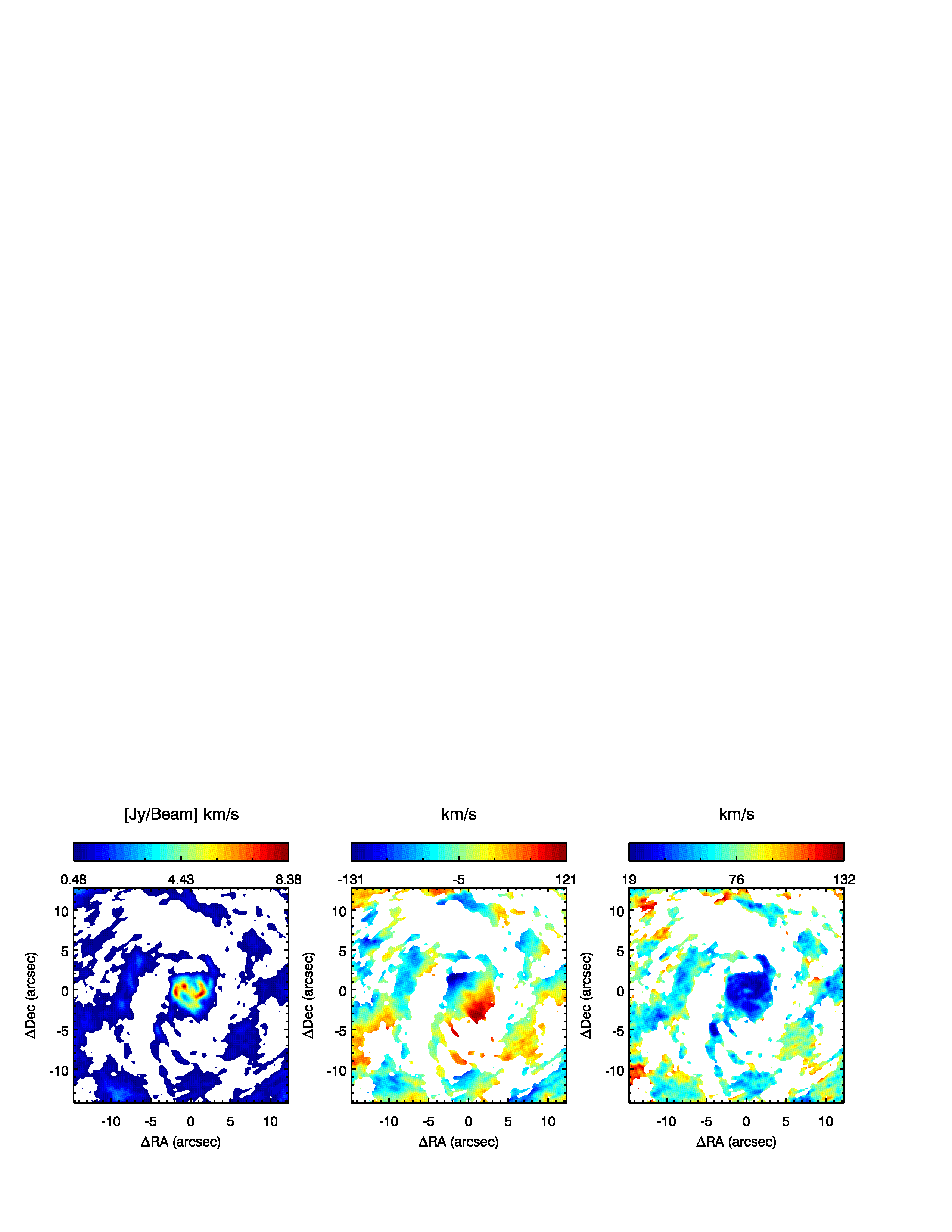}
  \caption{Moment maps of the \co2-1\ emission in \galaxy. Left to right panels show the 0th (integrated flux), 1st (velocity) 
and 2nd (velocity dispersion) moments. The top row shows the moment maps derived
from the highest spatial resolution maps (over a 12\arcsec\ $\times$ 12\arcsec\ FOV) 
to best emphasize the nuclear features. The moment 0 (left) panel has a r.m.s. of 0.17 Jy/beam \kms, and
pixels with flux density lower than 0.51 Jy/beam \kms\ (3$\sigma$) in the moment 0 image were `masked' in all panels of the row
by setting them to a value which results in a white color in the panel.  
The bottom row panels show the moments derived from lower spatial resolution (but higher signal to noise) maps
and show a larger 27\arcsec\ $\times$ 27\arcsec\ FOV to emphasize the larger scale spiral arms. 
The moment 0 (left) panel has a r.m.s. of 0.16 Jy/beam \kms, and
pixels with flux density lower than 0.48 Jy/beam \kms\ (3$\sigma$) in the moment 0 images were `masked' in all panels. 
}
  \label{momentsmaskedfigs}
\end{figure*}

In this work, we reanalyze the nuclear molecular and ionized gas kinematics using new ALMA and Gemini-GMOS/IFU data. 
We present new ALMA observations of \galaxy\ in the \co2-1\ emission line which covers the inner 
12\arcsec\ ($\sim$600 pc) at 1.3\kms\ channel spacing, i.e. a $\sim$2.6\kms spectral resolution. 
These new \co2-1\ observations are more sensitive allowing us to create datacubes at the intrinsic channel spacing of the observations
and have a higher image fidelity (due to the improved $uv$
coverage from the $\sim$32 antenna array) as compared to the previously published Cycle~0 \coo3-2\ observations. 
We compare the distribution and kinematics of molecular gas with that of ionized gas (specifically the \nii\ 6583\AA\ emission line) and stellar absorption lines observed with Gemini-GMOS/IFU at optical wavelengths. We argue that the kinematics can be best explained by a quenched spherical outflow in ionized gas, a decelerating outflow of molecular gas in the plane of the inner ($\sim$300 pc) disk (a scenario we favor over only bar-perturbed kinematics and streaming), and discuss  molecular gas streaming inflows to the nucleus. 

This work is structured as follows: In Sect.~\ref{obsersect} we present the observations and data processing. In 
Sect.~\ref{ressect} we present our results, including the morphology and kinematics of the ionized and molecular
gas and stars, a comparison with our outflow, bar-perturbation and streaming models, 
and a discussion of the results.
Finally, in  Sect.~\ref{concsect}, we present our summary and conclusions.


\section{Observations, Data Processing, and Software}
\label{obsersect}

We observed \galaxy\ with ALMA and Gemini-GMOS/IFU in order to obtain a comprehensive picture of the
morphology and kinematics of the 
molecular gas, ionized gas, and stars.
Images of \galaxy, illustrating the FOVs of our observations are shown in Fig.~\ref{largefieldfig}.

We observed \galaxy\ with ALMA as part of a survey of five nearby 
Seyfert galaxies during Cycle 2: \emph{project-ID 2012.1.00474.S} (PI: Nagar) 
originally approved for Cycle 1 but carried over to Cycle 2.
The observations of \galaxy\ were taken on June 29, 2014, using the 
ALMA Band 6 receivers on thirty-two 12-meter antennas.
Four spectral windows (SPWs) were used; two in
the lower sideband (LSB) and two in the upper sideband (USB).
Three of the SPWs were configured to cover the following lines at
relatively high channel spacing ($\sim$1.3\kms): 
\co2-1\ ($\nu_{obs}$ = 229.401922 GHz) ,
13CH3OH ($\nu_{obs}$ = 241.548041 GHz) and 
CS(J:5-4) ($\nu_{obs}$ = 243.728532 GHz). A fourth SPW was used 
in `continuum' mode to best detect any nuclear continuum emission. 
The SPWs were thus centered on 
229.415 GHz, 227.060 GHz, 241.554 GHz and 243.735 GHz, with bandwidths
of 1.875 GHz, 2.0 GHz, 1.875 GHz and 1.875 GHz, respectively, and 
spectral channel spacing of 1.27\kms, 20.53\kms, 1.22\kms, and  1.21\kms, respectively.
At these frequencies, the full-width half maximum of the 12~m primary beam is about 26\arcmin. 
Antenna baselines ranged from 17~m 
to 650~m, resulting in a typical synthesized beam of 0\farcs6 $\times$ 0\farcs5 
with a position angle (PA) of 25.3\degr.  
                                     
Observations were carried out in two continuous observation blocks, totalling 124~min.  
The nearby radiogalaxy J0519-4546 (PICTOR A)
was used as a phase, bandpass and flux-calibrator. 
Data were calibrated and imaged using CASA 4.2.1 \citep{casa}.
The \co2-1\ emission line was strongly detected over a velocity range of $\sim \pm 200$\kms, 
and we were able to map the CO line at the observed channel spacing of 1.3\kms. 
Thus, our effective spectral resolution (2.6\kms) is higher than the internal dispersion of a typical GMC.
At this spectral resolution, our highest spatial resolution maps 
(made with Brigg's weighting with the  \textit{robust} parameter set to $-$2) have 
a synthesized beam of 0\farcs52 $\times$ 0\farcs35 (beam PA= 13\degr).
The r.m.s. noise per channel in line free channels is $\sim$1 mJy/beam, and rises
by up to a factor 2 in channels with significant line emission.
Equivalent `natural weighted' maps (Brigg's  weighting with \textit{robust}=2) have 
a resolution of 0\farcs6 $\times$ 0\farcs5 (beam PA= 15.6\degr) and an r.m.s. noise per channel
of 1.2 mJy/beam in line free channels, rising by up to a factor of 4 in channels with significant line emission.
The task \textit{immoment} of CASA was used to create moment (integrated flux, velocity, and 
dispersion and skewness) maps from the above data cubes.

Gemini-GMOS observations of \galaxy\ were obtained on the night of 27th of September 2011 with
GMOS in IFU mode and using the R400\_G5325 grating in combination with the r\_G0326 filter 
(\emph{program ID: GS-2011B-Q-23}; P.I. Nagar). This grating yielded an intrinsic spectral resolution (FWHM) of 123\kms, which
was sampled on the CCD at $\sim$30\kms\ per pixel near the [NII] line. The total spectral coverage was from 5620\AA\ to 
6970\AA. 
The observations consisted of two adjacent IFU fields covering 7\arcsec\ $\times$ 5\arcsec\ each, 
resulting in a total spatial coverage of 7 $\times$ 10 arcsec. Six exposures of 350 seconds 
were obtained for each field, each slightly shifted in wavelength and position in order to correct for 
detector defects and fill in CCD chip gaps. 
The data was processed using specific tasks developed for GMOS data in the 
\textit{gemini.gmos IRAF}\footnote{http://iraf.noao.edu} package.

We use four software packages for obtaining velocities, velocity fields, and related parameters from
the datacubes or moment images. 
Ionized gas kinematics were obtained by fitting Gauss-Hermite polynomials and 
double Gaussians to the \nii\ 6583\AA\ emission line using a modified version of the 
\textit{profit}\footnote{http://w3.ufsm.br/rogemar/software.html} routine \citep{riffel2010}. 
The Gauss-Hermite polynomial fits were used to obtain total flux (moment 0), velocity 
(moment 1), and velocity dispersion (moment 2) maps over the full FOV. The nuclear stellar velocity 
and velocity dispersion was determined by using the 
\textit{Penalized Pixel Fitting (\textit{pPXF})}\footnote{http://www-astro.physics.ox.ac.uk/~mxc/software/} code 
\citep{capandems2004}, on the integrated (over our full FOV) spectrum of the galaxy, and using templates based on simple 
stellar populations (SSPs) from \citet{bruandchar2003}. 

We used a modified version of the \textit{Kinemetry}\footnote{http://davor.krajnovic.org/idl/} 
package \citep{kraj2006} to constrain the major axis and inclination of 
\galaxy\ via fits to the CO J:2-1 velocity field, and to determine the best
fit circular velocity field via fits to the moment 1 (velocity) maps of both CO J:2-1 and \nii.
This modified version uses an improved global optimization thereby yielding results that are less affected by the 
starting values and are more robust to missing pixels in the map \citep[discussed in detail in][]{venki2017}.
Some parameters such as the position angle and inclination of the galaxy can either be fixed or obtained on the fly \citep[see ][ for a more detailed description of the software and its features]{kraj2006}. 
On the first \kinemetry\ run we allowed both PA and inclination to vary with radius. The PA and inclination
were then fixed to their median values and a second run of \kinemetry\ was used to obtain the circular rotation map
and the coefficients of each circular velocity and perturbation term. 
Briefly, \kinemetry\ fits concentric elliptical rings to the velocity fields under the assumption that it is possible 
to define the latter such that data extracted along each ellipse can be described by a simple cosine law. 
Therefore, along each ellipse fitted to our velocity map, the program constructs a Fourier series as a function of 
azimuthal angle. 
When using \kinemetry\ we used six odd terms, i.e. $\rm \cos(n\theta$) and $\rm \sin(n\theta$) with n=1, 3, 5.

\begin{figure*}
\centering             
  \includegraphics[bb=7 14 619 270,width=0.8\textwidth,clip]{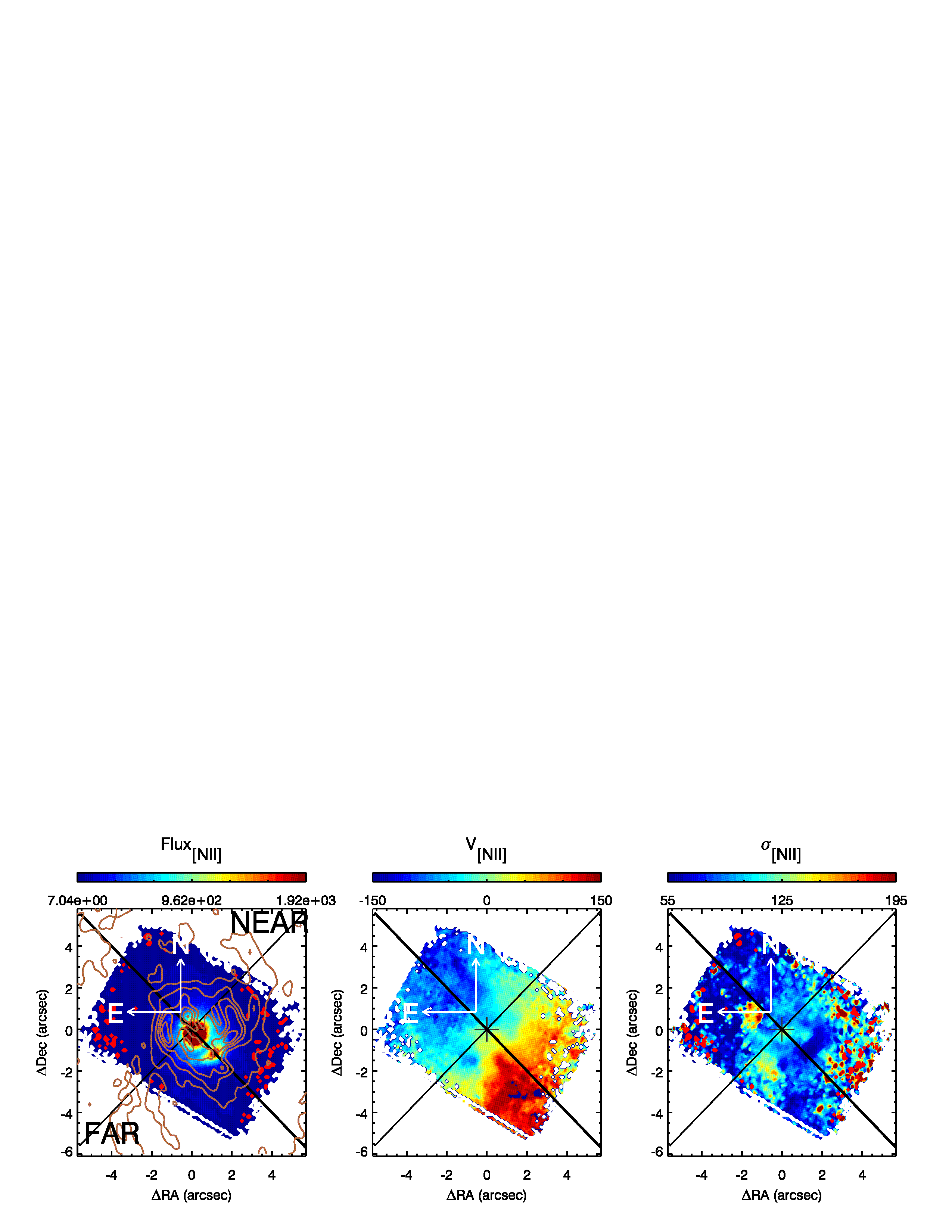}
   \includegraphics[bb=320 7 540 270,width=0.23\textwidth,clip]{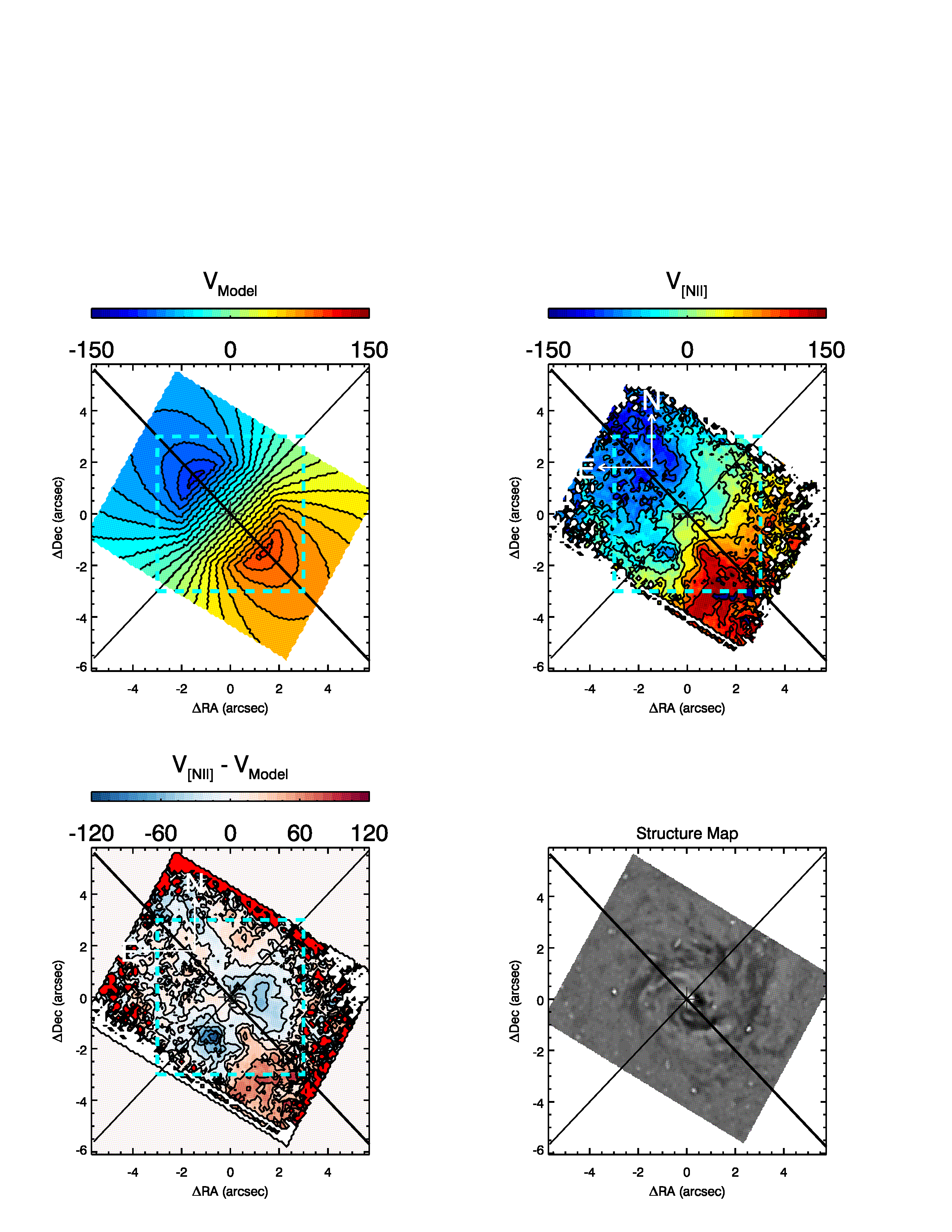}
   \includegraphics[bb=20 14 400 270,width=0.50\textwidth,clip]{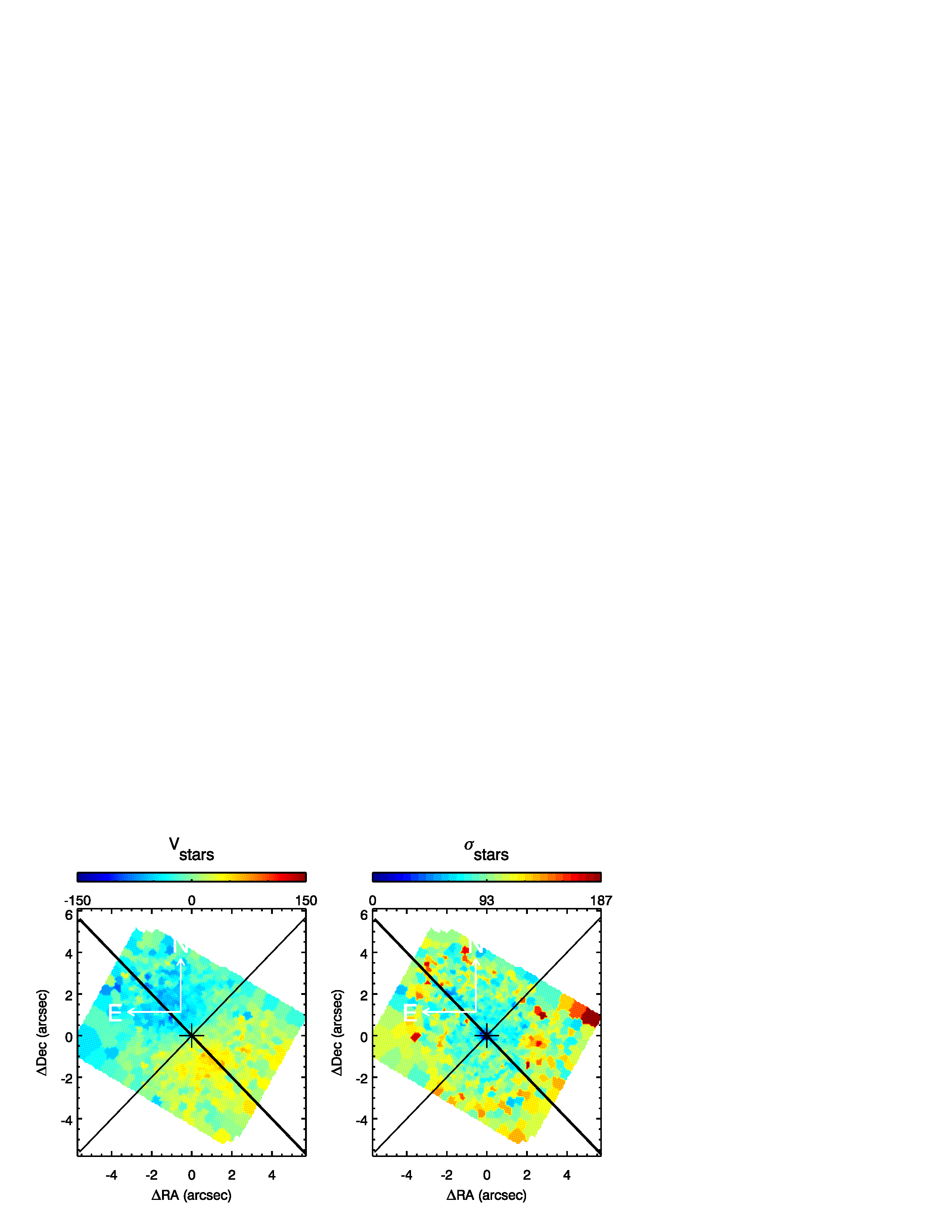}
  \caption{Top row: Moment maps of the \nii\ 6583\AA\ emission line in \galaxy. From left to right, the panels show the 0th (integrated flux), 1st (velocity) and 2nd (velocity dispersion) moments. 
The nuclear position (marked with a cross) was derived from the peak flux of the continuum in Gemini-GMOS datacube, and the
solid lines indicate the major axis (thick line) and the minor axis (thin line). 
In the moment 0 map (left panel), the brown contours show the integrated flux (moment 0) of the
\co2-1\ line.  
Bottom row: from left to right the structure map, and maps of the stellar velocity and stellar velocity dispersion. All maps have N to the top and E to the left (see compass). 
The structure map was created from a HST F606W filter on which unsharp masking was used to emphasize dust features. It is shown at the same size and orientation as the other panels.
  }
  \label{momentsgmosfigs}
\end{figure*}

\section{Results}
\label{ressect}
The molecular (CO; ALMA) and ionized gas emission lines (Gemini-GMOS/IFU) are detected at high
signal to noise out to the edge of the observed FOV.
The \nii\ emission line is detected in every pixel of the GMOS FOV at signal to noise ratios of
3 to 250 in the moment 0 maps. The CO line is detected in well defined structures which cover a fraction of the FOV: here the signal to noise ratio in moment 0 maps ranges between 8 and 35. 

\subsection{Observed Moment Maps: ALMA and Gemini-GMOS/IFU}
\label{momentssect}
Our ALMA 230~GHz continuum map shows only an unresolved nucleus 
and a few other weakly detected components.
We do not present or discuss these 230~GHz continuum maps further since
the sub-mm continuum morphology of the
nuclear region can be better appreciated in the 345~GHz continuum maps of C14
due to the dust emission being brighter at this frequency. 
The principal use of our 230~GHz continuum map is thus to set the position of the nucleus in the 
CO maps.
The extensive dust lanes in the nuclear region of \galaxy\ could cause a small
systematic offset between the nucleus and the location of 
the optical continuum emission peak. Since this systematic offset is most likely to be
significantly less than 0\farcs5 (see the structure map in the bottom panels of Fig.~\ref{momentsgmosfigs})
we here assume that the nucleus is coincident with the stellar continuum peak in the Gemini-GMOS datacube.

The moment 0 (integrated flux), moment 1 (velocity) and moment 2 (velocity dispersion) maps of the
CO J:2-1 line in \galaxy\ are shown in Fig.~\ref{momentsmaskedfigs}. 
The molecular gas in the nucleus of \galaxy\  has a clearly defined disk-like structure
in the inner 3\arcsec\ (\FPeval\fpresult{round(3*\scaleimg:0)}\fpresult~pc), even though this 
region is deficient in both atomic gas and H~II regions \citep{pence90,aguero2004,smajic2015}. 
Within this nuclear disk, the \co2-1\ traces a two-arm spiral structure in the inner 1\farcs7 (\FPeval\fpresult{round(1.7*\scaleimg:0)}\fpresult~pc); 
this spiral structure is also seen in near-infrared and optical images \citep{smajic2015} 
and in previous \coo3-2\ maps (C14).
This inner molecular spiral (in which the arms almost close into a ring) sprouts two 
more extended but fainter \co2-1\ spiral arms which extend out of the inner disk until roughly \FPeval\fpresult{round(2.8*\scaleimg:0)}\fpresult~pc 
(2\farcs8). 
These more extended spiral arms coincide with the dust lanes seen in 
HST images (Fig.~\ref{momentsgmosfigs}; see also C14). 
The CO velocity map shows velocities ranging over $\pm$ $\sim$140\kms.
While the disk in its inner 3\arcsec\ shows a predominantly rotational `spider' velocity diagram,
the velocities are asymmetrical, pointing to a warped inner disk or the presence of non-circular
velocities. Despite the common association of outer \hi\ disks to warped disks, there exists some evidence for the latter 
at parsec scales \citep{greenhill2003}. Warp scenarios at nuclear scales for molecular gas has been explored as, e.g. in \citet{schin2000} reporting that molecular gas could be warped or bar perturbed in the nuclear region, although without analyzing deeply the origins of thereof.
. The trailing pattern of spiral arms in the CO velocity maps 
agrees with that at larger scales in the right panel of Fig.~\ref{largefieldfig}, allowing us 
to assume that the near and far sides of the galaxy disk is to the NW and SE, respectively. 
The velocity dispersion map reveals a typical dispersion of 
$\sim$30\kms\ in the inner spiral arms (see also Fig.~\ref{velreswithfluxfigs}), 
with a high ($\sim$100\kms) velocity dispersion region $\sim$3\arcsec\ to the SW of the nucleus 
along the major axis. Note that this region does not correspond to the star-forming region 
which is clearly detected in the optical observations of \citet{smajic2015}. The nucleus of \galaxy\ shows 
a velocity dispersion of $\sim60$\kms\ (see also Fig.~\ref{velreswithfluxfigs}). 

The moment maps of the \nii\ line in \galaxy, obtained from the GEMINI-GMOS/IFU data, are shown
in the top row of Fig.~\ref{momentsgmosfigs}. Overall, these are roughly similar to those of the CO line.
The \nii\ moment 0 image clearly shows the bright \nii\ region to the SW, corresponding to the
optically-emitting star-forming region seen in \citet{smajic2015} and in agreement 
with a blue region seen to the SW of the nucleus in the CO velocity residual map. 
The velocity map shows kinematics consistent with rotation with velocities similar to those seen
in the CO maps.
Once more there are non-symmetrical velocity patterns closest to the nucleus:
note especially the excess of blueshifts seen on the far side of the galaxy disk $\sim$2\arcsec\ from
the nucleus. The map of the \nii\ velocity dispersion is more difficult to interpret.
The \nii\ dispersion is in general higher than that seen for CO, and the inner spiral
structure is not as clearly discernible as a higher dispersion region. The star-forming 
region to the SW has a dispersion of $\sim$50\kms\ in \nii\ (less than that in CO). 
The nucleus shows a velocity dispersion ($\sim$120\kms) significantly higher than that seen 
in CO and two regions $\sim$2\arcsec\ from the nucleus in the NE and S directions also show relatively 
high (150--180~\kms) velocity dispersions. These will be interpreted below in conjunction with the 
results of the two component fits to the \nii\ emission line.

The bottom row of Fig.~\ref{momentsgmosfigs} shows a structure map and the first two moments of the
stellar velocity field. The structure map was created by running the IDL routine `unsharp\_mask.pro' on a 
HST image taken through the F606W filter, in order to emphasize sharp changes in the image. While the highest
contrast dust arc is seen on the far side of the galaxy as expected, several strong dust features are also
visible on the near side of the galaxy. The stellar velocity map, derived from running \textit{pPXF} on a Voronoi
binned datacube (to achieve a minimum signal to noise of 25 in the continuum near the \oiii\ line in 
each spectrum), shows a clear rotation pattern. 
The stellar rotation velocities are significantly lower than those seen in the molecular gas. 
Since the map
is relatively noisy even after Voronoi binning we did not attempt to fit a PA and inclination to this velocity field using \kinemetry.
Visually, the PA appears consistent with the values we derive from our CO J:2-1 map; this is corroborated by our best fit rotation model to the stellar velocity field (see next section).

\subsection{Modeling the Observed Velocities: Rotation} 
\label{rotmodelsect}

 We used \kinemetry\ to analyze the \co2-1\ velocity field (see Sect.~\ref{obsersect}), both to constrain the
PA and inclination of the CO disk, and to constrain the relative contributions of circular rotation and perturbations. 
We assumed m=2 modes and thus use six (odd) Fourier decomposition terms. 
In the first run both the PA and inclination were allowed to vary with radius, and in the second run we fixed both to their median values from the first run. The Fourier decomposition coefficients of the best fit \kinemetry\ model are shown in Fig.~\ref{kinemetrycoefffig}. Here the $\cos\theta$ term represents the pure circular (rotation) velocity and the other terms are perturbations. In the innermost $\sim$1\arcsec, the $\sin\theta$ (radial) term is positive and dominates the pure rotation term (below we argue that this is best explained by a nuclear outflow) while the $\cos\theta$ (circular rotation) term dominates between $\sim$1 and 4\arcsec. Beyond 3.4\arcsec, the CO velocity field is sparse and the results of \kinemetry\ are thus less reliable. Nevertheless we note that the coefficient of $\sin\theta$ remains stable at $\sim$20\kms at radii beyond 1.4\arcsec, and the $\cos3\theta$ coefficient is significantly negative
between 4\arcsec and 5\arcsec\ which could signify an asymmetry about the minor axis. The other terms show relatively small amplitudes, and given that we sample
a very small range of radii (significantly less than the bar co-rotation radius) we are unable to reliably interpret their variations. To emphasize the changing reliability of these results with radius we plot two dashed vertical lines. To the left of these is the region with a 100\% of data coverage in the ellipse fitting (between 0 and 2.2\arcsec); between these there is a linear decrease from 100\% to a 50\% and to the right of these the data coverage of the ellipse fitting is lower than 50\%,  decreasing to 30\% at 3.4\arcsec and beyond. 
It is thus clear that beyond 3 to 4\arcsec\ the values of the coefficients are relatively unreliable due to the
sparseness of the velocity field, and we thus do not attempt to interpret, e.g., the fact that the s1 term remains
positive and almost constant. 

\begin{figure}
\centering         
  \includegraphics[bb=290 14 570 276,width=0.4\textwidth,clip]{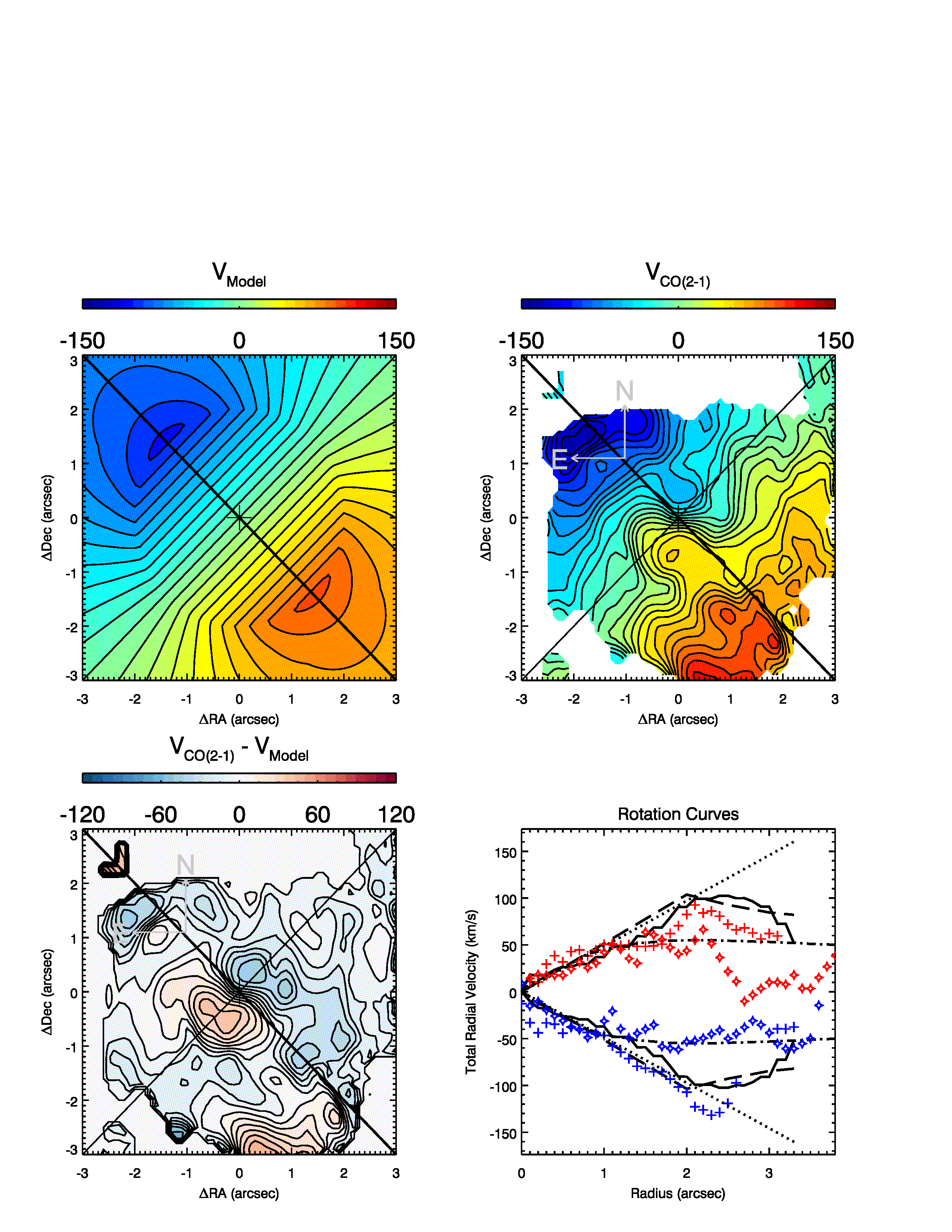} 
  \caption{Comparison of the (projected) rotation models and observed velocities extracted along the major axis, 
  in the central kpc of \galaxy. 
  The solid body model is shown with a dotted line, our ModC2014 model (see text) is shown with a dashed line,
  the circular (`gascirc') model obtained by the \kinemetry\ fit to the  
  \co2-1 velocity field is shown with the solid line, and the Bertola model fit to the stellar velocity
  field is shown with a dashed-dot line. 
  \co2-1\ and stellar velocities extracted along the major axis are shown with crosses and stars, respectively. 
  Blue is used for the NE (approaching)
  side of the galaxy, and red for the SW (receding) side of the galaxy.   
  The zero velocity on the $y$-axis corresponds to $\systvel$\kms; 
  at the nucleus, the velocity of the \co2-1\ line is offset
  from this by  $-$\off_2\kms;  a consequence of the asymmetric double-peaked profile
  of the nucleus, most likely caused by the effects of nuclear outflows and/or bar related perturbations.}
  \label{idlanrob-2fig}
\end{figure}

\begin{figure}
\centering         
\includegraphics[bb=0 0 1500 1200,width=0.4\textwidth,clip]{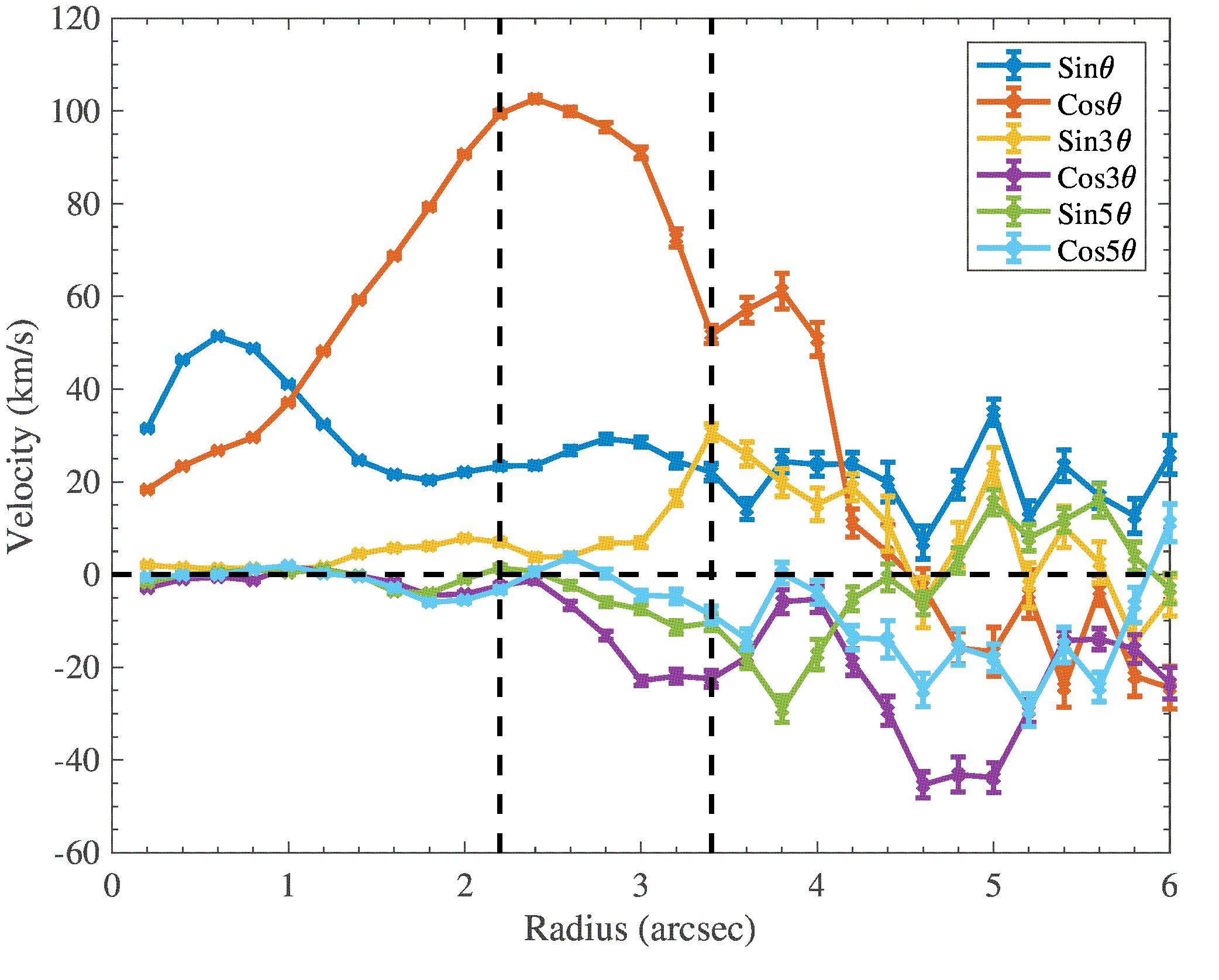}
  \caption{The amplitudes of the Fourier components obtained from the \kinemetry\ analysis of the CO 
  velocity map,
  as a function of distance from the nucleus. The solid red line represents the coefficient of the $\cos\theta$ term (pure circular rotation in a disk) and the other solid lines show the additional components (associated to perturbations) in the Fourier decomposition, following the colours specified in the inset. Only odd Fourier components were used.
  The vertical lines delineate radii at which we have abrupt changes in the fraction of pixels at a given
   radius which have values in the velocity map. At radii smaller than the left vertical line, this fraction is 1. 
   Between the two vertical lines the fraction drops linearly from 1 to 0.5, and beyond 4\arcsec\ the fraction
is relatively steady at $\sim$0.3.
  The horizontal dashed line delineates zero velocity.}
  \label{kinemetrycoefffig}
\end{figure}

We fit the observed stellar velocity field with a `Bertola' model rotation curve \citep[Eq.~2;][]{bertola91}.
This model uses six parameters: the maximum amplitude of the rotation curve \textit{A}; the radius at which this maximum amplitude is achieved \textit{c};  a \textit{p} factor which drives the slope of the rotation curve at larger radii (p=1 gives a flat rotation curve at large radii and p$>$1 gives a decreasing rotation curve at large radii, emulating a finite total mass in the disk), the position angle of the major axis; the inclination of the galaxy, and the systemic velocity. Of the six parameters of the model we fixed the \textit{c} parameter (by visual inspection of the radius at which the rotation amplitude reached its maximum) and the inclination (the same value obtained running \kinemetry\ on the CO velocity field). The best fit parameters obtained were \textit{A}=200\kms, \textit{p}=1.5, PA=45\degr\ (same as derived by \kinemetry\ on the CO velocity field), and a systemic velocity offset of $-$5\kms\ with respect to that used in this work (\systvel\kms). This rotation model is shown with a dash-dotted line in Fig.~\ref{idlanrob-2fig}, and is later (Sect.~\ref{barpersect}) used in the bar perturbation analysis.

The asymmetries in the observed CO and \nii\ velocity fields are best appreciated once
ordered rotational motions are subtracted out. 
We use three rotation models for our analysis of the observed gas velocity maps: all
models use a major axis P.A. of 45\degr\ and an inclination of i=33\degr, derived
from running \kinemetry\ on our CO J:2-1 velocity field.
These three models differ in the parametric form of the circular rotation velocity with radius: 
(a) solid body rotation with parameters: 
\begin{equation}
\label{rotcurvmodel}
\rm Vc_{radial}=\left(S_{rot}*r\right) *cos(\phi)*sin(i). 
\end{equation}
where $\rm S_{rot}$ is equal to \FPeval\fpresult{round(312.13/\radbulge:1)}\fpresult~[\kms pc$^{-1}$],
 r and $\phi$ are the polar coordinates in the 
velocity map, and i is the galaxy inclination; 
(b) An empirical axisymmetric rotation model (hereafter, ModC2014)
based on the nuclear rotation curve derived by C14, but
with a gradual decrease in circular velocity beyond 2\arcsec, roughly following the results from the
\kinemetry\ fit to the \co2-1\ velocity field. Recall that C14 derived the nuclear rotation curve by
using their observed \coo3-2\  velocity field - specifically by minimizing the residual 
(observed $-$ model) velocities - at small radii, and literature \ha\ velocities at larger radii
(the black line in their Fig. 9). 
Since the nuclear CO kinematics are highly perturbed and the velocity field is relatively sparse (especially beyond 2\farcs5), it is not clear that a \kinemetry\ fit or a minimization of residuals will produce a reliable circular rotation model. In fact the circular velocity model obtained by applying \kinemetry\ to our CO velocity field, 
and that from fitting a Bertola model to the stellar velocity field, are both significantly different from the  C14 rotation model (Fig.~\ref{idlanrob-2fig}): 
the most significant difference is a decrease in the rotation velocities beyond 2\arcsec. 
Since a major function of the rotation model in the following sections is to emphasize asymmetries in the observed velocity field, we create a smooth rotation model 
which follows C14 (and the solid body model above) in the inner 2\arcsec\ and decreases (to reflect the
\kinemetry-derived and Bertola models)  at larger distances (dashed lines in Fig.~\ref{idlanrob-2fig}). 
This ModC2014 model is used to produce residual (observed $-$ model) velocity maps which better emphasize deviations from circular rotation as compared to using the C14 rotation model; and 
(c) the gas circular rotation model (specifically the variable ‘gascirc’)  
obtained by running  \kinemetry\ on the CO J:2-1 velocity map.

A direct comparison of the rotation models considered by us in the inner kpc of \galaxy\  is 
shown in Fig.~\ref{idlanrob-2fig}.  
The CO J:2-1 velocities extracted along the major axis (crosses) show several differences from the
solid body and ModC2014 models. First, the velocities are not axisymmetric with the blue-shifted
velocities (to the NE) larger than the red-shifted ones (SW) between 1--2.5 arcsec from 
the nucleus. Second, both blue and red shifted sides show wiggles with larger velocities in the
inner 0\farcs5 and relatively small velocities at distances $\geq$2\farcs5 arcsec from the nucleus. The circular gas velocity fit
of \kinemetry\ to the \co2-1 velocity field (solid line) is by definition an axisymmetric fit; it
 well follows the solid body rotation model until 2\farcs4 (except for an excess of velocities
in the inner 0\farcs5 (below we argue that this is due to a nuclear outflow) after which it shows
 rotation velocities slightly lower than ModC2014. 

On the other hand, the stellar velocities along the major axis (stars in Fig.~\ref{idlanrob-2fig}) 
presents two stages: the blue and red shifted sides are quite similar until $\sim$2\arcsec: in the inner arcsec it is 
completely consistent with our ModC2014 model. Beyond 2\arcsec, on the NE (blueshifted) side the stellar velocities flatten 
at $\sim$60\kms\ while on the SW (redshifted) side the stellar velocities reach a similar peak but then decrease to almost
 zero velocity at 2.8\arcsec, after which they gradually increase again to $\sim$40\kms.

For the three models - solid body rotation, ModC2014, and \kinemetry-derived - the residual (observed minus model)
\co2-1\ velocity maps reveal similar asymmetries in the inner 2.6\arcsec. The differences in the residual maps are seen at greater radii:  the \kinemetry\ model 
undersubtracts the observed velocities while the ModC2014 and solid body rotation models oversubtract the observed velocities. 
The \kinemetry\ model results in the best and most symmetrical residual velocity map for \co2-1, but does not work well for the GMOS/IFU data. For GMOS-derived ionized gas velocity fields, the inner region is 
well subtracted but a radii greater than 2\arcsec\ the velocities are not well subtracted 
as a consequence of the very low model velocities; i.e., rotation velocities in ionized and stellar lines do not decrease at radii greater 2.6\arcsec\ as in the case of \co2-1. Given that we are interested primarily in identifying deviations from axis-symmetric rotation in the innermost region, rather than accurately predicting the true rotation curve, unless
otherwise stated, we consistently use the ModC2014 model for all (ALMA and GMOS) kinematical analysis in this work.

\begin{figure*}
\centering         
  \includegraphics[bb=40 120 400 490,width=0.45\textwidth,keepaspectratio]{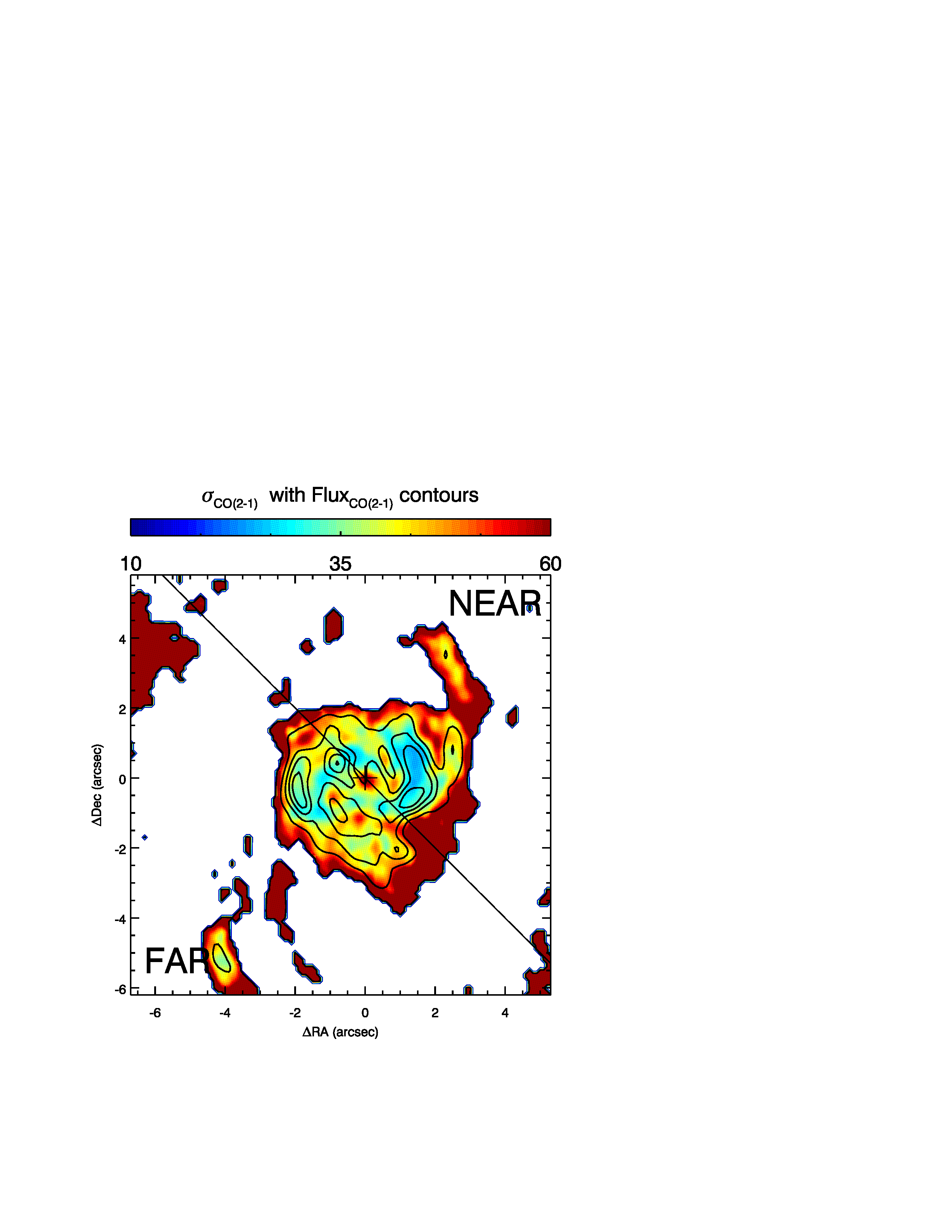} 
  \includegraphics[bb=40 120 400 490,width=0.45\textwidth,keepaspectratio]{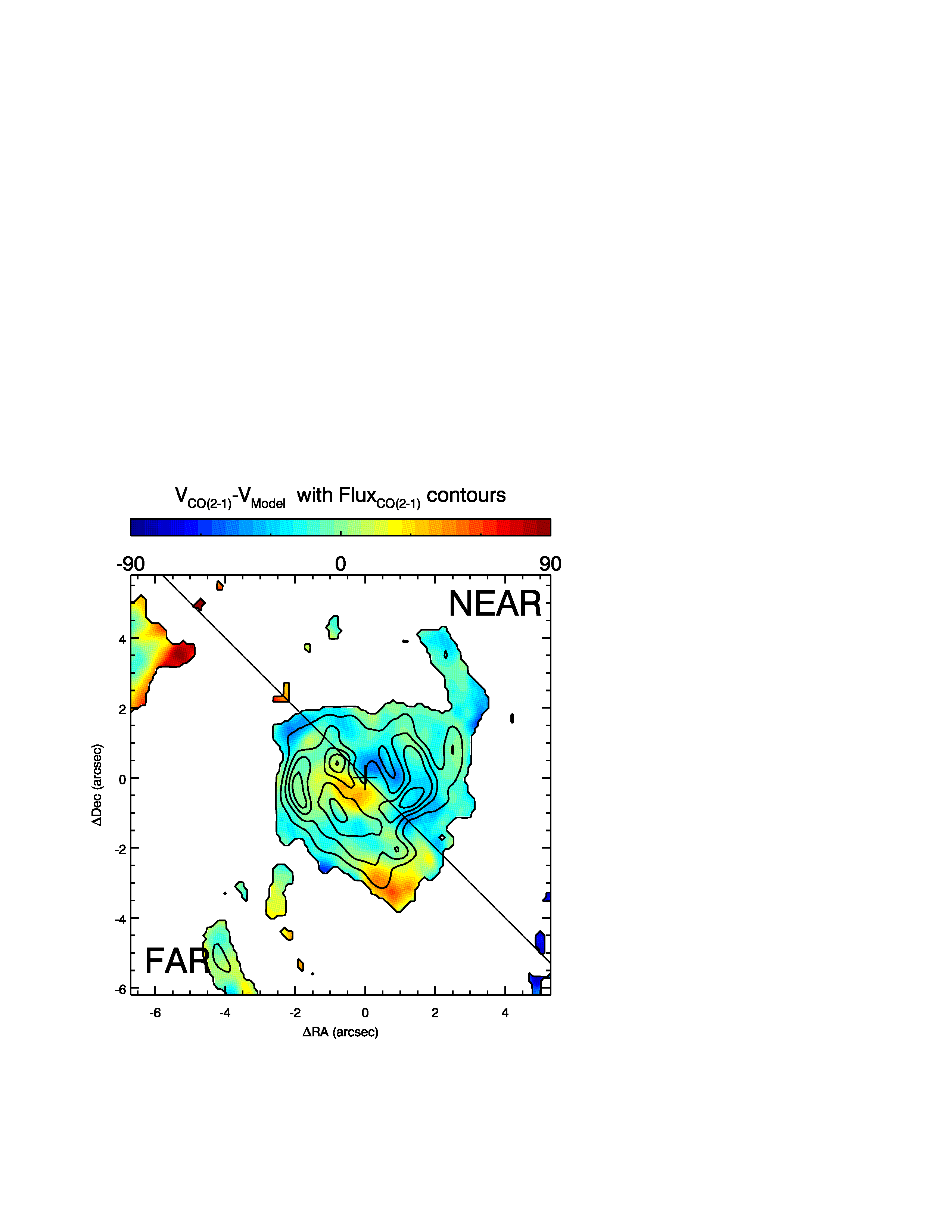} 
  \includegraphics[bb=40 120 400 490,width=0.45\textwidth,clip]{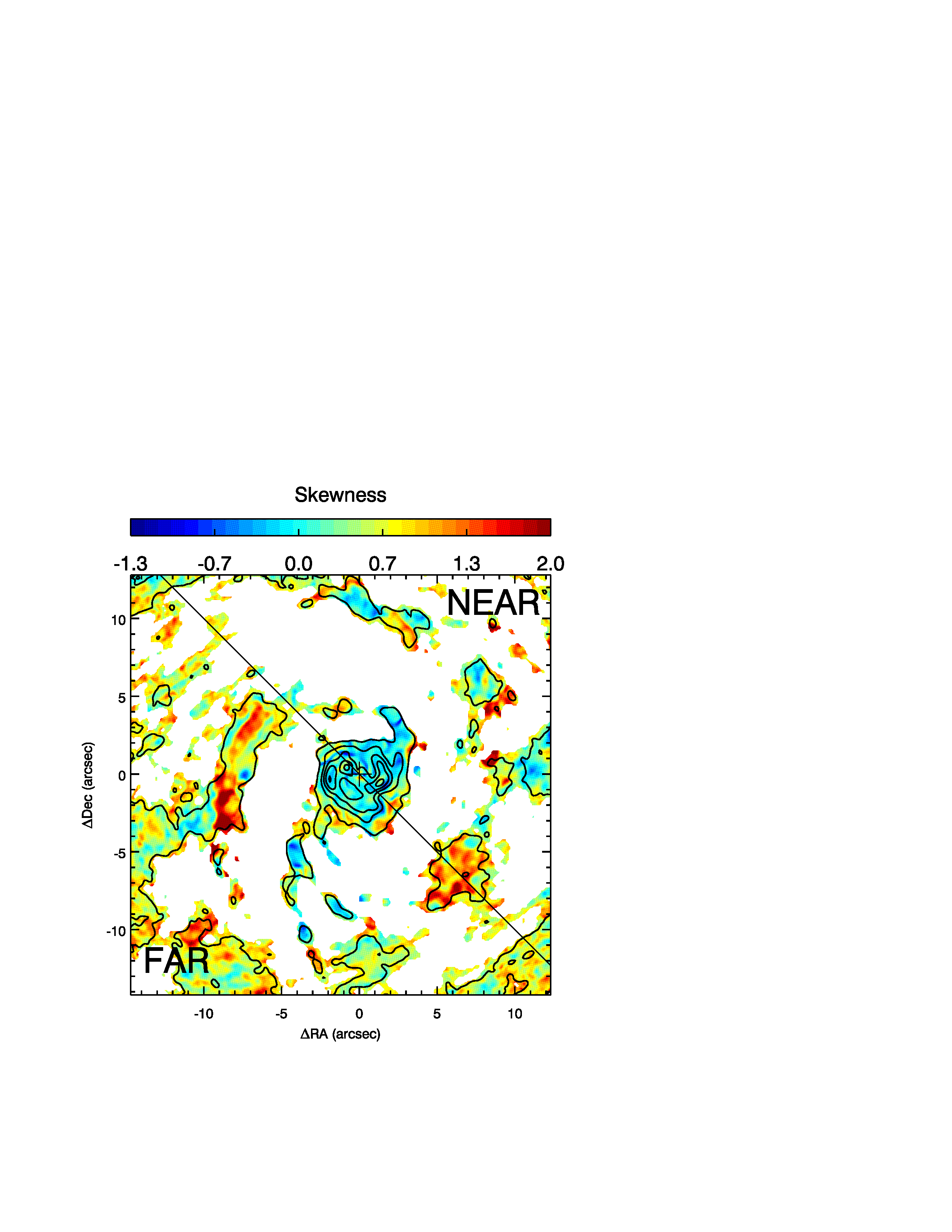}
  \includegraphics[bb=40 120 400 490,width=0.45\textwidth,keepaspectratio,clip]{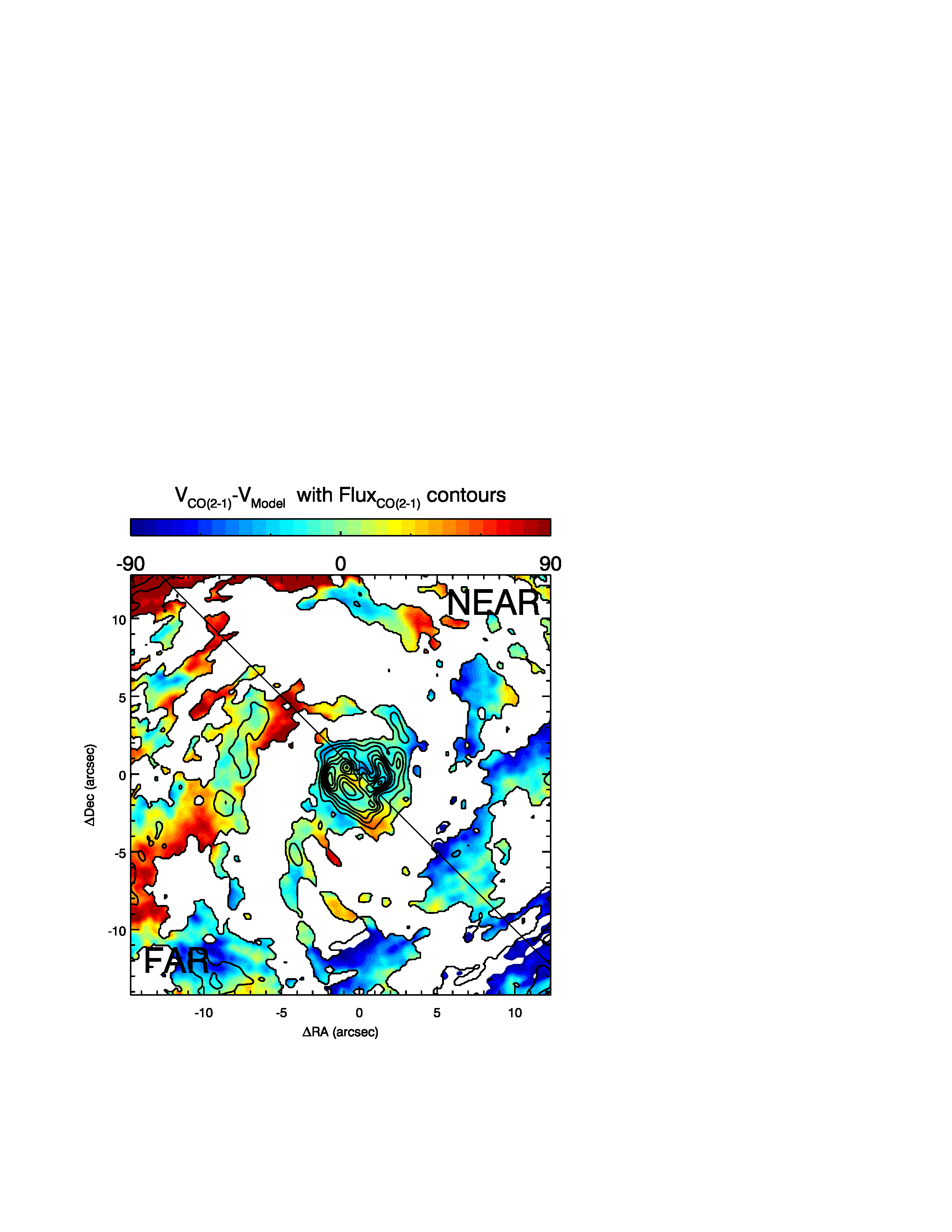} 
  \caption{\textbf{Top:} Maps of the \co2-1\ velocity dispersion (left) and velocity residuals after subtraction
   of the ModC2014 rotation model (right panel) for the inner FOV. Contours in both panels show the moment 0 (integrated flux) map of the \co2-1\ emission, ranging from 0.4~mJy/beam \kms\ to 5.4~mJy/beam \kms. 
  \textbf{Bottom left:} moment 3 (skewness) map of the \co2-1\ emission for the larger FOV, shown in color following the color bar: 
          blue colors represent spectra with excess emission towards the blue side of the weighted mean 
          velocity. The CO Moment 0 (total flux) map is overlaid with black contours. 
          \textbf{Bottom right:} As in the top right panel but for the larger FOV. 
        Pixels with velocity less than $-$90 \kms\ in the right panels are shown in white. Contours in bottom panels show the moment 0 (integrated flux) map of the \co2-1\ emission, ranging from 0.4~mJy/beam \kms\ to 8.4~mJy/beam \kms.
}
  \label{velreswithfluxfigs}
\end{figure*}

\begin{figure*}
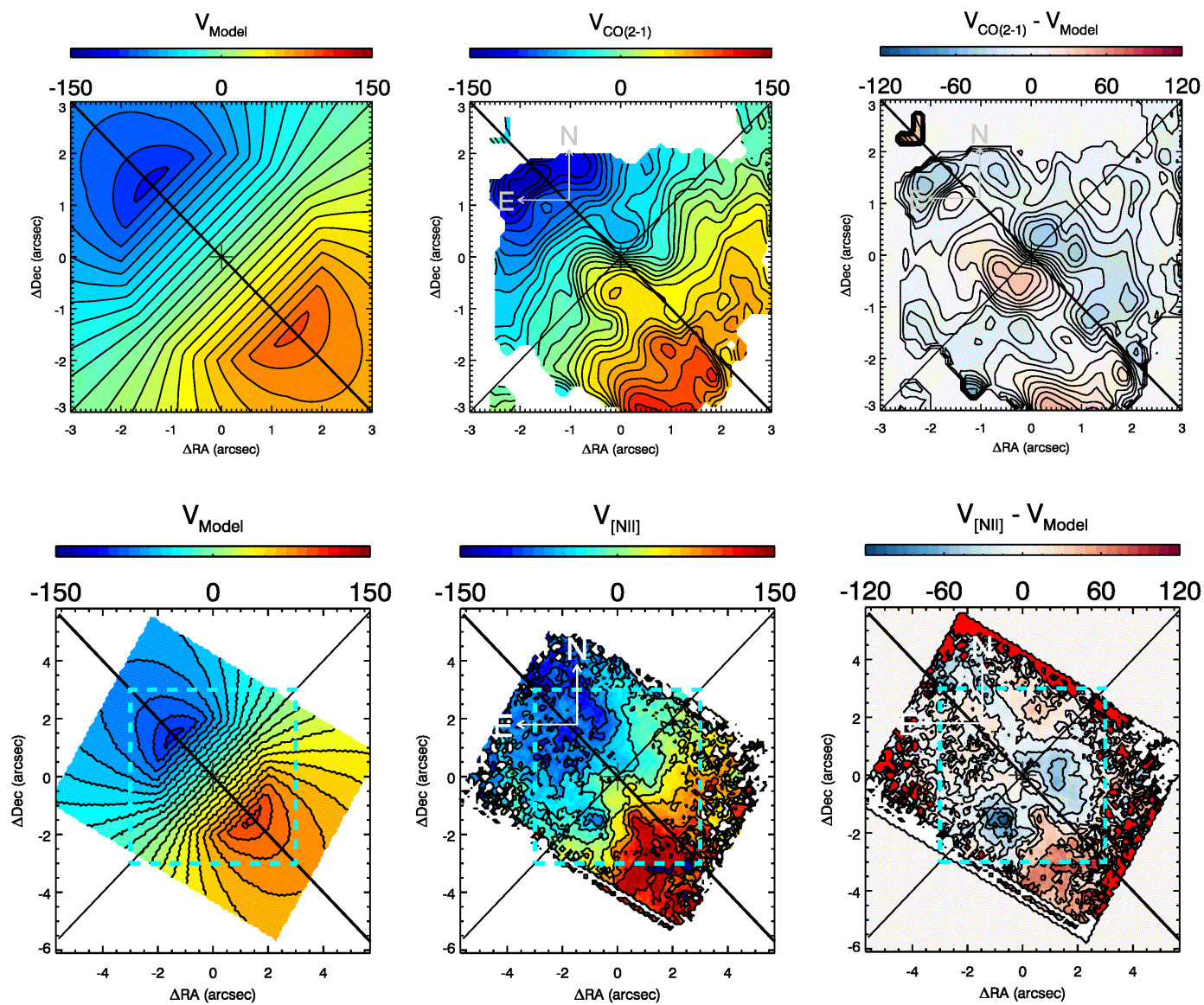

\centering        
  \includegraphics[bb=7 320 274 625,width=0.33\textwidth,clip]{idl-analysis_ngc1566.png} 
  \includegraphics[bb=293 320 560 625,width=0.33\textwidth,clip]{idl-analysis_ngc1566.png}
  \includegraphics[bb=7 14 274 319,width=0.33\textwidth,clip]{idl-analysis_ngc1566.png}
  \includegraphics[bb=28 326 255 640,width=0.33\textwidth,clip]{GMOSkinanalysis_GMOS_ngc1566.png}
  \includegraphics[bb=322 326 549 640,width=0.33\textwidth,clip]{GMOSkinanalysis_GMOS_ngc1566.png}
  \includegraphics[bb=28 14 255 320,width=0.33\textwidth,clip]{GMOSkinanalysis_GMOS_ngc1566.png}
  \caption{The nuclear velocity field of \galaxy\ in the \co2-1\ (top row) and \nii\ 6583\AA\ (bottom row)
  emission lines, in contours and color following the color bar above each panel.
  The left column shows the expected radial velocity field from the ModC2014 model (see text), 
  the middle column shows the observed radial velocity, and the right column shows the
  residual velocities (observed $-$ model).  
  A cyan dashed squares were drawing inside bottom panels for showing the FOV of top panels. 
  In these figures we use a systemic velocity of $\systvel$\kms and
  the nuclear position (marked with a cross) is determined by the 230~GHz continuum emission peak. 
  The major (PA=45\degr) and minor axes are shown in solid lines. 
  One arcsecond corresponds to \scaleimg\ pc.
}
  \label{zoominfigs}
\end{figure*}

The residual (after subtraction of the rotation model) velocity field of the \co2-1\ emission line
is shown in Figs.~\ref{velreswithfluxfigs} and \ref{zoominfigs}. 
The departures from pure rotation are now clearer, especially in the inner 3\arcsec. 
The largest deviations are (a) blue and red shifted clumps $\sim$1\arcsec\ to the NW and SE of
the nucleus; (b) red spiral arms $\sim$4\arcsec\ to the N and SE with the latter less redshifted, and
(c) a diffuse clump some redshifted $\sim$3\arcsec\ to the SW of the nucleus.
The blue (residual velocity $\sim$50\kms) clump $\sim$1\farcs5 to the SW of the nucleus 
along the major axis 
in our CO residual velocity map (Fig~\ref{zoominfigs}) marks the location 
of the star-forming region noted by \citet{smajic2015}: recall that this region is discernible
in our \nii\ moment maps, both as a  
a high flux region in the \nii\ moment 0 map, and a 
relatively low velocity dispersion region in the \nii\ moment 2 map. Further, this is also the region found to have a velocity gradient in its \oiii\ emission line \citet{dasilva2017}.  Disturbances produced by the star forming region could explain why the CO velocity here does not  agree with the expectation from pure 
rotational. 

\subsection{Modeling Observed Velocities: Molecular Outflows?}
\label{outflowsect}

The \co2-1\ residual velocity map in Fig.~\ref{zoominfigs} shows excess blueshifts
to the NW (near side) and redshifts to the SE (far side) of the galaxy, around 1\arcsec\ from
the nucleus along the minor axis. This is the expected
signature of outflows within the plane of the galaxy disk. This feature (but at lower
spatial resolution, and at significantly lower spectral resolution and image fidelity) was noted by
C14 and \citet{smajic2015}, but they argued that the small velocities seen in their 
\coo3-2\ 
maps made an outflow scenario unlikely. \citet{smajic2015} have shown that residuals in the nuclear kinematics
of the \h2\ are consistent with outflows along the minor axis, but that these residuals are also easily explained 
by deviations introduced by the density waves of the nuclear spiral;
non-circular orbits, e.g. a closed elliptical orbit with axes not parallel to one of the symmetry axes (minor or major) can produce residual velocities \citep{smajic2015}. 

We nevertheless argue for the presence of a nuclear outflow (which may of course 
co-exist with other bar-, warp, or spiral-related perturbations) based
on the following reasons:
(a) the presence of a nuclear outflow in \galaxy\ is not unexpected as previous studies have claimed kinematic
and morphological evidence for the presence of outflows in the NLR \citep{schm-kinn96,davies2016,dasilva2017},
which are most likely to intersect the disk given the observed geometries,  and in the larger scale disk \citep{pence90}.
Note also the evidence of a blueshifted velocity in \oiii 5007\AA\  near the star-forming region $\sim$1\farcs5 SW of the 
nucleus discussed above which \citet[][see their Fig.~20]{dasilva2017} interpreted as a consequence of contamination
from an AGN outflow; we also detect this blue residual in our CO residual (observed $-$ rotation model) velocity
map  (Fig.~\ref{zoominfigs}); 
(b) the unresolved nuclear aperture shows a double peaked profile with Full Width at Zero Intensity (FWZI)
$\sim$200\kms\ (Figs.~\ref{aperturesfig} and~\ref{nuclearspectfig}), 
higher than that seen in the lower fidelity maps of C14.
If these velocities are attributed to an outflow, the fact that opposite velocities are observed
on each side of the nucleus implies that the outflow axis is not aligned with our line of sight. Large
angles to the line of sight are unlikely as this would imply extremely high true outflow velocities. An outflow
in the plane of the disk would imply an outflow with velocities up to 180\kms.
Conversely, attributing these observed velocities along the minor axis to other perturbations in the plane of the disk, 
requires  radial velocity perturbations of $\sim$80--100\kms\ in a nuclear region where the 
intrinsic (undisturbed) rotation velocities are expected to be $\lesssim$40\kms. As we will show in Sect.~\ref{barpersect},
our modelling of the bar-related perturbations does not reliably produce both the morphology and the large perturbations seen in the observed velocity field; 
(c) the pv diagram along the minor axis (bottom right panel of Fig.~\ref{pv1fig}) not only shows
the high-velocity components ($\pm$60--90\kms) in the nuclear aperture but also lower brightness emission which
connects these high-velocity components to the zero velocity components seen at r$\sim$1.8\arcsec\ on both
sides of the nucleus. To the N (negative offsets in the bottom right panel of Fig.~\ref{pv1fig}) the decrease in 
velocity in the inner arcsec is clearly seen, and to the S (positive offsets) the decrease is more
clearly seen in the r$\sim$1-2\arcsec\ range.
(d) the pv diagrams show velocity deviations which are consistent with
radial outflows in the plane of the disk over several PAs (and not just the minor axis) and over
apertures at distances of several synthesized beams from the nucleus (see Sect.~\ref{pvsect}); 
(e) as discussed in the next section, the kinematics of the emission line gas in the optical 
(from GMOS-IFU) are consistent with 
a nuclear spherical (or bipolar) outflow, which makes the interpretation of a related molecular outflow less surprising.
In summary, we support the presence of a nuclear outflow, without ruling out the presence of additional bar- or spiral-related 
perturbations (Sect.~\ref{barpersect}). Other scenarios, e.g., a warped disk or non-coplanar disks 
\citep[e.g.,][]{wong2004,gar-bur2014} cannot be constrained by us due to the limited resolution, the sparse velocity field, and the lack of a reliable circular rotation model for the galaxy.

This posited molecular outflow is most likely primarily in the disk of the galaxy for the following reasons:
(a) outflows outside the disk are often related to nuclear jets \citep[e.g.:][]{morganti2013,sakamoto2014}, but
there is no clear evidence for radio-traced jets and outflows in the nucleus of \galaxy. The potential 
radio extension in PA 10\degr\ (Sect.~\ref{introsect}) is not aligned with our posited outflow axis. 
Further, apart from the
Seyfert 1 classification there are no data (e.g., maser disks) to constrain the orientation of the central engine; 
(b) outflows are also often seen perpendicular to the plane of the disk, especially in the case of starburst 
driven winds
    \citep[eg.,][]{veilandrub2002,veilleux2005,leroy2015}. However, for such a polar outflow the blueshifted (redshifted) emission would be seen towards the far (near) side of the galaxy disk, the opposite of that seen in \galaxy; 
(c) a large (e.g. >30\degr) opening angle for the outflow, often the case in radiation-pressure-, jet-, and starburst-driven
    outflows will produce a large observed velocity dispersion due to the varying projection angles of the outflowing gas to the line of sight. 
    In \galaxy\ we see a relatively low \co2-1\ velocity dispersion (FWHM of $\sim$30\kms) in the posited
    nuclear outflow components (see Figs.~\ref{velreswithfluxfigs} and \ref{nuclearspectfig}). A spherical outflow
    scenario can be rejected as this would produce a large velocity dispersion centered on zero velocity, under the assumption of optically-thin emission in \co2-1. Note that in Sect.~\ref{outflowgmossect}, we argue for the 
    presence of spherical outflow in the ionized gas, which does not contradict our claim that the molecular outflow is in the disk; 
(d) the posited nuclear outflow has a limited extension, and an apparently decreasing velocity, both of which argue for
    deceleration of the molecular gas in the high-density medium of the disk; 
(e) the higher molecular gas density in the disk of the galaxy will make this component more easily detectable in short integrations, as compared to more diffuse molecular gas outside the galaxy disk.
In summary, while the molecular outflow could have a larger opening angle (and indeed be isotropic) we appear to 
be preferentially detecting this component within the galaxy disk.

Is the posited outflow AGN or starburst-driven? As mentioned above, there is no clear evidence of a radio jet, and
so any AGN-driven outflow would most likely be due to radiation-pressure. We note that the highest velocities in the outflow are detected at the position of the nucleus and not towards the star-forming knot 1\farcs5 to the SW. 
Several authors have presented photometric and kinematic data which argue against the likelihood of a starburst-driven
outflow in \galaxy: \citet{davies2016} used diagnostic diagrams of \oiii/\hb\ to demonstrate that there is no significant contribution from
star-forming regions in the nucleus and it is the radiation pressure from the AGN which dominates in the inner scales.  \citet{smajic2015} present similar diagnostic diagrams using Molecular Hydrogen (\h2) and report an AGN domination for the nuclear region inside 1\arcsec\ and a relatively modest SFR ($\rm \sim8\times10^{-3}$ [M$_{\odot}$yr$^{-1}$]) in the inner 3\arcsec,
implying a relatively low star-forming efficiency given the observed  gas reservoir. Both the galaxy-wide SFR 
\citep[$\sim$4.32 $\rm M_{\odot}yr^{-1}$;][]{gruppioni2016} and the SFR surface density
\citep[$\sim$0.033 $\rm M_{\odot}yr^{-1}kpc^{-2}$;][]{hollyhead2016} are relatively low, and thus star formation is not
expected to drive a nuclear outflow  \citep[e.g., ][]{cicone2014}.

As seen in the pv diagrams of Fig.~\ref{pv1fig}, a model which sums
the  
ModC2014 model and our empirically derived  outflow model (black solid lines in the 
pv diagrams; see Sect.~\ref{pvsect}) provides a much better fit (as compared to a pure rotational model) to 
the inner 2\arcsec\ in the pv diagrams at all PAs.
Note that our synthesized beam of $\lesssim$0.5 arcsec well resolves the central
4 arcsec of the galaxy (e.g. Fig~\ref{zoominfigs}). 
Nevertheless there are several specific features which cannot be fit only with the model of
radial outflows (in the galaxy disk) plus rotation, e.g. the apparent morphological double
structure of each 
outflow lobe (Fig.~\ref{zoominfigs}) and the pv diagram in PA=75\degr\ (middle right panel of Fig.~\ref{pv1fig}) where 
observed velocities 1.5--2 arcsec from the nucleus to the E are not well fit by the model. 
These are discussed in Sect.~\ref{pvsect}. 

In the residual velocity map of CO (top right panel of Fig.~\ref{zoominfigs}) the blue- and red-shifted lobes 
1\arcsec\ from the nucleus 
have a double-peaked morphology. The largest velocity deviations are along PAs of  100\degr\ and 140\degr, i.e.
straddling the minor axis.  
How can this be explained? Do streaming motions into the
nucleus along PA $\sim$120\degr\ create the valley between the two peaks? Further, why is it that almost the entire
near side of the galaxy (NE) has a blue residual, while almost the entire far side of the galaxy (SW)
has a red residual in the top right panel of Fig.~\ref{zoominfigs}? This is not due to the use of an
incorrect major axis PA (varying the major axis PA does not change these features). Effectively,
the NE and SW sides of the galaxy are not axisymmetric in their rotation (see, e.g., Fig.~\ref{idlanrob-2fig}). 

\subsection{Modeling Observed Velocities: Ionized Gas Outflows?}
\label{outflowgmossect}
The \nii\ line velocity map was (obtained from the Gauss-Hermite fit version of 
\emph{profit.pro} after subtracting the broad \ha\ line emission).
The \nii\ velocity residual map constructed using the same rotation model
used for the \co2-1\ line at first glance appears morphologically similar to 
its \co2-1\ counterpart (bottom panels in Fig.~\ref{zoominfigs}). However the P.A. of the major axis for
the \nii\ rotation (especially to the SW) appears to be $\sim$15$\degr$ smaller than 
that used in our model.  Further,
the \nii\ residual velocity map clearly shows blue residual velocities at $\sim$1--2\arcsec\ on 
both sides of 
the nucleus (with the largest velocity deviation to the SE), in contrast to the \co2-1\ 
residual velocity map which clearly
shows opposite colors on each side of the nucleus (interpreted above as the sign
of a nuclear molecular outflow in the disk). 

In the \nii\ velocity residual map, the blue region on the near side of the galaxy 
(NW of the nucleus) is in rough agreement with the
equivalent blue region in the \co2-1\ residual map, and thus would be consistent with the 
outflow scenario posited for the \co2-1\ data. 
On the other hand, the blue region in the \nii\ velocity residual map on the far side of the galaxy
(SE of the nucleus) is located 
along the minor axis at $\sim$1\farcs8 from the nucleus: this does not match the feature seen
in the \co2-1\ residual map which is closer to the nucleus ($\sim$1\arcsec) and
redshifted. Note that this blue SE feature in the \nii\ residual velocity map
is roughly cospatial with a region of high dust extinction
(structure map in Fig.~\ref{momentsgmosfigs}). A red region closer 
to the nucleus matches with the red region to the SE seen in the CO residual map, however, this consists in redshifted radial velocities around 10\kms\ which is too low to be considered significant 
as a part of an outflowing gas. There is another small region to the SW at $\sim$2\arcsec\ which is redshifted and around 50\kms\ but in spite of having a similar dynamic range, it does not match with the redshifted outflow region to the SE in the ALMA residual map ($\sim$1\farcs5), being placed too close to the star-forming region.

\begin{figure}
\centering            
 \includegraphics[bb=70 10 520 630,width=0.5\textwidth,clip]{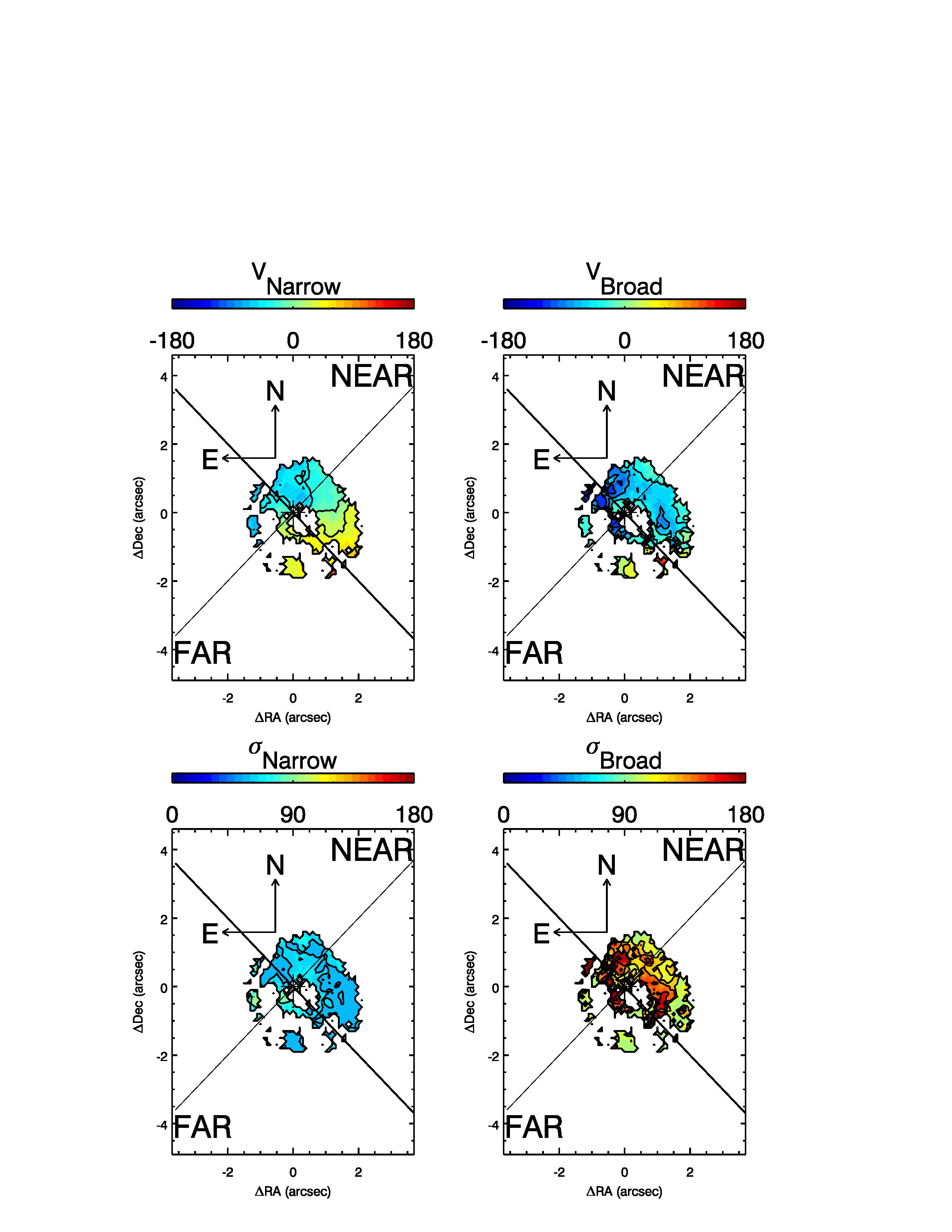}
  \caption{The velocity and dispersion maps of the \nii\ emission line obtained from the two-Gaussian fit version of \emph{profit.pro}. 
Top panels: from left to right, the velocity and velocity dispersion maps of the narrow and broad components, respectively. Bottom panels: the velocity dispersion maps of  the narrow (left panel) and broad (right panel) components. 
  }
  \label{gmosbluestcomponentfig}
\end{figure}

The \nii\ emission in the inner $\sim$2\arcsec\ is double-peaked, and thus cannot be well fit with a single Gauss-Hermite polynomial. 
We, therefore, used the two Gaussian
fit version of \emph{profit.pro} to search for potential independent velocity components in the \nii\ emission line. Meaningful two-component fits were obtained in part of the inner 3\arcsec\ radii nuclear region and the resulting velocity maps are shown in Fig.~\ref{gmosbluestcomponentfig}.  
These maps include only pixels for which
a double component fit produced a meaningful result. For regions where the two-component fit was not possible, the single component fit remains valid. Note that in the latter case the single component fit shows primarily the equivalent of the narrow component but in a few regions the velocity dispersion of the single component fit is similar to that of the broad component of the two-component fit.  The two-component fit is mainly obtained near the nucleus and on the near side of the galaxy. On the far side (and a small nuclear region to the W on the near side), the regions in which a two component fit is not possible coincide well with the regions of high dust extinction. 
The `narrow component' shows velocity dispersions ranging between 60 and 90\kms\ (bottom left  panel of the figure) and the velocity field of this component mainly follows that expected from our rotation model. 
The second component, which we refer to as the `broad component',  
shows velocity dispersions of $\sim$140--160\kms\ (bottom right panel of the figure) and does not appear to participate in
regular rotation. 

The strong blue residual region to the SE in the single component \nii\ residual velocity map is not
fit with a double component.  
The weaker blue residual region to the NW in the single component \nii\ residual velocity map
is now seen to be blue in its broad component only; in its narrow component this region follows regular rotation. 
Note that both blue knots do not correspond to regions of high dust columns in the
structure map. 

Based on the velocity field of the broad component of the two component fit to \nii, 
the velocity field of the high dispersion ($\gtrsim$120~\kms) areas in the single 
component fit to \nii\, and the distribution of the nuclear dust,
we postulate the presence of an expanding sphere of ionized gas, i.e. a spherical ionized 
outflow, for the reasons given below.
With the presence of dust (dominantly in the plane of the galaxy disk) we would 
preferably see emission from the hemisphere 
in front of the galaxy and moving towards us, i.e. blueshifted radial velocities. 
In the absence of dust one would expect a large dispersion and a median velocity close to
systemic. 
Note that given the relatively low inclination of \galaxy\ (33\degr), dust in the inner 
2\arcsec\ ($\sim$100~pc) of the disk produce an almost equal extinction of light from the bulge 
for both the `near' and `far' side of the galaxy disk. This is clear in the structure map
where dark dust lanes are seen on both the near and far sides in the inner $\sim$2\arcsec. 
Only at larger radii are the structures of the 
dust lanes more prominent on the near side of the galaxy disk.
In the case of \galaxy\ we do not obtain a two-component fit in areas where the structure
map implies marked dust lanes (non-intuitively these are on the far side of the galaxy disk in the
innermost arcsec) and 
find large blue shifts in the broad component of the two-component fit on the near side 
of the galaxy disk
in areas where the structure map shows less marked dust lanes. 
Further, the maps derived from the single component fit to the \nii\ line (Fig.~\ref{momentsgmosfigs}) 
show two regions of high dispersion, about 2\arcsec\ from the nucleus to the SE (the blue knot
referred to above) and to the NE. These two regions effectively correspond to the broad component,
and are also blue in their velocity. Thus we effectively see blue velocities in the broad component in almost every compass direction implying a  spherical outflow in the inner $\sim$2\arcsec\ which is visible to us primarily in areas of lower dust extinction. 
Similar kinematic signatures, and thus interpretations, were observed and used 
in previous IFU studies, e.g. in the nearby Seyfert galaxy NGC~2110 \citep{schnorr2014b}, 
and in some radio-quiet quasars \citep{liu2013}. Nevertheless, despite all these signs, 
we are not neglecting a potential presence of a bipolar outflow. Namely, it might be reasonable to think that what we see as a spherical outflow might be instead ionized gas ejected from the nucleus in opposite directions but we notice just a part of them as a consequence of a poor resolution in the GMOS data ($\sim$0.5--0.6\arcsec). 

We note that our postulation of this spherical outflow of ionized gas does not contradict our postulation of cold molecular gas outflows in the galaxy disk. The cold molecular gas outflows are expected to be preferentially detected in the plane of the disk for two reasons: a low density CO outflow would be optically thin and its profile would thus be centered on zero velocity and the significantly larger abundance of molecular gas in the disk as opposed to above the disk, makes the disk molecular gas much easier to detect.
    
\subsection{Observed Position-Velocity (pv) Diagrams: ALMA}
\label{pvsect}
\begin{figure*}
\centering            
  \includegraphics[bb=40 120 390 494,width=0.33\textwidth,clip]{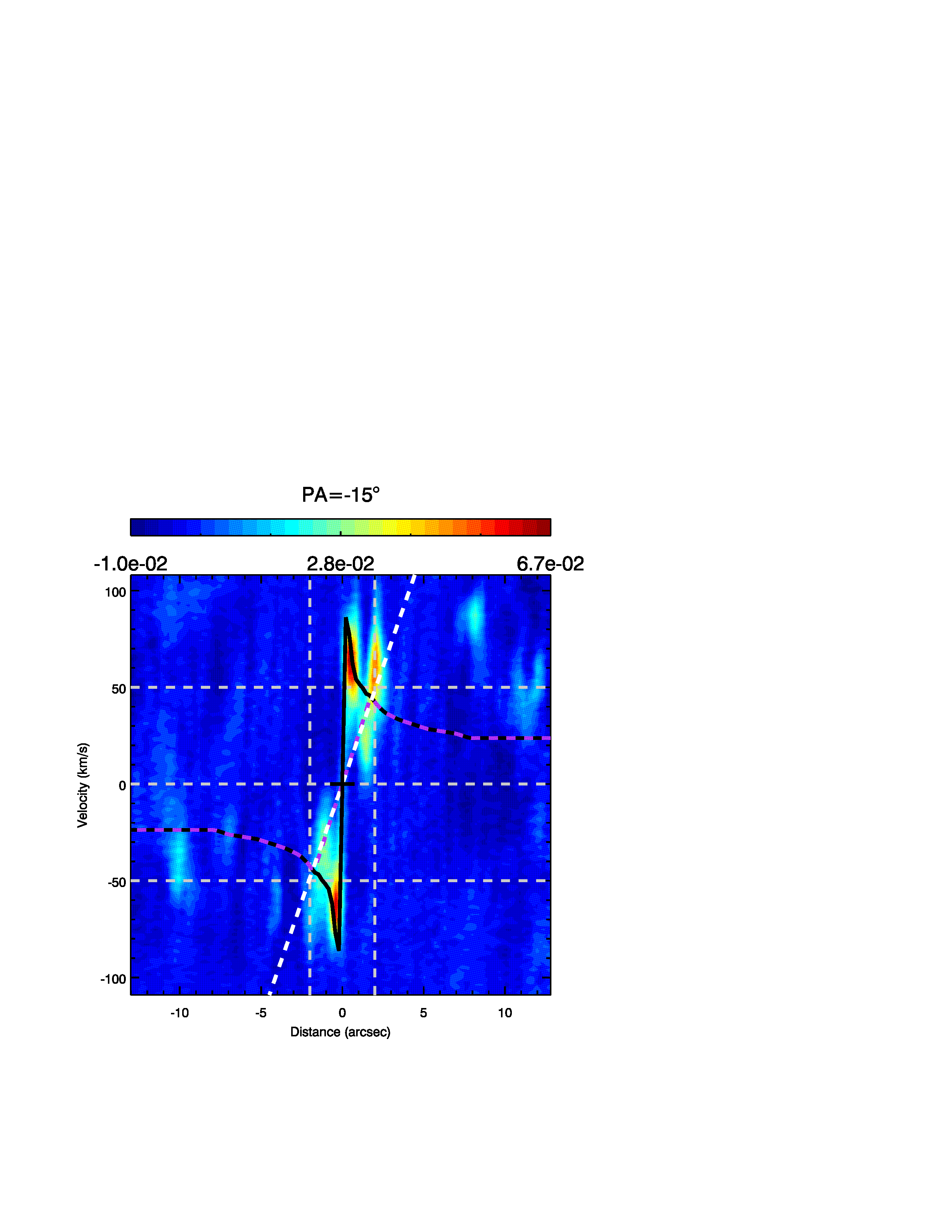} 
  \includegraphics[bb=40 120 390 494,width=0.33\textwidth,clip]{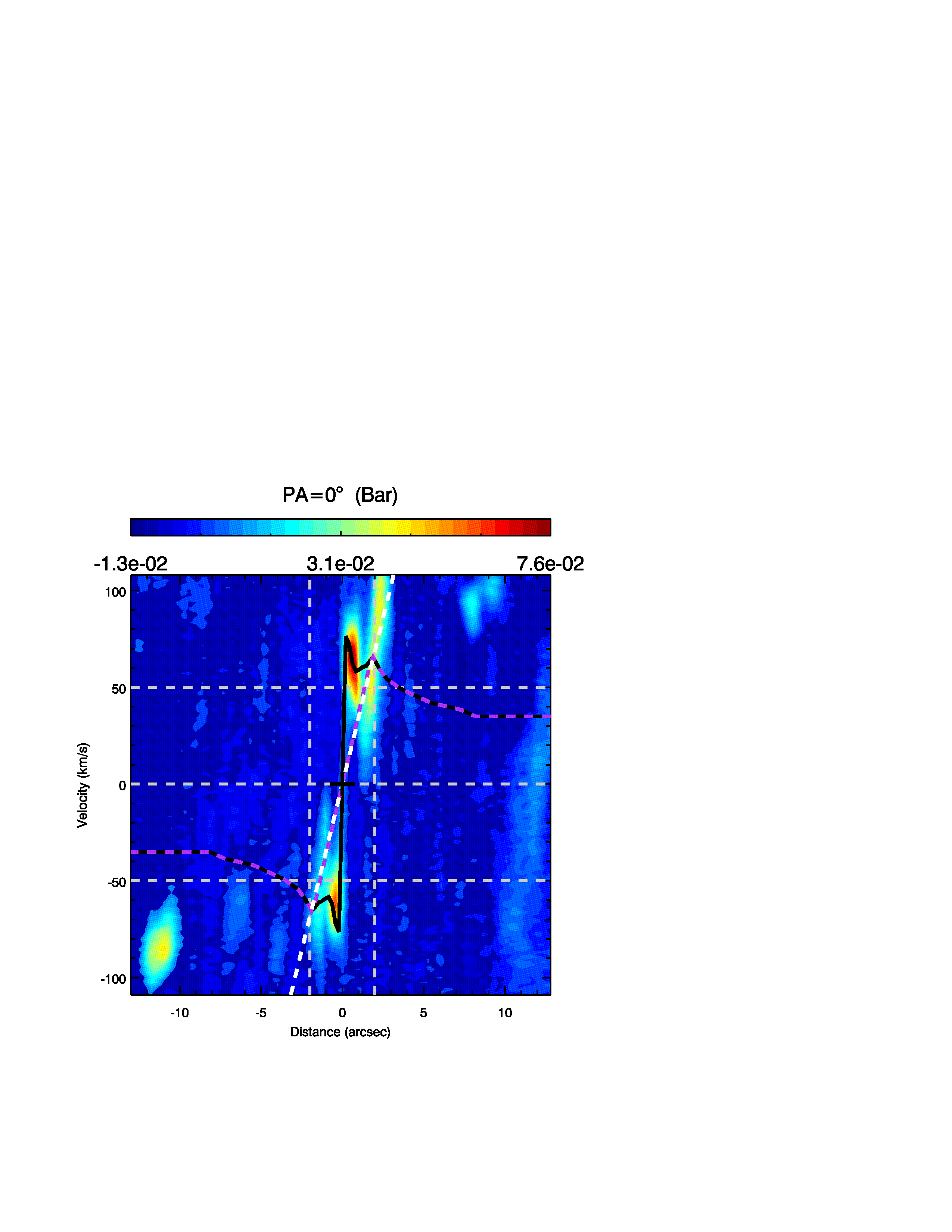} 
  \includegraphics[bb=40 120 390 494,width=0.33\textwidth,clip]{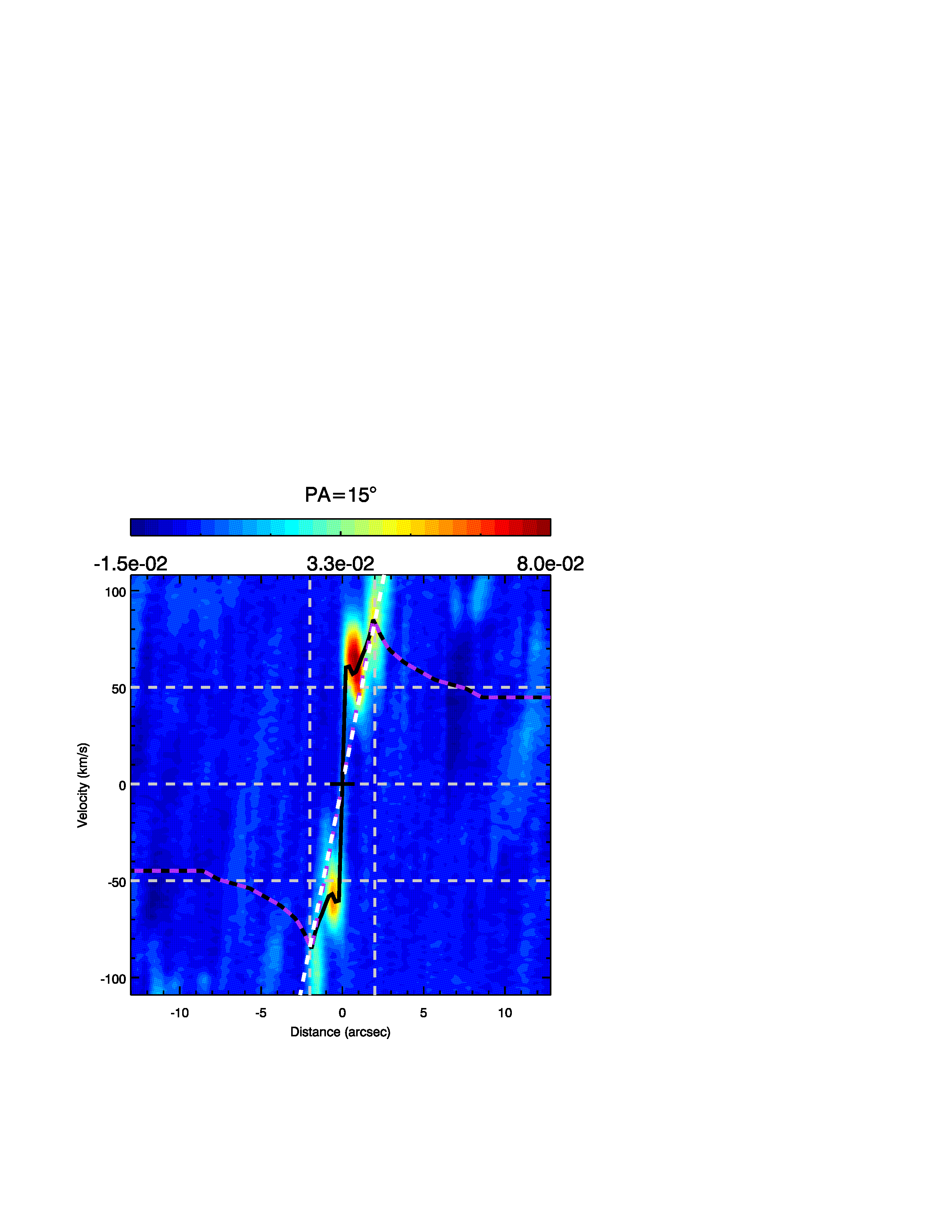}
  \includegraphics[bb=40 120 390 494,width=0.33\textwidth,clip]{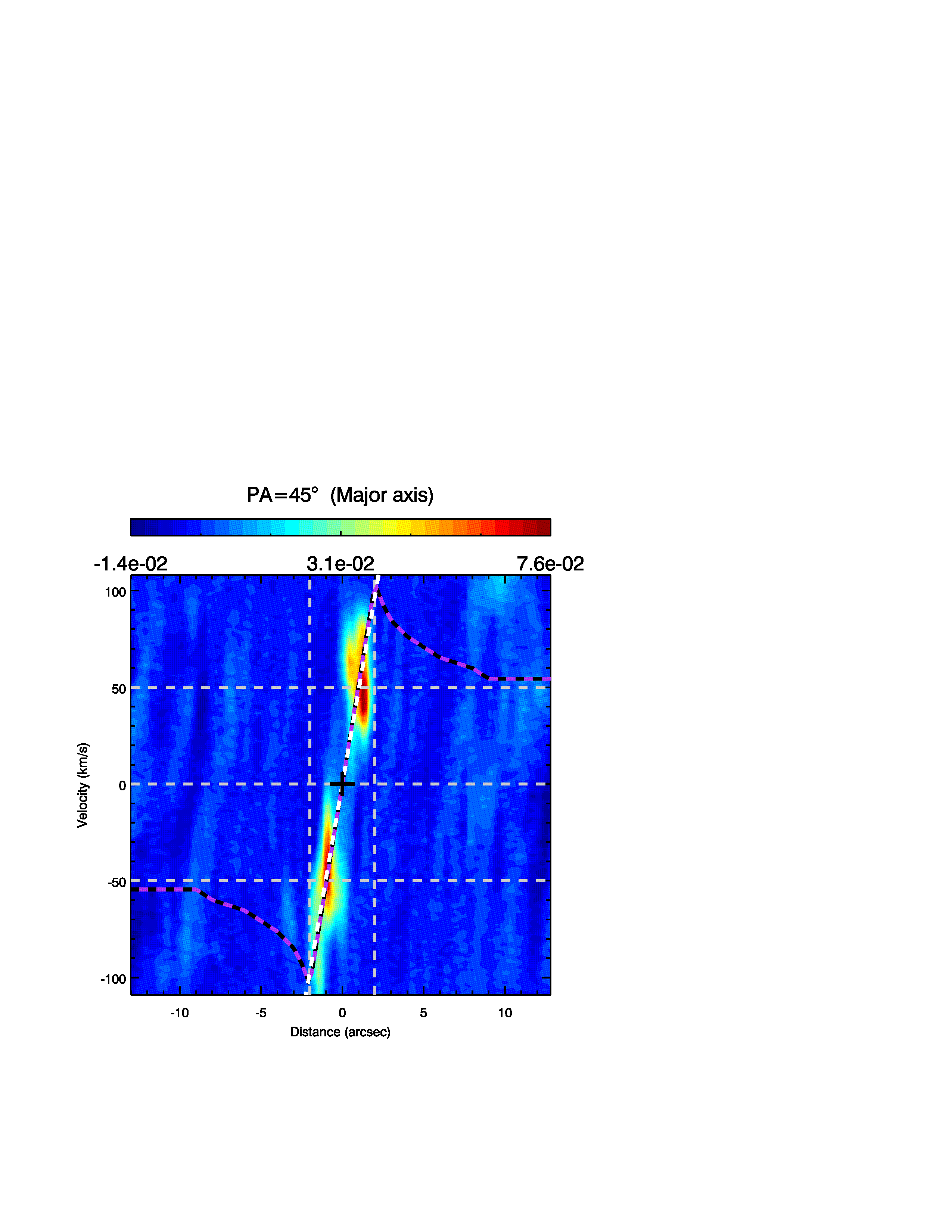} 
  \includegraphics[bb=40 120 390 494,width=0.33\textwidth,clip]{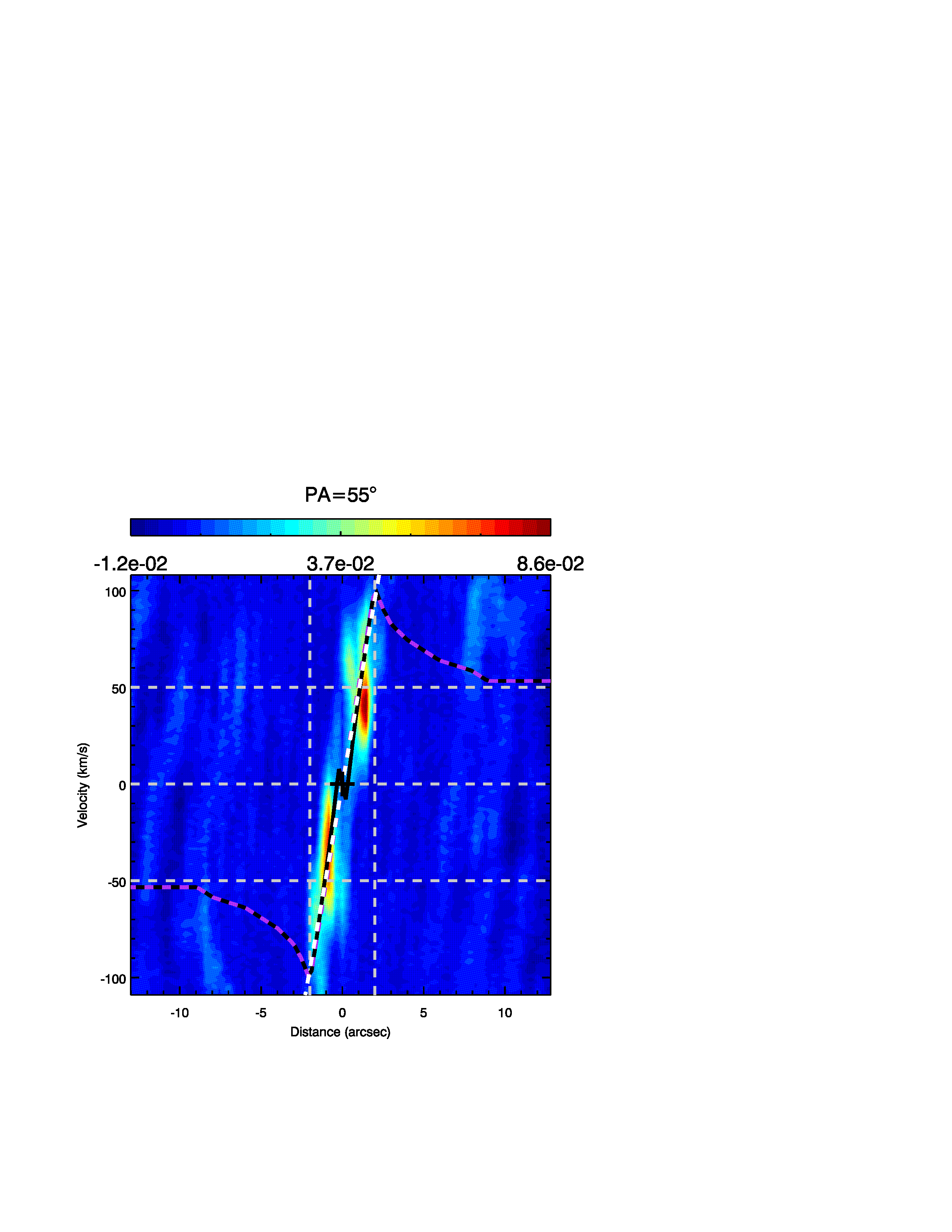}
  \includegraphics[bb=40 120 390 494,width=0.33\textwidth,clip]{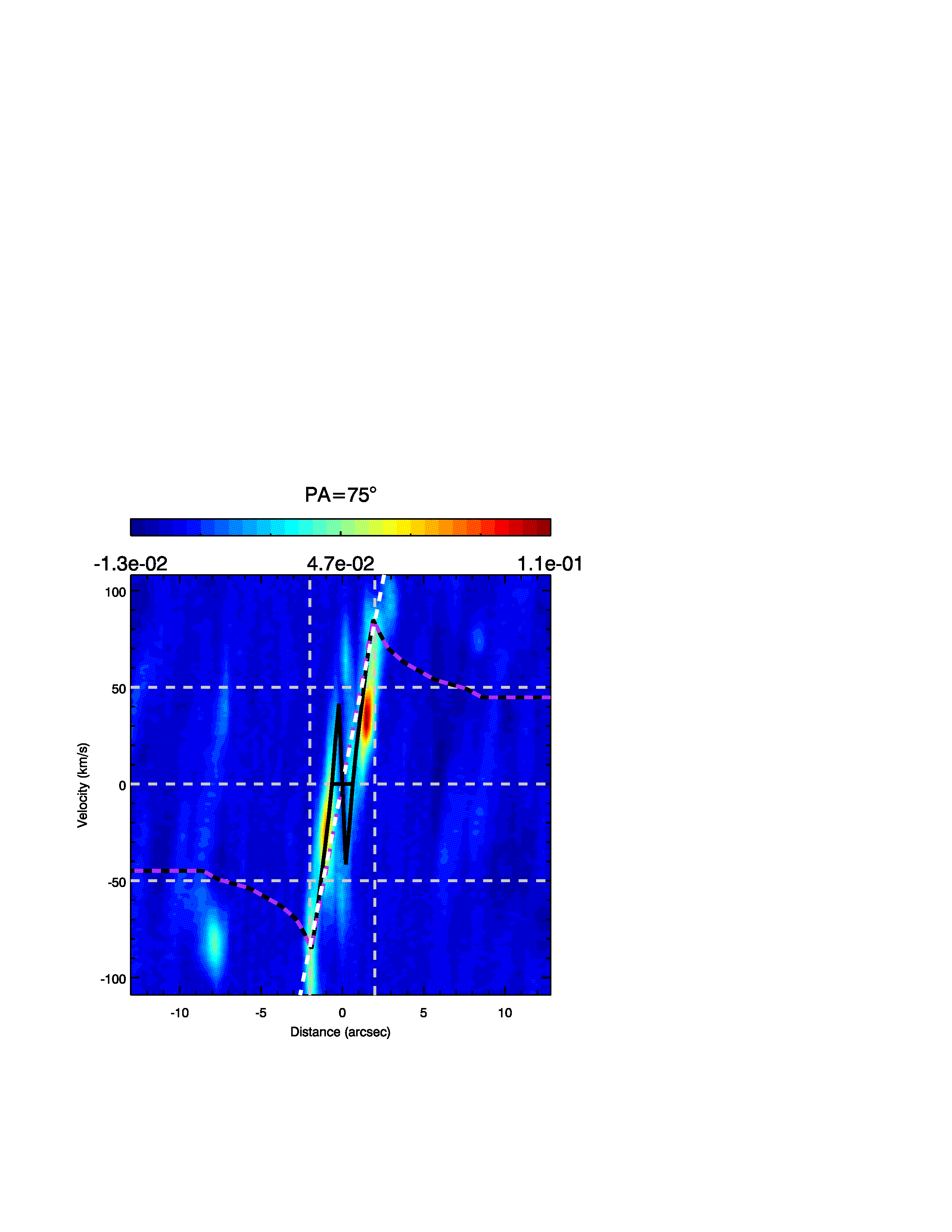}
  \includegraphics[bb=40 120 390 494,width=0.33\textwidth,clip]{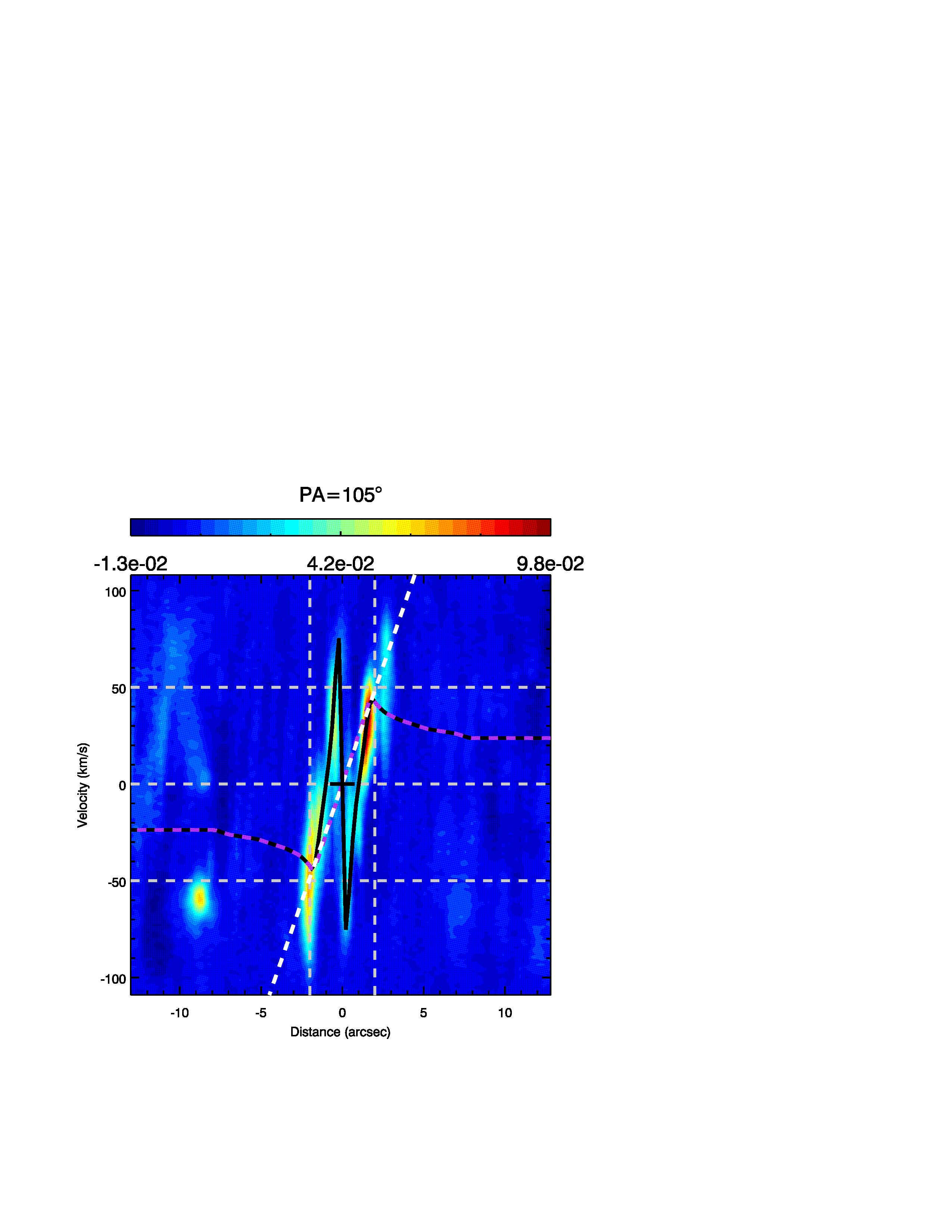}
  \includegraphics[bb=40 120 390 494,width=0.33\textwidth,clip]{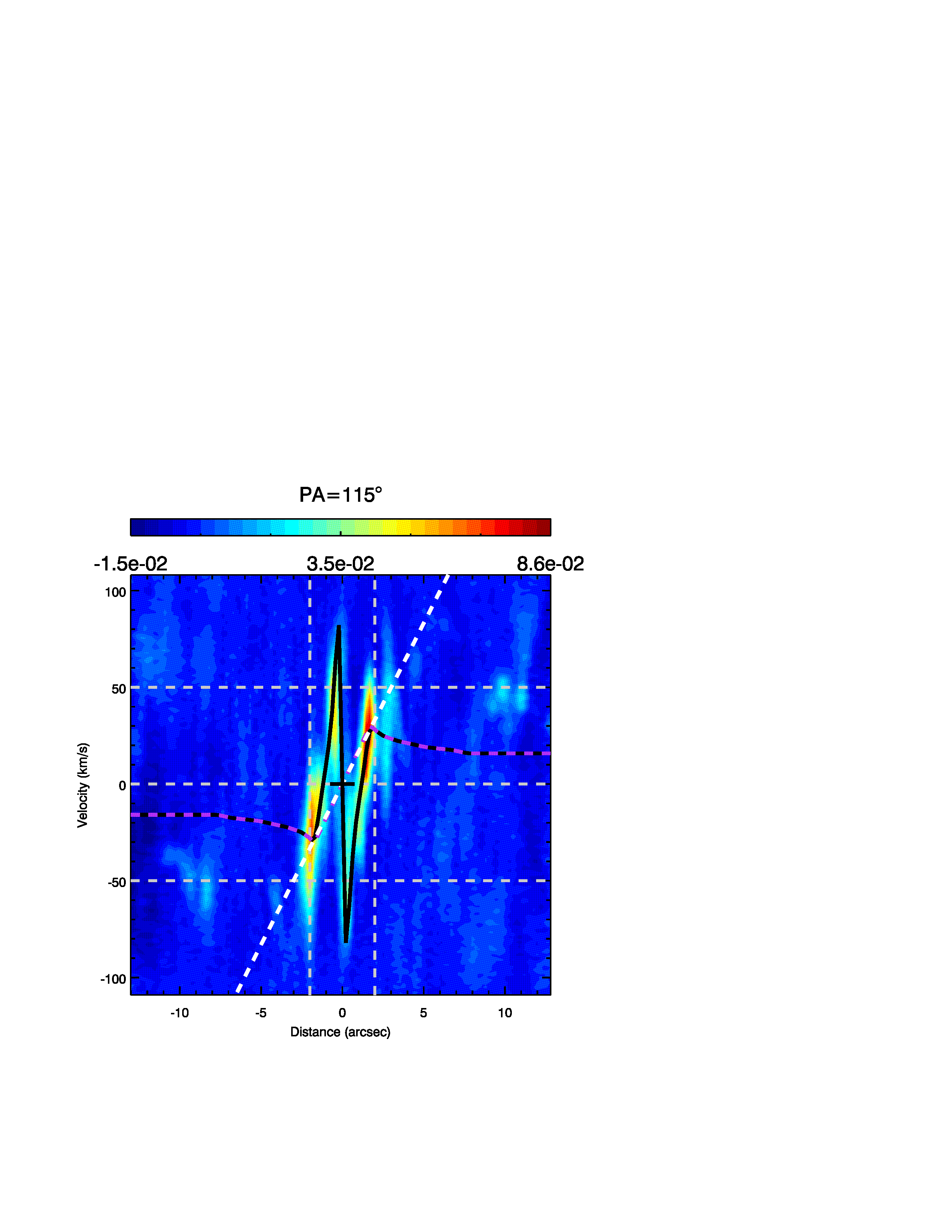}
  \includegraphics[bb=40 120 390 494,width=0.33\textwidth,clip]{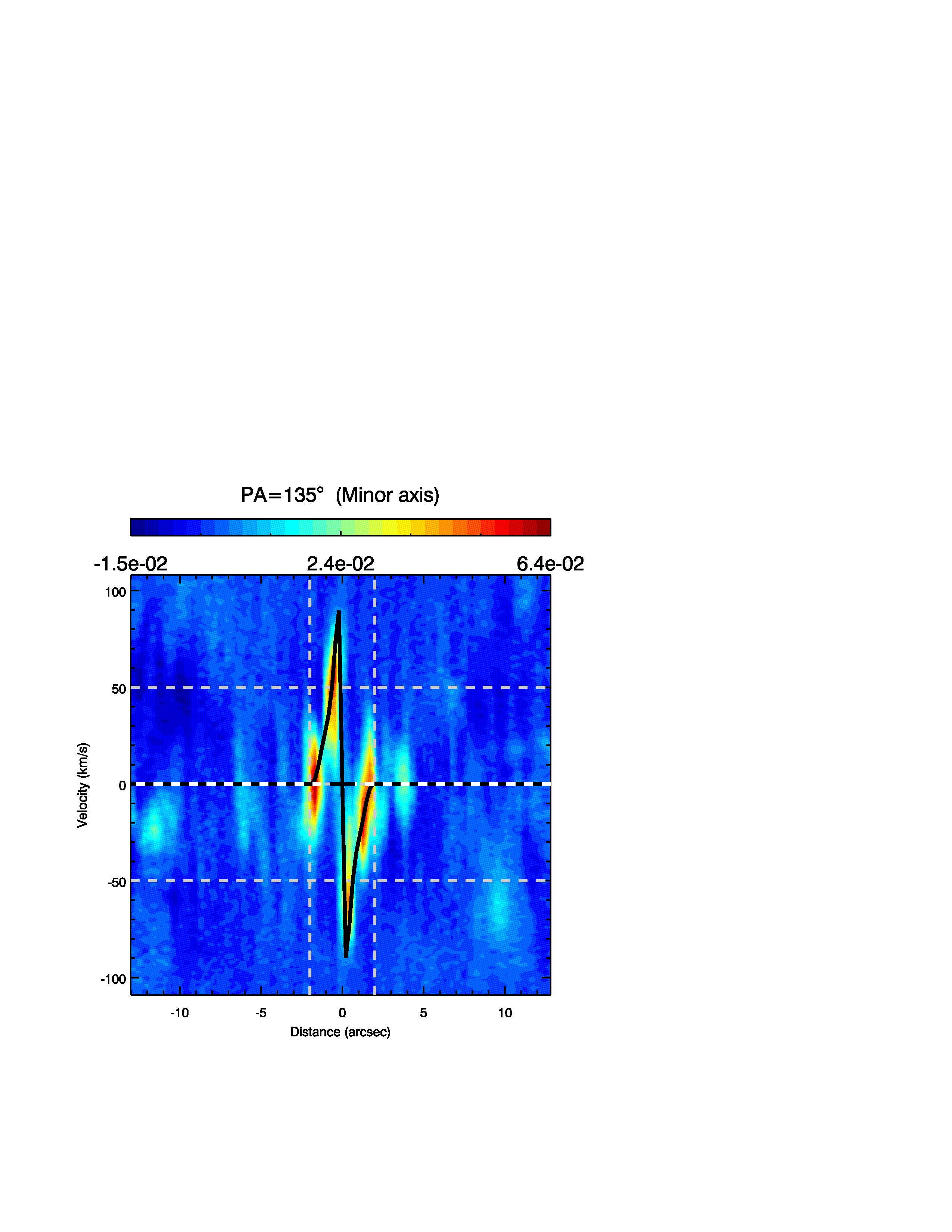} 
\caption{Position-Velocity diagrams of the \co2-1\ emission in \galaxy\ along several PAs  
are shown in color following the color bar above each panel.
The PA of the `slit' over which the pv diagram was extracted is indicated above each panel, as are the PAs 
corresponding to the major and minor axis of the galaxy, and of the large scale bar.
Negative offsets on the $x$-axis correspond to the PA listed above the panel, i.e. positive offsets
are along the 180\degr\ plus the listed PA. 
The black cross indicates the position of the 230~GHz continuum peak
(presumed to be the galaxy center) and the systemic velocity of the \co2-1\ line
(\systvel\kms).
To guide the eye horizontal and vertical lines delineate
$\pm$2 arcsec from the nucleus and $\pm$50\kms\ from the systemic velocity.
The dashed white and purple lines are the solid body rotation model and the  
ModC2014 rotation model, respectively. 
The solid black line shows the expectation of adding our outflow model to the ModC2014 model (see Sect~\ref{outflowsect}). 
The pv diagrams, were created from a `hybrid' datacube: 
the inner $\sim$12\arcsec\ $\times$ 12\arcsec\ square region 
centered on the nucleus is taken from a high resolution (Briggs weighting, Robust=$-$2) 
map with spatial resolution 0\farcs52 $\times$ 0\farcs35 and an r.m.s. 
noise of 0.1 mJy/beam per channel (up to 2 mJy/beam per channel in channels with strong signal), while the rest of the cube is from a higher signal to noise
(Briggs weighting, Robust=2) map with 
spatial resolution 0\farcs6 $\times$ 0\farcs5 and an r.m.s. noise of 1.2 mJy/beam per channel (up to 4 mJy/beam per channel in channels with strong signal). 
  }
  \label{pv1fig}
\end{figure*}

In this section we present position-velocity (pv) diagrams for the \co2-1\ line. These pv
diagrams were extracted from the ALMA datacubes using a slitwidth of  0.2\arcsec; they 
are thus limited in spatial resolution by the intrinsic resolution of the images ($\sim$0\farcs4).
Fig.~\ref{pv1fig} shows the pv diagrams along the major axis, minor axis, the large scale bar,
and other relevant PAs including PA=115\degr\ (20\degr\ 
from the minor axis) for a direct comparison with the pv diagram of CO and \ha\ in Fig.~3 of \citet{aguero2004}. 
In their figure, the `Blueshifted Knot' which they interpret as inflow motion is clearly seen in our \co2-1\ pv diagram. 

In each pv diagram we have overlaid the predictions of solid body rotation 
(white dashed line), the ModC2014 model (purple dashed line),
and the sum of the ModC2014 model with our outflow model (solid black line).  
The solid body rotation and ModC2014 are 
essentially the same over the inner $\pm$3 arcsec, after which the latter flattens in velocity.
Our outflow model was derived as follows: we used the 
pv diagram along the minor axis (bottom right panel of Fig.~\ref{pv1fig}) to measure the
radial velocity of the brightest CO emission at a given distance from the nucleus 
on both the NW (positive offsets in the pv diagram) and SE sides. 
The absolute values of these velocities as a function of distance from the nucleus were then interpolated and deprojected (assuming that
the outflow is in the disk) to construct a function of outflow velocity vs. position.
As seen in the bottom right panel of Fig.~\ref{pv1fig} the redshifted `outflow' velocities provided 
more constraints closest to the nucleus and the blueshifted
velocities provided better constraints at slightly larger distances. The final outflow model
starts with outflow velocities in the disk of \maxoutflow\kms\ at the (unresolved) nucleus and decreases monotonically 
to zero velocity 2\arcsec\ from the nucleus. 

First concentrating on the pv diagram along the minor axis in Fig.~\ref{pv1fig}, we see
that the NW side of the minor axis shows a clear
deceleration in velocities when going from 1\arcsec\ to 2\arcsec\ from the nucleus, while the SE
side the bright emission at 1.5--2\arcsec\ (the inner spiral arm) shows a larger velocity dispersion 
which does not clearly vary with distance. However, this SE side shows a clearer decrease in velocities
between 0 and 1\arcsec. A similar scenario has been reported in NGC 1068 for both hot and cold molecular gas \citep[][respectively]{barbosa2014,gar-bur2014}; outflowing nuclear molecular gas, with outflow velocity
decelerating from 200 \kms\ to 0, accumulates in an off-centered ring 100 pc from the nucleus 
\citep{barbosa2014}. 

For the pv diagram along PA=0, which is aligned with the large scale bar in \galaxy,
we see (1) gas consistent with outflows, which is well fit by our outflow model;
(2) gas which is almost in rotation 1--2\arcsec from the nucleus on either side, but showing a 
steeper rotation curve which reaches zero velocity at a position offset from the center. 

For the pv diagram along the major axis (PA 45\degr), several velocity components can be seen.
These include gas in rotation, and some contamination from the disk outflow component, since 
at the nucleus, the `slit' (limited by the spatial resolution of our ALMA observations)
expectedly picks up the gas outflowing along the minor axis and other angles.
Moreover, gas in the inner spiral to the NE is preferentially redshifted 
and gas in the inner spiral to the SW is preferentially blueshifted, i.e. 
both spirals show $\sim$40\kms\ (in projection) deviations towards values of zero velocity: the most obvious 
interpretation of this is that gas originally in circular rotation is slowed down on hitting 
the ends of the nuclear molecular gas `bar'. This loss of momentum could potentially result in inflows. Similar
velocity offsets are also seen at slit PAs offset 10\degr\ from the major axis 
(e.g. PA=55\degr; Fig.~\ref{pv1fig}), and is very dramatic 
on the SW side in the pv diagram with slit PA=75\degr, at the point where an outer spiral pattern breaks 
off from the inner spiral pattern.  

Molecular gas in the inner spirals always show a large 
velocity dispersion (around $\sim$80\kms).
The inner spiral to the NW (about 1 arcsec from the nucleus) always show 
velocities which are bluer than that expected from rotation or rotation+outflow. This is 
clearly seen in all pv diagrams which intersect this arm (e.g. PAs 45\degr, 55\degr, 
75\degr). The opposite inner spiral (that to the SE) shows the opposite, i.e. velocities 
redder than expected from rotation and rotation+outflow (e.g. most obvious in the PA 
105\degr\ and 115\degr\ pv diagrams) but this velocity offset is not as well defined as 
in the case of the NW arm. Given that the NW arm is mainly on the near side of the 
galaxy disk and the E arm is on the far side, this is what would be expected from a 
streaming outflows along the spiral rather than streaming inflows! We speculate that
these structures are absorbing the momentum of the nuclear outflow and thus heating up and expanding. 

Another region which consistently shows large differences from the rotation+outflow model
is the double cavity (between the nuclear bar-like structure and the inner spiral arms)
to the NW and SE, for offsets of 1 to 2 arcsec from the nucleus. When the slit passes through 
these cavities the pv diagrams (especially those at PA=0, and PA=$-$15\degr) show a characteristic
pattern which can be explained by slower than rotation velocity between 1 and 2 arcsec (but
increasing in the correct sense), and a large dispersion of velocities (all larger than those 
expected from rotation) at 2 arcsec. 

While the predictions of our rotation plus outflow model are in general consistent with 
the position velocity data at different PAs especially in the inner 2\arcsec, at PAs close 
to 0\degr\ (see the pv diagrams at PA=0\degr,$-$15\degr\ and 15\degr) 
one can see significant differences between the data and models in the inner 1\farcs5. 
Here we clearly see a  
component of gas which follows a steep velocity gradient decreasing to zero velocity
at a distance of 1\farcs5 from the nucleus on both sides. 
One potential explanation for this anomalous rotation is an inner 
counter rotating gas disk with major axis in PA$\sim$0, 
fed by gas inflowing along the large scale bar. 
This possibility is motivated by observational evidence that
bars are an efficient pathway for transporting gas from galactic scales to nuclear scales in both active 
and inactive barred galaxies \citep{sakamoto99,crenshaw2003,regandteub2004,sheth2005}.
Alternatively, these features are a consequence of perturbations due to the bar, as
discussed in the next section.   

\subsection{Modelling Observed Velocities: Bar perturbations}
\label{barpersect}

\begin{figure*}
 \centering
 \includegraphics[bb=50 120 395 450,width=0.4\textwidth,clip]{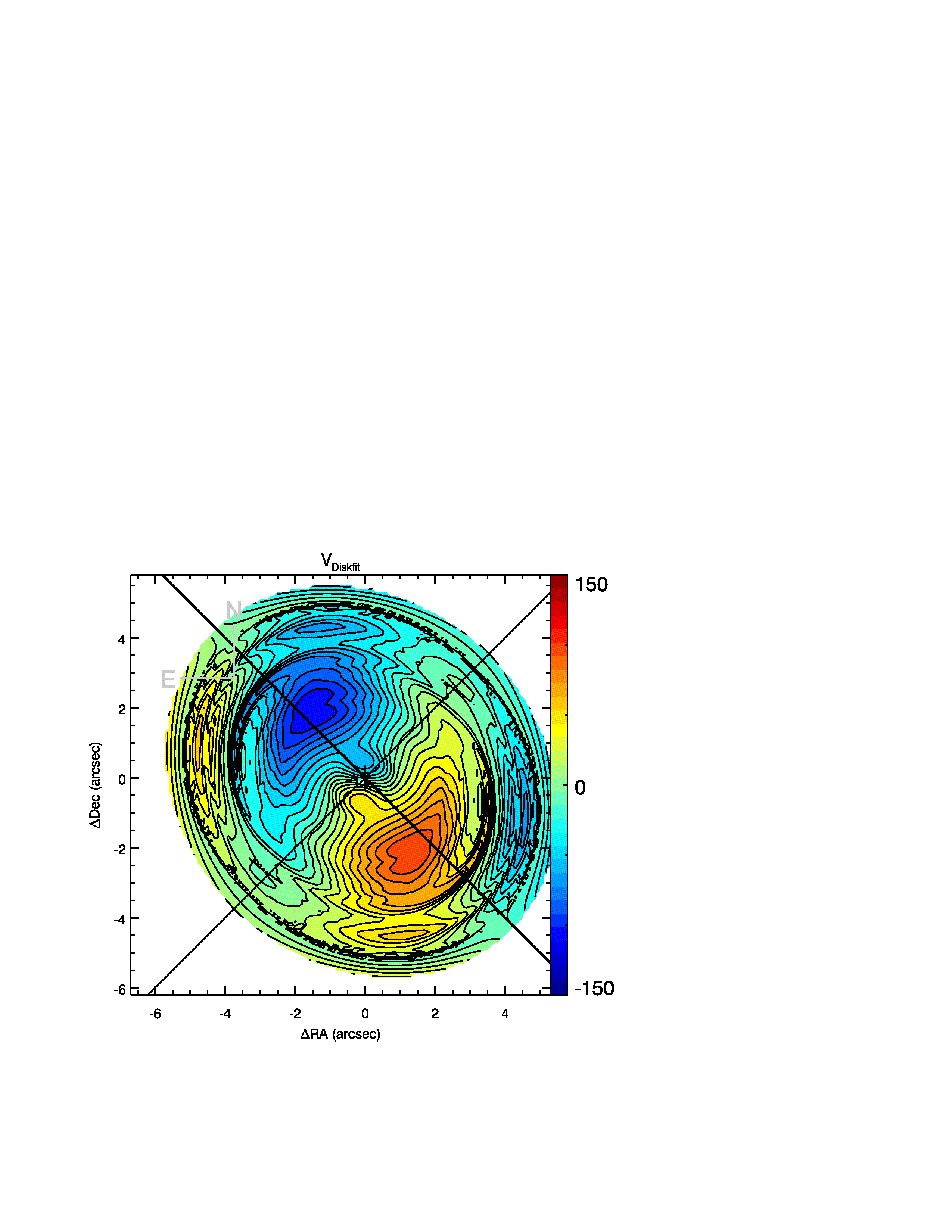}
 \includegraphics[bb=50 120 395 450,width=0.4\textwidth,clip]{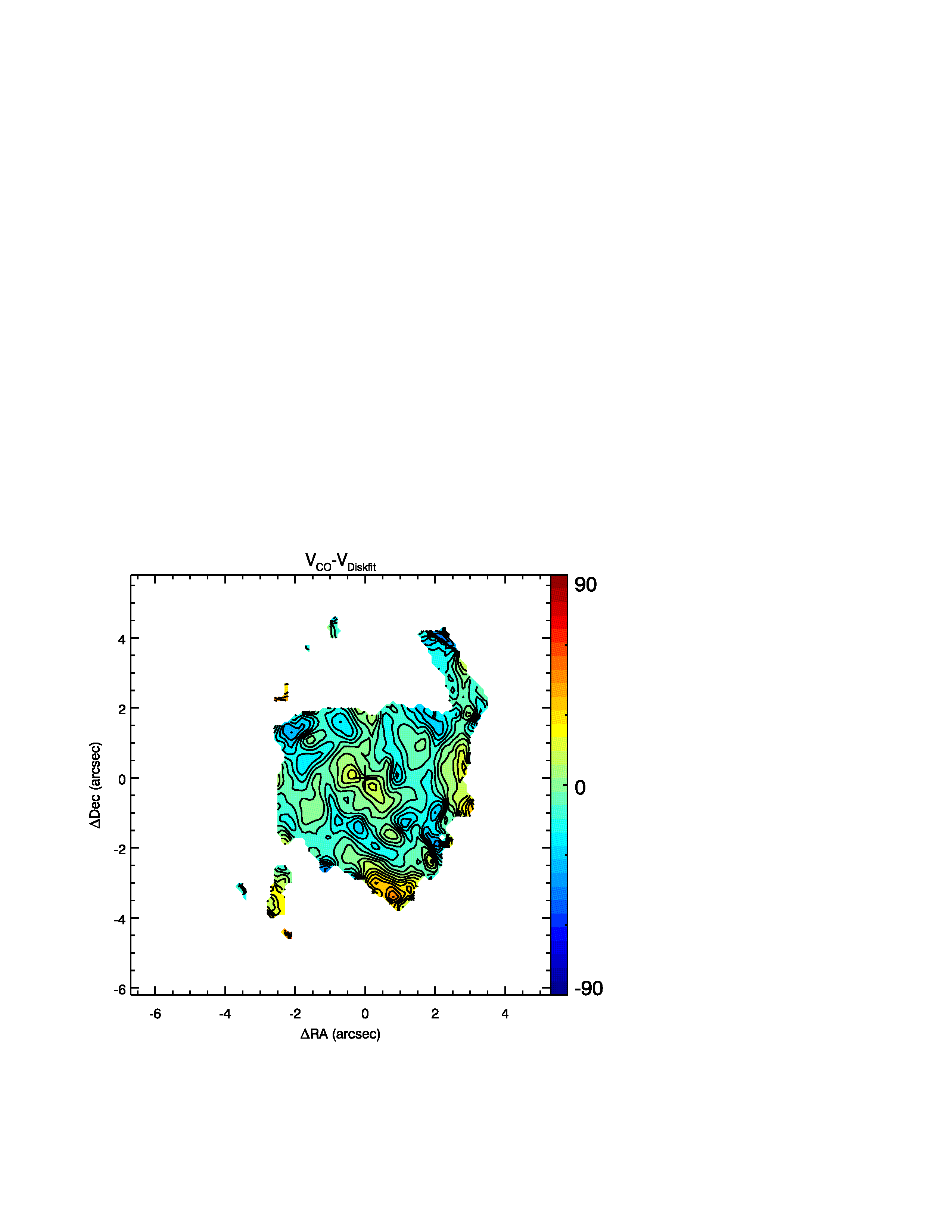}
 \includegraphics[bb=50 120 395 450,width=0.4\textwidth,clip]{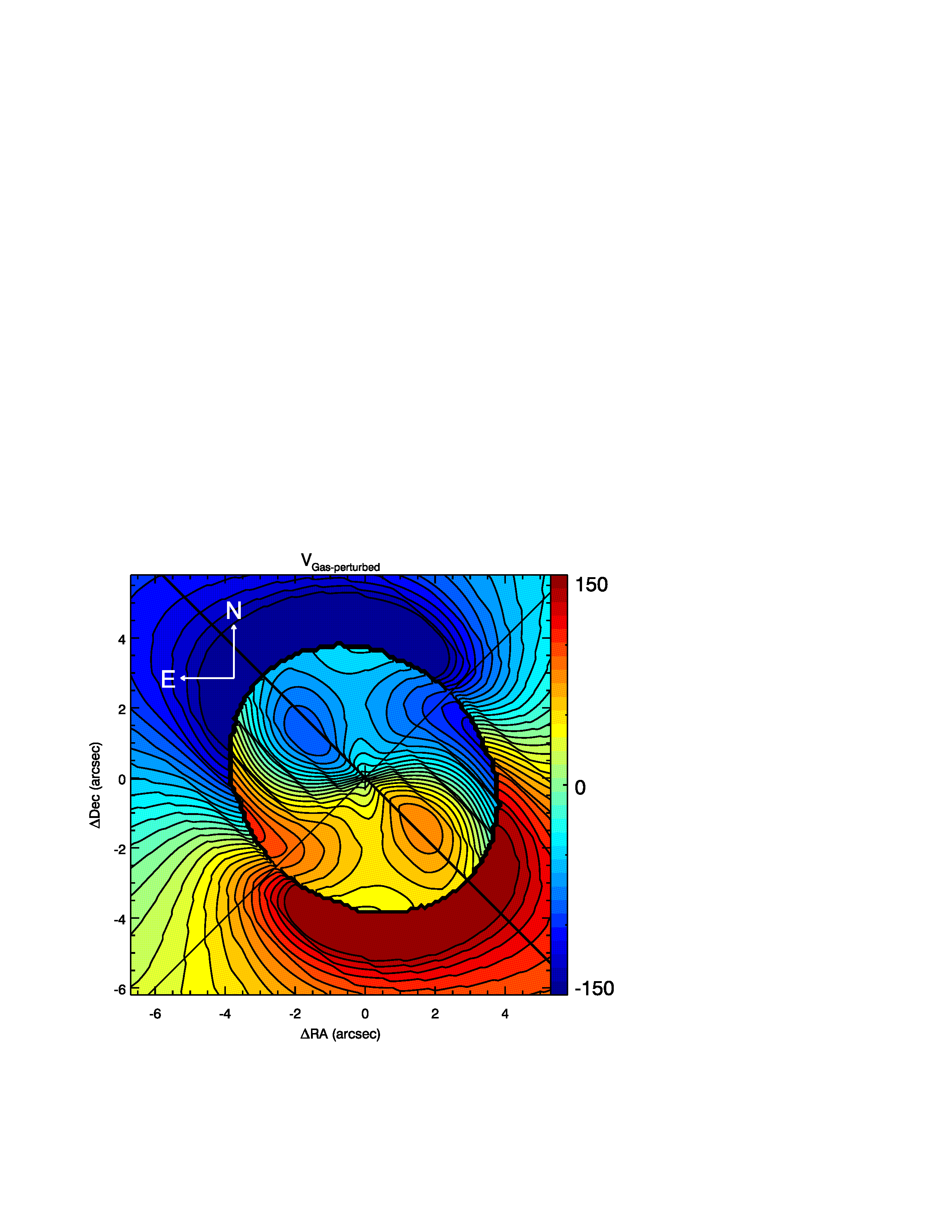} 
 \includegraphics[bb=50 120 395 450,width=0.4\textwidth,clip]{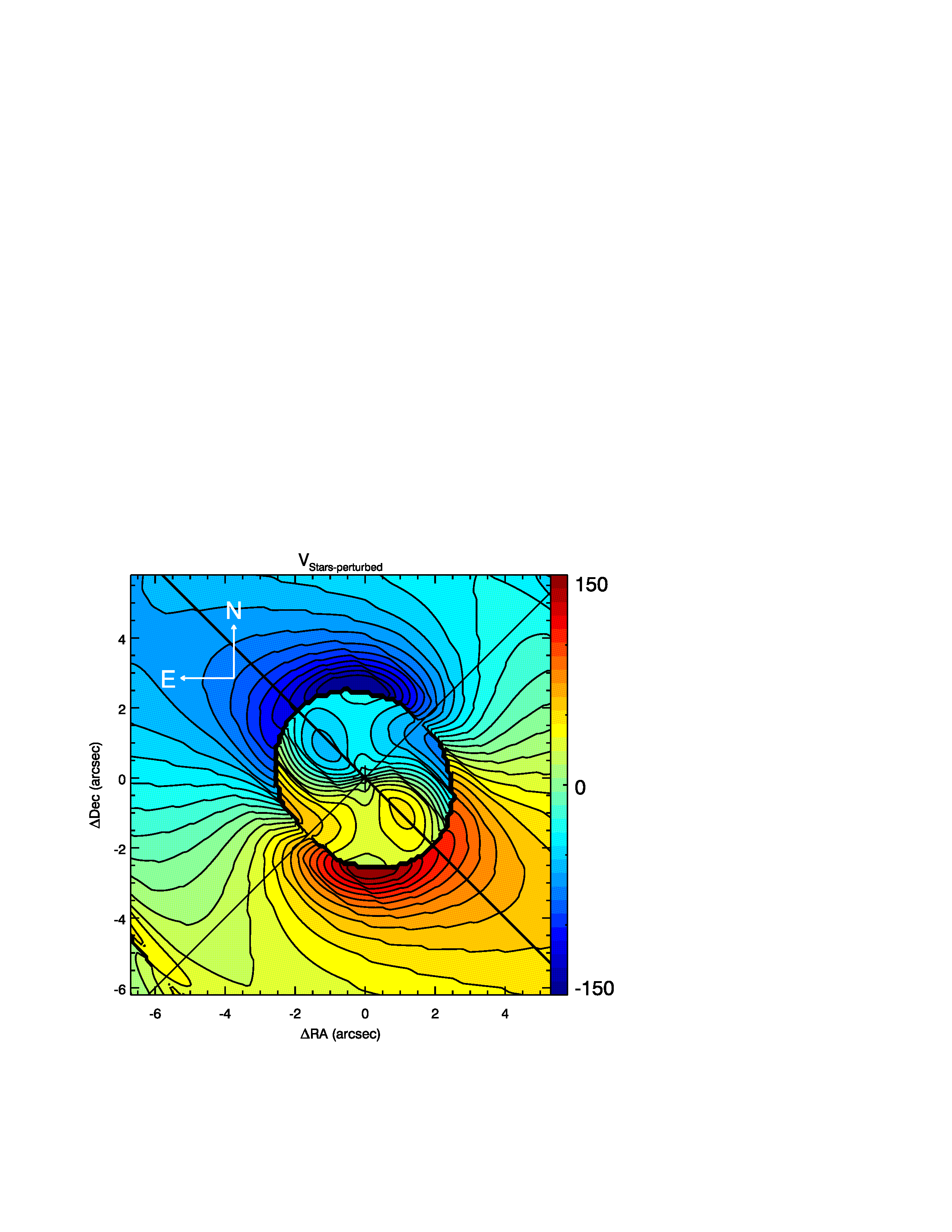}
   \caption{\textbf{Top:} \textit{Diskfit} model, with perturbations from a bar at PA  0\degr, fit to the CO velocity field  (left panel) and the resulting residual CO velocity field (observed $-$ model)  (right panel). \textbf{Bottom:} illustrative 
   bar-perturbed velocity fields resulting from our linearized epicyclic perturbation models (see text) and using 
   $\lambda=0.2$ and $\rm \Omega=120 km s^{-1} kpc$. The left (right) panel shows the results when setting the intrinsic rotation
   curve as that from C14 (our Bertola model fit to the stellar velocity field).  
  }
  \label{barmodsandresfig}
\end{figure*}

Velocity perturbations due to bar(s) are believed to play an important role in
fueling the SMBH and in triggering nuclear star formation. 
It is well known that \galaxy\ hosts an intermediate strength nuclear bar 
with radius $\sim$1.7 kpc in PA$\approx$0 
\citep[][C14]{hackandschwei83,pence90,mulchaey97,aguero2004,dicaire2008,comeron2010,kendall2011}, 
which could be largely responsible for the velocity perturbations seen in 
the molecular gas. C14 used torque maps to show that the asymmetries
in the velocity field of the nuclear molecular gas are predominantly produced by the bar.
They also briefly explore estimates for the bar pattern speed. 
However, they did not make a detailed kinematic analysis of the bar-produced perturbations.

To analyze the effect of bar-produced perturbations we use both \textit{Diskfit} \citep{diskfit2007} and our own
Fourier component decomposition software \citep{finlezphd2017} 
based on the the linear perturbation analysis described in \citet{wong2004,fathiphd2004}.

The \textit{Diskfit}\footnote{http://3w.physics.queensu.ca/Astro/people/Kristine\_Spekkens/diskfit/} 
package can be used to fit both the image and the velocity field of a galaxy. In imaging mode, an
input image is fit with one or more of a bulge, disk, and bar, resulting in estimates of the relative flux
and morphology (ellipticity, brightness profile, and PA) of each component.  
In velocity mode, \textit{Diskfit} models asymmetric rotation-dominated
velocity fields using a combination of tangential and radial perturbations to a fitted circular velocity
model. 
We fit our \co2-1\ velocity field using \textit{Diskfit} considering only m=2 potential perturbation
(i.e. bars) modes
and  using the galaxy nuclear position, galaxy PA, galaxy inclination and bar PA as fixed values 
(mm continuum peak position, 45\degr,33\degr, and 0\degr, respectively). 
The best fit model obtained by \textit{Diskfit}, and the velocity residuals (observed $-$ Diskfit model) are shown in the top panels of Fig.~\ref{barmodsandresfig}.
The best fit model from \textit{Diskfit} differs from our toy rotation model (Fig.~\ref{zoominfigs}, top left 
and Sect.~\ref{rotmodelsect}) in that the apparent rotation axis moves to a slightly smaller
PA in the inner 4\arcsec, the inner 1\arcsec\ shows twisted isophotes, and there is a resonance
at $\sim$4\arcsec, which mainly falls in a region where we do not have observed velocities due to
low signal to noise. 

The residual velocity map obtained after subtracting the \textit{Diskfit} (Fig.~\ref{barmodsandresfig}
top right panel) shows smaller deviations from  systemic as compared to the velocity residual 
made from our rotation-only model (top right panel of Fig.~\ref{zoominfigs}), especially to the SE of
the nucleus, and in general in the inner arcsecs. However, the \textit{Diskfit}
model still does not attain the highest velocities seen in the inner arcsecs. We note that
\textit{Diskfit} only allows us to change basic photometric parameters of the galaxy, e.g., disk PA and inclination and bar PA,
and the input observed velocity field. Since all these are relatively well defined for \galaxy, we are unable to further
fine-tune the results of \textit{Diskfit}.

To better illustrate the differences between the \textit{Diskfit} model and the observed velocity
field, we plot the  \textit{Diskfit} model (orange lines)
on the observed pv diagrams at several relevant PAs (Fig.~\ref{pvbarmodfig}). We immediately note that
the best-fit \textit{Diskfit} model was derived from the velocity field (intensity weighted average
velocity at each spatial pixel) rather than the full datacube, so that comparing the model directly to the
pv diagram is not really fair. Instead it is more correct to compare the model (orange lines) with the
velocities from the  moment 1 map (intensity weighted velocity; black dashed lines in the figure).
While the \textit{Diskfit}
model slightly overpredicts the velocities seen along the major axis (top right panel), and the pattern
of the velocities seen along the minor axis (bottom middle panel), it fails to predict (by a factor $\sim$2)
the large peak velocities seen along the minor axis (PA=135\degr), or along PAs 0\degr\
and $-$15\degr. That is, the bar perturbations are unable to explain the $\sim$90\kms\ radial velocities
seen in the inner 1\arcsec\ to the NW and SE of the nucleus.

\begin{figure*}
 \centering
 \includegraphics[bb=75 130 440 475,width=0.7\textwidth,clip]{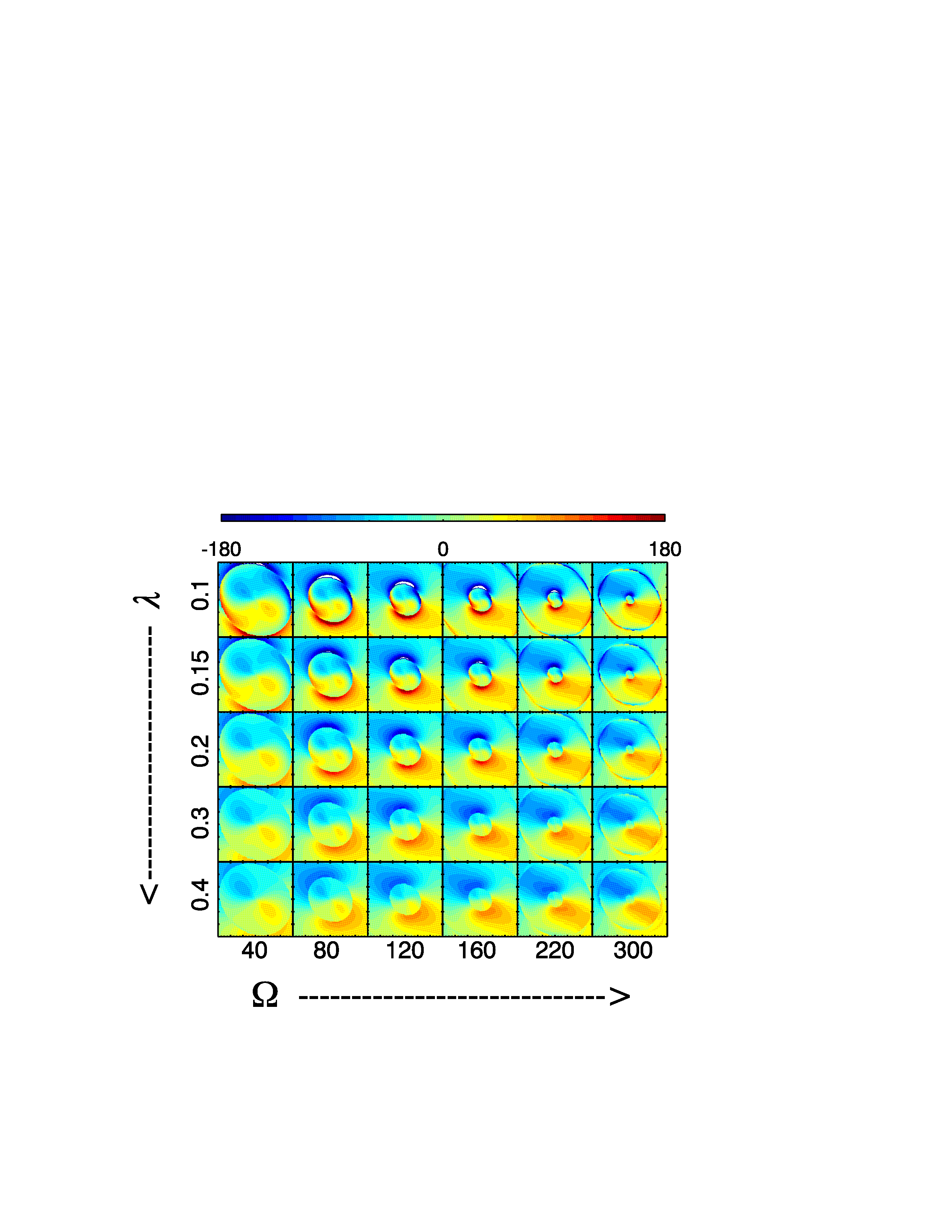}
   \caption{Bar-perturbed velocity fields obtained with our epicyclic perturbation models applied to our
   best fit Bertola stellar rotation curve (see text),
on varying the bar pattern speed (left to right; $\Omega$ in units of \kms\ kpc$^{-1}$) 
and the dimensionless damping parameter (top to bottom; $\lambda$). The FOV of each image is 12\arcsec\ $\times$ 12\arcsec 
and major tick marks are shown every 1\arcsec. All velocities follow the same color bar shown
on the top. 
}
  \label{barpertmodfig}
\end{figure*}

While \textit{Diskfit} models the observed velocity field with a base rotation model plus perturbations 
in radial and tangential velocities (one component each in the case of m=2 modes), 
it does not use (or at least does not provide details to the user) a physically-based model with, 
e.g., a mass-based rotation curve or a fixed bar pattern speed. 
We thus additionally model the observed CO velocity field with linearized epicyclic perturbations
produced by a bar
\citep[for details see, e.g.,][]{wong2004,fathiphd2004} applied to a physically derived
rotation curve (an exponential disk whose mass is constrained by near-IR photometry)
for a  given bar pattern speed ($\Omega$), damping factor (associated to a frictional force; $\lambda$), and bar PA and ellipticity \citep{finlezphd2017}. 
Our code is based on the algorithms proposed in \citet{franx94}, \citet{wong2004} 
and \citet{fathiphd2004}. Note that we specifically use the `m=2' potential
(relevant for bars) which introduces changes in the 1st and 3rd harmonic coefficients \citep{schoenmakers97},
and that this perturbation analysis is valid only for `weak' bars, i.e. when the bar potential does not dominate
the disk potential).

\begin{figure*}
\centering            
   \includegraphics[bb=40 120 390 494,width=0.33\textwidth,clip]{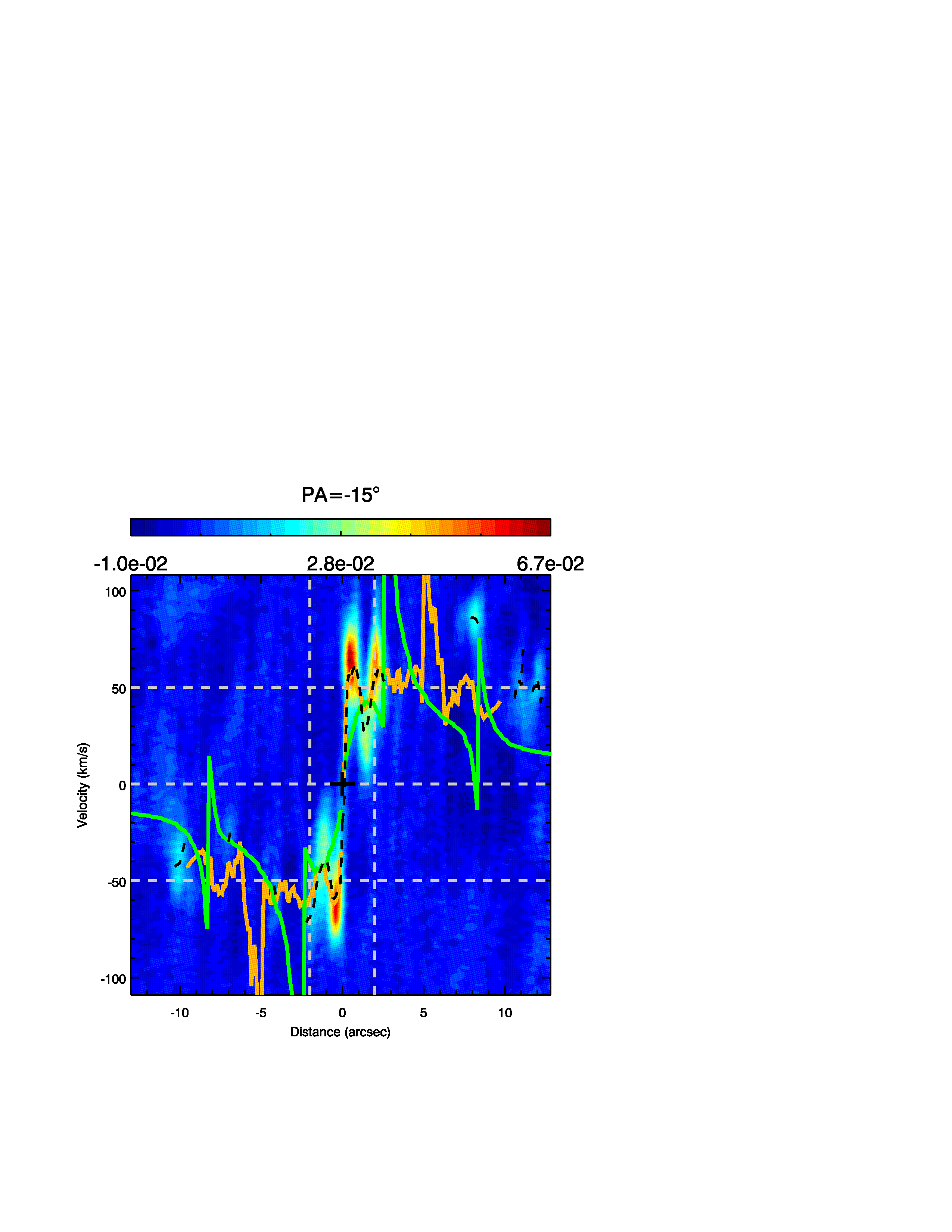} 
   \includegraphics[bb=40 120 390 494,width=0.33\textwidth,clip]{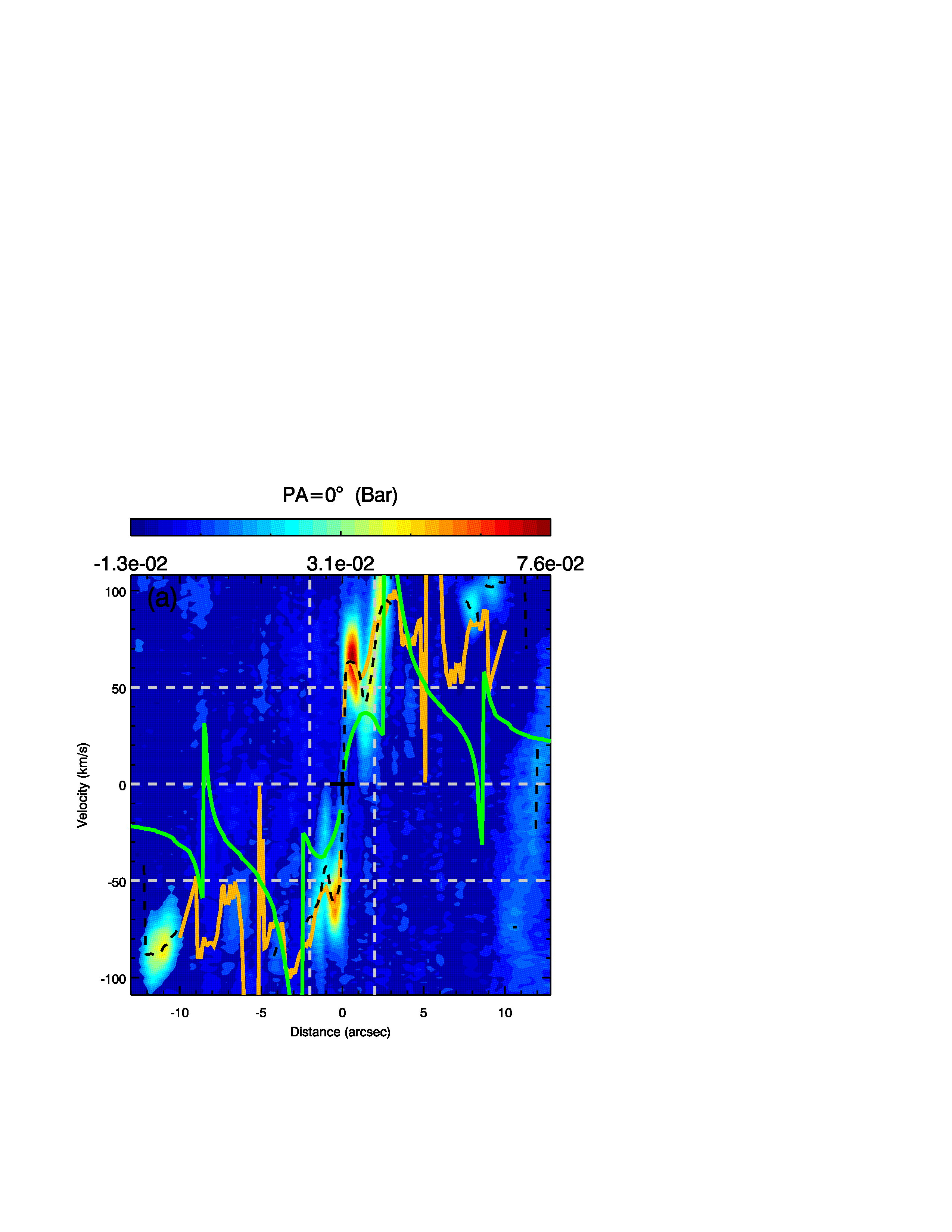}
   \includegraphics[bb=40 120 390 494,width=0.33\textwidth,clip]{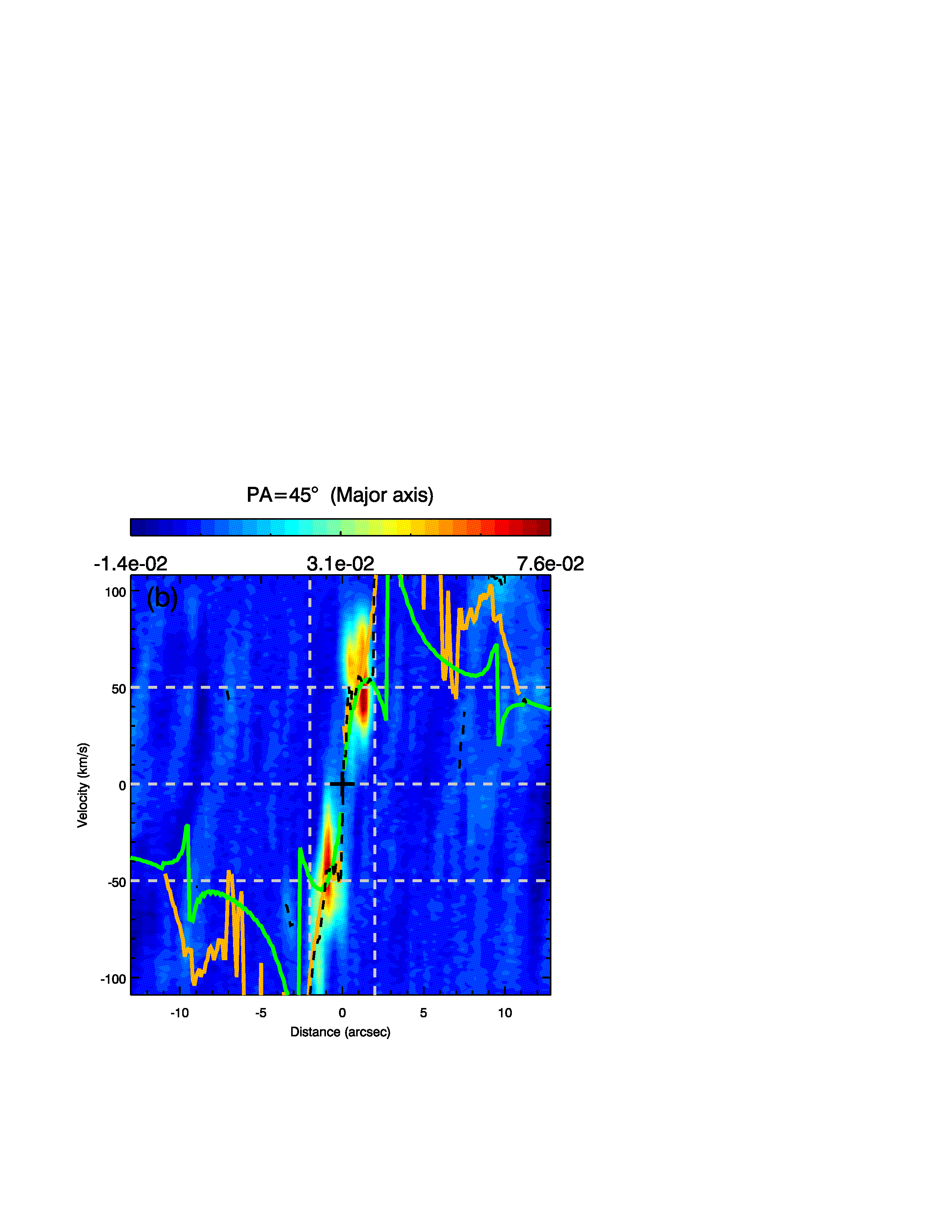}
   \includegraphics[bb=40 120 390 494,width=0.33\textwidth,clip]{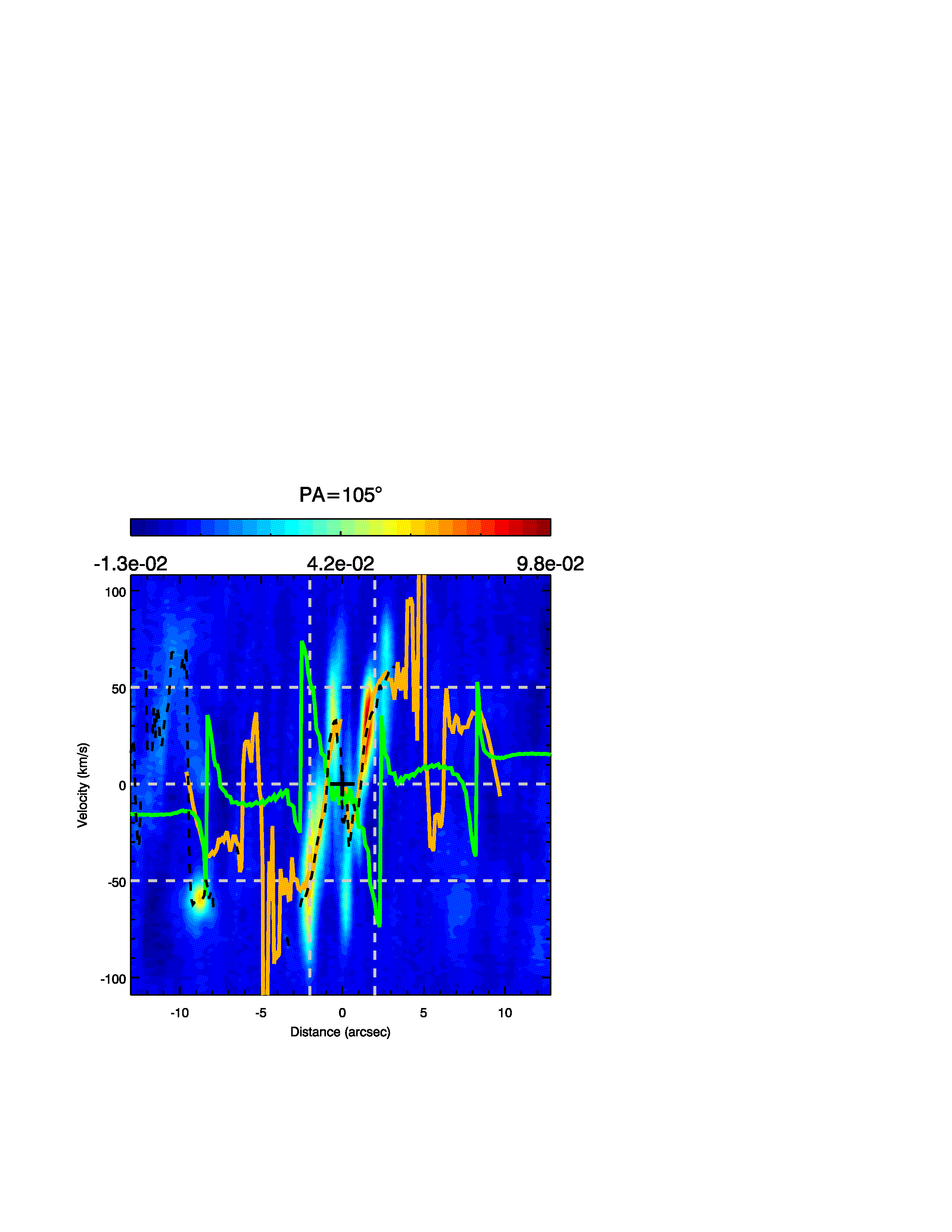}
   \includegraphics[bb=40 120 390 494,width=0.33\textwidth,clip]{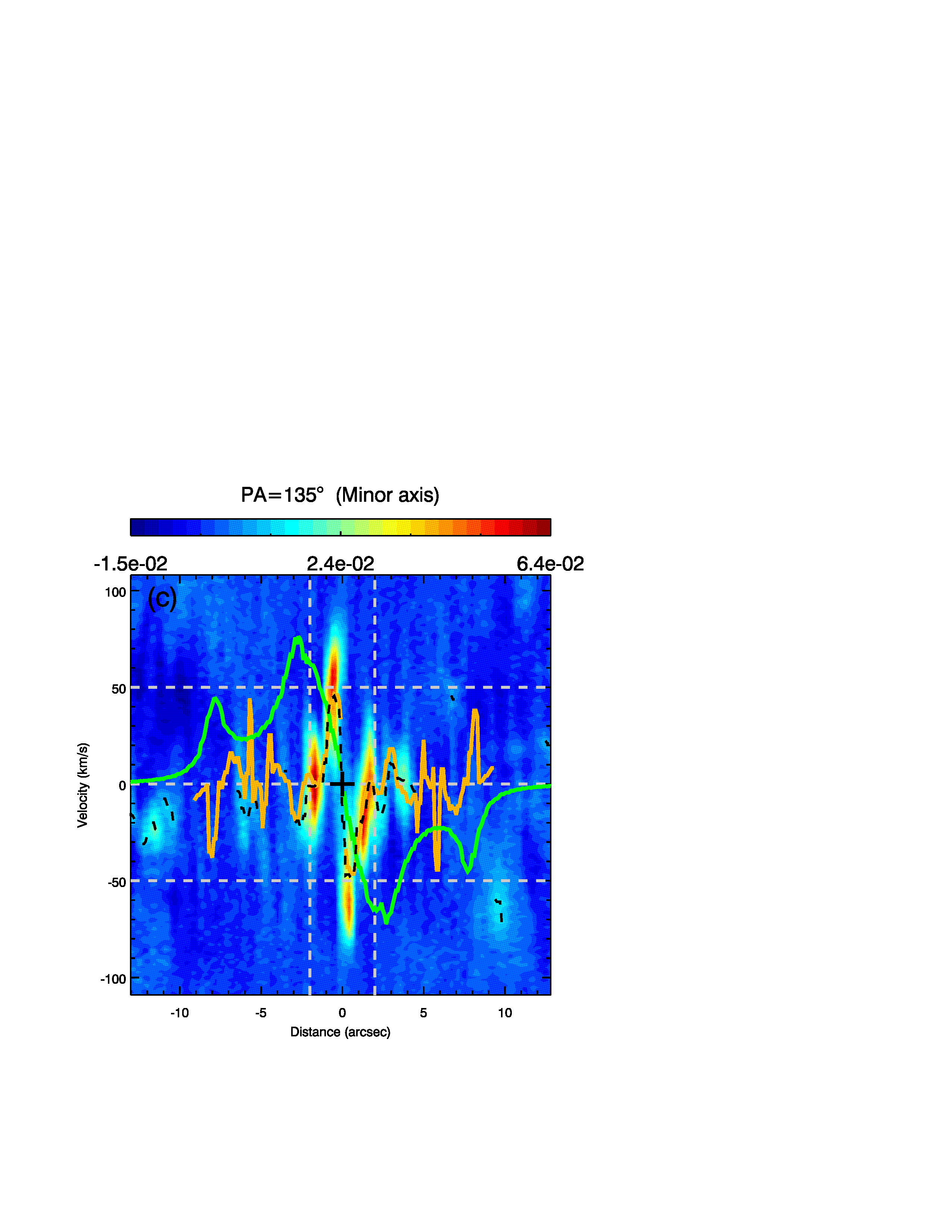}
   \includegraphics[bb=40 120 390 494,width=0.33\textwidth,clip]{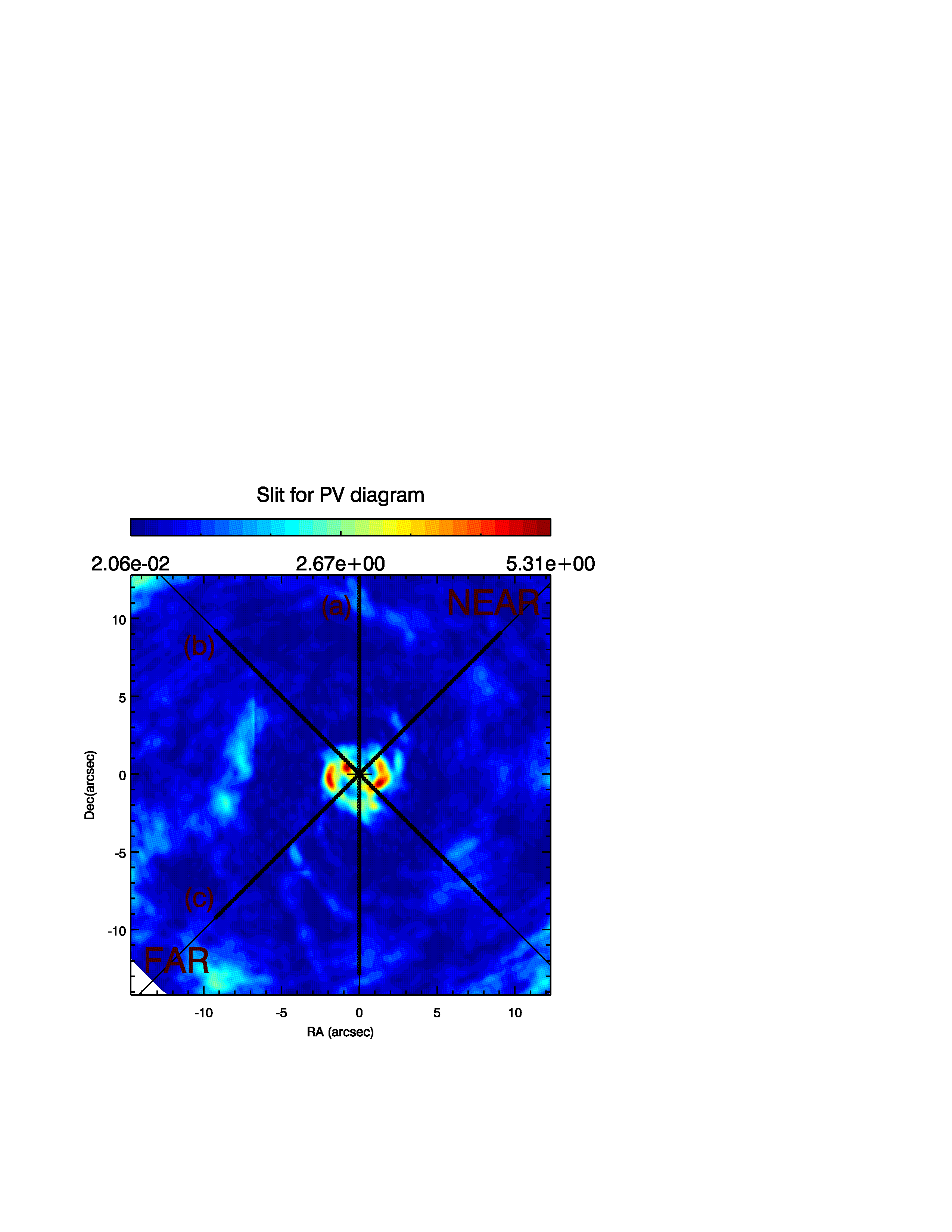}  
  \caption{Position-Velocity diagrams, as in Fig.~\ref{pv1fig}, but here the overplotted
lines show the velocity of the corresponding pixel in the CO moment 1 map (dashed black lines; i.e. the flux-weighted
average velocity at a given pixel), the best-fit velocity field from \textit{Diskfit} (solid orange line), and for
comparison the $\lambda$=0.2 and $\Omega$=$\rm 120 km s^{-1}kpc$ model from our linear perturbation analysis (bottom right panel
of  Fig.~\ref{barmodsandresfig}; solid green line). 
To better trace the highest velocity
components in the CO moment 1 (velocity) map, we used a cutoff of  8mJy/beam/channel ($\sim4\sigma$) to
create the moment 1 map which is overplotted here and used as the input map to \textit{Diskfit}. The PA of the `slit' along which the pv diagram was extracted is marked above each panel. The bottom right panel shows the moment 0 map of \co2-1\ together with the positions of the slits used to create the pv-diagrams in the upper row panels. 
  }
  \label{pvbarmodfig}
\end{figure*}

We used \textit{Diskfit} to decompose an IRAC 3.6\micron\ image of \galaxy, obtained from NED, 
into bulge, disk, and bar components.
The galaxy and bar PAs were fixed and other parameters allowed to vary. We further assume a 
constant mass to light (M/L) ratio for all three components. We find that the disk contains 58\% of the total mass 
(light) of \galaxy\ with a bar to disk mass ratio of 0.7. 
Given the total mass of \galaxy\ derived by \citet{sheth2010}, the disk mass is $2.2\times10^{10}[M_{\odot}]$. 
Alternatively,  the $3.6\mu$ disk luminosity with a   
$M/L_{3.6\mu}$ ratio of 0.47 \citep{mcgaugh2014} implies a disk mass of $\rm 4.4\times10^{9}[M_{\odot}]$.  
An exponential disk with these total masses  was then used to
derive a first-guess intrinsic (i.e. before bar perturbations) 
axisymmetric rotation curve \citep[details in ][]{finlezphd2017}. The disk mass was then slightly adjusted
(to $\rm 6\times10^{9}[M_{\odot}]$) 
in order to better fit (by eye) the model rotation curve of the CO in the inner disk (i.e. ModC2014)
or $\rm \sim 2\times10^{9}[M_{\odot}]$ to agree with our best-fit
Bertola model \citet[][Eq.~2]{bertola91} to the stellar velocity field. 
Note that these masses are an order of magnitude lower than that predicted by \citet{korchagin2000} ($\rm 1.78\times10^{10}[M_{\odot}]$). 

We then ran our linear perturbation code, in m=2 mode, using as inputs the intrinsic axisymmetric rotation
curve(s) derived above, the PAs of the galaxy disk and bar, and the bar ellipticity ($\epsilon=0.42$), the latter derived from our \textit{Diskfit} decomposition. 
The bar pattern speed ($\Omega$) and the damping factor ($\lambda$) were allowed to vary. The resultant model velocity fields for a range of values of $\Omega$ and $\lambda$, when using the best-fit Bertola model of the stellar velocity field, are 
shown in Fig.~\ref{barpertmodfig}. The most notable effect of varying $\Omega$ is the change in the radii of the
resonances. Gas orbits change abruptly when 
crossing these resonances; the effect of increasing damping (increasing $\lambda$) is to smooth
out these large swings in the orbits. Most of the panels in Fig.~\ref{barpertmodfig} show the characteristic
`butterfly' pattern  expected from bar perturbations. However, for this butterfly
pattern to fall within the central $\sim$2\arcsec\ as observed, i.e. to explain the innermost high velocity features, 
one requires extremely high 
($\sim$300\kms\ kpc$^{-1}$) bar pattern speeds. 
Alternatively, the intrinsic rotation curve requires to 
rise slower or flatten at lower velocities.
We must note that the uncertainty in the distance to \galaxy\ (see Sect. 1) plays a significant role in 
the bar pattern speeds used here. 
If a distance of 20~Mpc is used for \galaxy\ then the bar pattern speeds we list here would halve, so that
less extreme bar pattern speeds could replicate the observed resonances.  
In any case, even if the resonance radii are matched, the pattern of the model velocities are significantly 
different from
the observed CO velocity field (and the larger scale \ha\ velocity field from \citet{pence90}: specifically 
at higher pattern speeds the strongest perturbations inside the inner resonance are in PA $\sim$100\degr, 
offset from the PA of our posited outflow, and beyond the inner resonance the kinematic axis of is highly curved,
starting at PA $\sim$0 and then curving to the observed PA of the galaxy. 

For illustration, we compare the predictions of the perturbation model which uses 
the Bertola best fit model to the stellar velocity field as the intrinsic rotation curve,
and parameters $\Omega$ = 120 [\kms kpc$^{-1}$] and  $\lambda$ = 0.2
(the model shown in the third row, third column of Fig.~\ref{barpertmodfig}) with our 
observed pv diagrams in Fig~\ref{pvbarmodfig}. 
While this model does not well fit the observed
velocity field, it uses a pattern speed argued for in C14 (based on corotation placed at the bar end)
and a damping parameter within the range of values
typically invoked for other well studied galaxies \citep[between 0--0.5][]{wada94,fathi2005}, and is thus
a good reference point.

As a further illustration, the bar-perturbed velocity fields for $\Omega$ = 120 $\rm [km~s^{-1}kpc]$, 
$\lambda$ = 0.2, and 
for both options of the intrinsic rotation curve (gas- and stellar-rotation curve models)  are 
shown in bottom panels of Fig.~\ref{barmodsandresfig}. 
The model which uses an intrinsic rotation curve similar to that of the  gas (bottom left)
exhibits a resonance at $\sim$4\arcsec, similar to that obtained by \textit{Diskfit} (left top), but presents less pronounced 
nuclear distortions as compared to the 
 \textit{Diskfit} model.  
Using the slower rising Bertola (stellar velocity) model as the intrinsic rotation model (bottom right) 
changes the position of the resonance to  ($\sim$2.6\arcsec), but also gives lower velocity distortions along the
minor axis, or rather the higher velocities seen in the observed velocity field ($\sim$1\arcsec) are further out in the model ($\sim$3\arcsec): 
therefore, to spatially matching these velocity distortions 
requires higher bar pattern speeds or a slower rise in the intrinsic axisymmetric rotation curve.
Apart from the mismatch in resonance radii, these two panels also clearly illustrate the mismatch between
the observed and modelled velocity fields noted above; specifically, the misalignment of kinematic axes inside the resonance
(related to the posited outflow), and the large curvature in the kinematic axis beyond the inner resonance.  

Overall, we are unable to convincingly fit the observed CO kinematics with perturbations produced by the large
scale bar.
Our linearized epicyclical bar perturbation models, which use realistic values for the intrinsic rotation curves and the bar 
pattern speed (with the caveat of the uncertainty in the distance to \galaxy), are able to reproduce the 
amplitudes of the inner perturbations. However, the resonances are produced further out than observed, and
the velocity changes are not as sharp as observed.  Higher bar pattern speeds, perturbations by an inner bar in
a different PA, or different intrinsic rotation curves, would be required. 
 \textit{Diskfit} reproduces reasonably  many of the  observed features in the velocity map, and at first
 glance provides a reasonable explanation for the perturbations observed, even if the amplitude of these perturbations is not
 as high as observed. However, we are wary of the results of \textit{Diskfit} for two main reasons. First, \textit{Diskfit}
 does not provide feedback on the underlying physical parameters of the resultant model, and thus, e.g., we are unable to 
 evaluate whether the bar pattern speed used is physical and second, we have applied \textit{Diskfit} to about a dozen galaxies
 for which we have disturbed optical emission line kinematics over the inner 5" of the galaxy and almost always found relatively good fits (Schnorr-Muller, priv. communication), even though our detailed multi-component analysis either found the perturbations to be due to bars \citep{schnorr2017a} or outflows and/or streaming inflows \citep[most other cases, eg.,][]{schnorr2017b}. In fact, \citet{diskfit2007} obtained a good fit to the velocity field in NGC2976, but to conclude that the perturbations were due to the bar, they confirm their existence at the PA predicted by the model, according previous photometry. This consistently good performance of \textit{Diskfit} makes it more difficult to believe that the fits are truly consistent and physically motivated 
rather than  empirical best fits to distorted velocity fields.
We emphasize that we are not stating that bar-related perturbations do not exist in the velocity field, rather we argue that bar-related
perturbations are not the unique and dominant driver of the observed nuclear perturbations in the CO velocity maps, and it is most
likely that the nuclear perturbations are produced by an AGN-driven outflow.

\subsection{Modeling Observed Velocities: \co2-1\ Streaming?}
\label{streamsect}

\begin{figure*}
\centering            
  \includegraphics[bb=20 310 274 558,width=0.3\textwidth,clip]{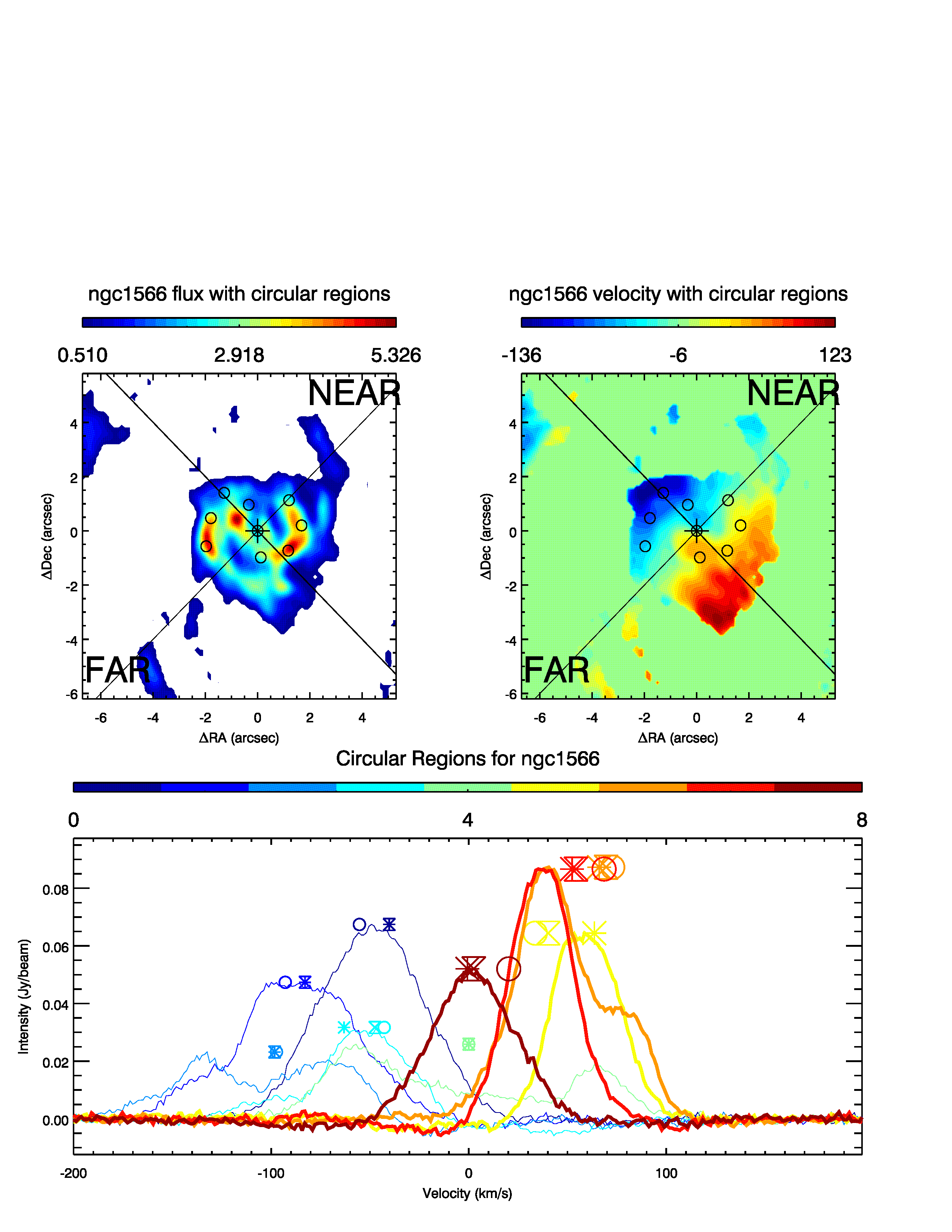}
  \includegraphics[bb=7 14 570 310,width=0.69\textwidth,clip]{circularapertures_stream50_ngc1566.png}
  \caption{CO spectral profiles in apertures along the inner spiral arms. The left panel shows the location of
  each circular aperture (0\farcs2 in radius) overlaid on the CO flux map.
  For reference, the center of the galaxy, 
  major axis, and the near and far sides of the disk, are indicated. Apertures are numbered 0 to 8 with aperture 0 being the farthest aperture on the E arm, aperture 4 the nuclear
  one, and aperture 8 the farthest aperture on the W arm. 
  The right panel shows the extracted \co2-1\ spectra; 
  thick lines are used for the spectra corresponding to apertures 
  from the W arm (Apertures 5 to 8), and different colors are used for each spectrum, following
  the color bar on top of the panel.
  Symbols with the corresponding color (plotted at the $y$ value of the peak flux density
  of the spectrum) denote the radial velocities expected in that aperture
  for our rotation model (hourglasses), our outflows + rotation model (asterisks), and our rotation plus radial streaming inflow model (open circles). In some apertures, adding outflows and/or streaming inflows does not change the predicted radial velocity; this is a result of projection effects and/or the fact that our outflow model has zero velocity beyond $\sim$2\arcsec\ from the nucleus. 
  }
  \label{aperturesfig}
\end{figure*}

The presence of putative outflows and/or bar related perturbations (previous sections) 
makes it difficult to search 
for signatures of streaming motions in the inner few arcsec in velocity maps or even 
residual velocity maps (e.g. Fig.~\ref{zoominfigs}). 
That is, since the velocity maps show the intensity
weighted average velocity of the spectrum corresponding to each spatial pixel, they
are likely dominated by the 'outflow' signature in the inner $\sim$ 3\arcsec.
Further, when the intensity of the gas in rotation dominates that of the gas in inflow, velocity
residual maps will not show strong indications of the inflow. 
It is thus important to examine the velocity profile of
each pixel or aperture in order to separate outflows, rotation, and streaming inflows.
Ideally, one requires a complete velocity field to analyze the azimuthal average
of radial gas velocities at each radius. While this is often possible in the case of ionized gas,
molecular gas is often detected only over a limited 
range of azimuths at each radius. In the case of \galaxy, our CO velocity maps are 'complete' out to 
a radius of $\sim$3\arcsec, beyond which the velocity filling factor is $\sim$5--40\%.
Under the assumption that these detected CO regions dominate
the CO flux at their respective radii, the detected clumps or arms can still be used to constrain
the presence of  streaming flows.
Given the above, we model streaming inflows with a very simple toy model in which 
the inflow is assumed to have a
constant radial inflow velocity (which we fix to \streamvel\ after initial inspection of the results). 
This velocity is then projected and added to the projected radial velocity expected from
our rotational model. 
We first examine the spectral profiles in apertures along the inner spiral arms
(left panel of Fig.~\ref{aperturesfig}).
The \co2-1\ spectra extracted from these apertures are shown 
in the right panel of Fig.~\ref{aperturesfig}, together with the average radial velocities
expected from our models of rotation, outflows, and streaming (and combinations thereof). 

\begin{figure*}
\centering            
  \includegraphics[bb=20 310 274 558,width=0.3\textwidth,clip]{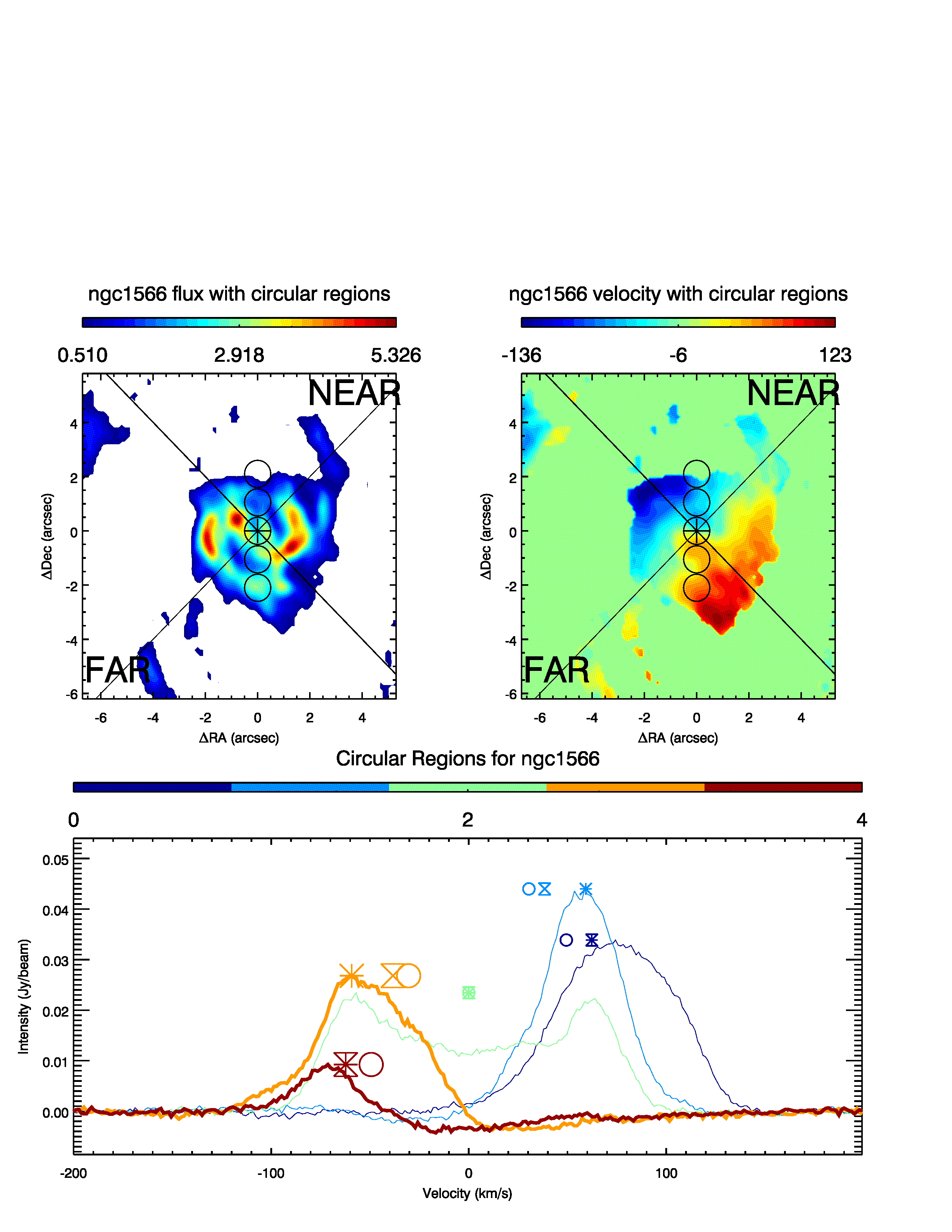}
  \includegraphics[bb=7 14 570 310,width=0.69\textwidth,clip]{PA0_stream50_ngc1566.png}
  \caption{Same as Fig.~\ref{aperturesfig}, but for apertures oriented along the  bar. Apertures are numbered 0 (southernmost) to 4 (northernmost), and apertures here are each 0\farcs5 in radius.
  }
  \label{aperturesfigpa0}
\end{figure*}

Analyzing the \co2-1\ spectra in the right panel of Fig.~\ref{aperturesfig} 
we see the following: 
(a) The nuclear spectrum (light green) is clearly double peaked: the peaks at 
$\rm V_{rad}\approx$65\kms\ are attributed to putative nuclear outflows 
(previous sub-sections), and the highest velocities seen are $\rm V_{rad}\approx$100\kms.
Note that there is a plateau of \co2-1\ emission at lower velocities, potentially from
gas rotating in the disk (recall that our rotation models predict velocities of
$\rm V_{rad}\approx$0--50\kms within this aperture). 
(b) The off-nuclear apertures show profiles with varied shapes and widths, and the off-nuclear apertures which 
intersect the galaxy major axis
clearly show multiple velocity components.
(c) For the W spiral arm (spectra plotted in thick lines) 
there is a large mismatch between the spectral profiles and the expectations
from our rotation model only (hourglass symbols in Fig.~\ref{aperturesfig}).
In order of increasing distance from the nucleus along the W spiral arm, 
gas in the first aperture rotates faster than predicted; 
the gas profile in the second aperture has a strong red shoulder at a velocity
consistent with rotation, while the profile peak is offset $\sim$30\kms\ to the blue;
gas in the third aperture rotates slower than predicted;
and gas in the farthest aperture is centered at zero velocity since the aperture
lies on the minor axis of the galaxy. 
(d) For the E spiral arm (spectra plotted in thin lines) 
the profiles are more centered on the predictions of our rotation only model.
In order of increasing distance from the nucleus along the E spiral arm, 
gas in the first aperture (which includes the edge of the strong CO knot $\sim$1\arcsec\ 
from the nucleus to to NE) lies close to the prediction of the rotation model but a
clear blue shoulder is seen;
gas in the second aperture shows a clear double-peaked profile with the expected rotation velocity
lying in the middle of the two peaks;
gas in the third aperture also shows a broad blue shoulder. 
(e) Including our decelerating outflow model (i.e. using the predictions of rotation plus outflows; 
asterisks in the figure) the model predictions change 
significantly only for the two off-nuclear apertures closest to the nucleus. Here the aperture 
to the S (solid yellow spectrum in the figure) fits the prediction satisfactorily, i.e. 
as if almost all gas is in outflows, 
but to the N (cyan spectrum), while the aperture profile shows an extra blue wing in the correct velocity
direction for outflows the magnitude of the offset does not fit well with our model, and the
gas seems to be dominantly in rotation rather than outflows. 
(f) Using a model which sums our streaming toy model to pure rotation
    (open circles in the figure), 
    we see that the apertures in the W arm are inconsistent with the predictions of
    streaming inflows:  the mismatch between the spectrum peak and the prediction increases 
    when changing from rotation only to rotation plus radial inflow. 
    In the E arm, however, the profiles are in general as consistent with the streaming inflow+rotation
    model as with the rotation only model, i.e. while the profile peaks are consistent with
    rotation, the prominent shoulders on these profiles are roughly consistent with streaming inflows. 
    This is also clearly seen in the skewness map (Fig.~\ref{velreswithfluxfigs} where almost the full
    inner spiral arm structure shows a blue 'skewness', independent of being on the near- or far-side of the
    galaxy disk.

We also show the spectra in apertures along the PA of the large scale bar in Fig.~\ref{aperturesfigpa0};
here we use larger apertures (0\farcs5 radius) to obtain a higher signal to noise.
Here the peak of the spectral profiles are consistent with the predictions of rotation+outflow
and a shoulder is seen roughly at the predicted velocity of rotation plus radial streaming
inflows. 
 
To test for streaming inflows along spiral patterns further from the nucleus, we also examined 
the spectra from apertures along the spiral patterns SE to S of 
the nucleus which connect to the inner spiral arm discussed above, and spectra in 
apertures tracing spiral structure to the N and NW of
the nucleus, which connect to the W inner spiral.
We do not show these spectra as
the essential results can be seen in the skewness map of the \co2-1\ line (Fig.~\ref{velreswithfluxfigs},
left panel):
here blue (red) colors represent pixels where the CO spectral profile is skewed in the sense of having
excess emission towards the blue (red) of the weighted mean velocity at that pixel. To avoid
contamination by noise the skewness was calculated using the spectral profile down to 10\% of the
peak flux. If most of the gas follows regular rotation and a smaller fraction of gas participates in a radial streaming inflow then we expect blue (red) skewness on the far (near) side of the galaxy. 
The inner 2\arcsec\ shows profiles skewed consistently to the blue.  
show that the gas is mainly consistent with rotation.  
Two spiral arm sections at $\sim$10\arcsec\ from the nucleus (roughly at the putative location of the  Inner Lindblad Resonance (ILR) of the large-scale bar  \citep[][C14]{comeron2010}) 
to the SSE and to the N show a skewness consistent with inflows. All spectra in these arms show non symmetric profiles which are likely from multiple velocity components. In fact velocity differences between the peak of the profile and the shoulders to the red match well the predictions of the offset (from rotation) velocity expected from streaming inflows.

\subsection{Molecular mass in the inner kiloparsec}
\label{molmasssect}
Molecular gas mass is typically estimated from the CO luminosity \citep{solbout2005}
using
M$_{\rm mol}~[M_{\odot}] = \alpha_{CO} \times$\lprimeco\ where:
\begin{equation}
 \label{solomoneq}
L^{\prime}_{CO}= 3.25 \times 10^{7} \times S_{line}\Delta\nu~\frac{D_{L}^{2}}{(1+z)^{3}\nu_{obs}^{2}}.
\end{equation}
Here, \lprimeco\ has units of $\rm K~km~s^{-1}~pc^{2}$,
$S_{line}\Delta\nu$ is the integrated flux density of the CO J:1-0 line in $\rm Jy~km~s^{-1}$,
$\rm D_{L}$ is the luminosity distance in Mpc, z is the redshift, and $\rm \nu_{obs}$ 
is the observed frequency in GHz. 
There remains significant debate on the value of \alphaco, and we use 
the value \alphaco$\rm =4.3~[M_{\odot}~(K~km~s^{-1}~pc^{2})^{-1}]$ as suggested by 
\citet{bolatto2013} for the Galaxy and other nearby spiral galaxies which are not extreme starbursts.
Note that \lprimeco\ 
is directly proportional to the surface brightness in K units, and therefore the 
\lprimeco\ ratio of two CO J transitions gives the ratio of their 
surface brightness temperatures. Furthermore, \lprimeco\ is constant for
all J levels if the molecular gas emission comes from thermalized optically-thick regions, 
i.e., the brightness temperature and line luminosity are 
independent of J and rest frequency for a given molecule \citep{solbout2005}. 

Since we observed the \co2-1\ line, we require to convert $\rm L^{\prime}_{CO J:2-1}$ to
$\rm L^{\prime}_{CO J:1-0}$, a conversion which depends on the physical conditions of the
gas.
\citet{bajaja95} have observed the CO J:1-0 and CO J:2-1 lines in \galaxy\ at low resolution using the SEST 
telescope, and C14 have presented ALMA \coo3-2\ observations of \galaxy.
The nuclear CO J:2-1 / CO J:1-0 intensity ratio found by \citet{bajaja95} is 1 in temperature
units: as expected from thermalized optically-thick gas.
C14 compared the \citet{bajaja95} CO J:2-1 integrated flux densities with their 
ALMA-derived \coo3-2\ integrated flux densities and inferred 
that the latter was missing some flux; they thus also assumed that the gas is
thermalized and optically-thick in their calculation of molecular gas masses. 

To further constrain the observed flux ratios, we downloaded the 
\coo3-2\ data of C14 from the \emph{ALMA Science 
Archive}\footnote{https://almascience.nrao.edu/aq/} and created a Moment 0 map. 
For thermalized optically-thick gas one expects that the integrated flux density
(in units of Jy~\kms) of each CO J transition varies as $\nu^{2}$.
We find that in the inner $\sim$3\arcsec\ radius (the inner disk),
the integrated CO J:2-1  flux density ($\rm 251\pm25~[Jy~km~s^{-1}]$) 
is 1/2.3 times that of \coo3-2. 
Over a larger field, $12\arcsec\times12\arcsec$ in size,
the total flux density of the CO(J:2-1) line ($\rm 406\pm41~[Jy~km~s^{-1}]$) 
is half that of \coo3-2.
We thus, as in C14, assume thermalized optically-thick gas, i.e. 
$L^{\prime}_{\rm CO J:2-1}$/$L^{\prime}_{\rm CO J:1-0}$ = 1.
Under this assumption, using a value of \alphaco\ listed above, and $\rm D_{L}$ = \distangal\ Mpc,
the molecular gas mass in the inner spiral arms
is $(6.6\pm0.7)\times 10^{7} [M_{\odot}]$,
and the total molecular gas mass in the central $12\arcsec\times12\arcsec$ region is 
$(1.1\pm0.1)\times 10^{8} [M_{\odot}]$.

\begin{figure}
\centering            
  \includegraphics[bb=40 120 390 460,width=0.4\textwidth,clip]{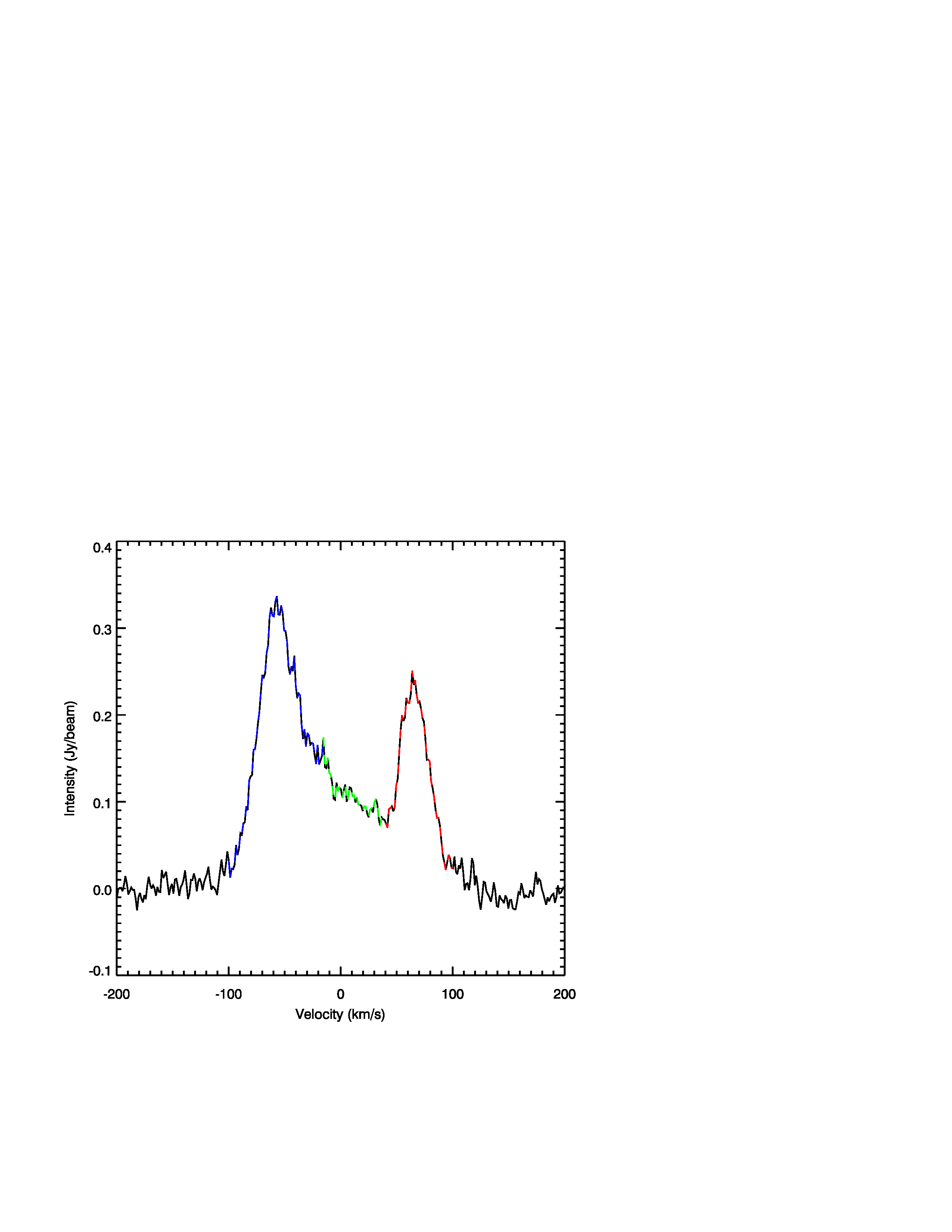}
  \caption{The CO J:2-1 spectrum (black) in a circular aperture centered on the nucleus with radius 0\farcs2. 
The spectrum was divided by eye into three parts: the pure outflow components (blue and red) and an intermediate velocity region (green; see text) between the blue and red outflow peaks.
  }
  \label{nuclearspectfig}
\end{figure}

Estimating the molecular gas mass in the nuclear outflows of \galaxy\ is more difficult. Recall
that outflowing molecular gas is clearly detected out to $\sim$2\arcsec\ (Sect.~\ref{outflowsect}),
and perhaps out to 5\arcsec\ (Sect.~\ref{streamsect}).
To estimate the mass and momentum of the outflowing gas in the unresolved nucleus we extract
the nuclear CO J:2-1 emission spectrum in a circular
aperture of 0\farcs2 in radius (similar to our synthesized beam area; Fig.~\ref{nuclearspectfig}).
This spectrum appears to be made of three distinct components: 
a strong blue Gaussian representing emission from the blue outflow ($(2.0\pm0.2)\times10^{5}~[M_{\odot}]$), 
a weaker red Gaussian representing emission from the red outflow ($(1.0\pm0.1)\times10^{5}~[M_{\odot}]$)
(Sect.~\ref{outflowsect}) and an intermediate velocity region (green line in the figure; radial
velocities between $\pm$40\kms; $\rm (0.8\pm0.1)\times10^{5}~[M_{\odot}]$). 
This intermediate velocity region
could originate in (a) gas in a spherical outflow in the nucleus since variation in the projection
angles to the line of sight will give emission at all velocities; (b) unresolved emission from
gas in solid body rotation-only; (c) emission from dispersion-dominated gas in the nucleus. 
With the masses calculated above, and using the (deprojected) mean flux-weighted velocity of each of the 
above components under the assumption that the outflows are in the plane of the disk 
we can estimate the outflow momentum in the unresolved nucleus, obtaining 
$\rm (-10.5\pm1.5)\times10^{6}~[M_{\odot}$ \kms] and $\rm (6.6\pm0.9)\times10^{6}~[M_{\odot}$ \kms] 
for the blue and red outflows in the inner (unresolved; $\lesssim$\FPeval\fpresult{round(0.4*\scaleimg:1)}\fpresult~pc) nucleus, respectively. 

Since the outflows extend beyond the inner (unresolved) nucleus we can 
extend this outflow mass analysis to larger scales. From the bottom right panel of 
Fig.~\ref{pv1fig}, we can distinguish that the putative outflows along the minor axis show two stages:  
(1) an initial stage between radii of 0 to $\sim$1\arcsec\ which show velocities close to the
peak velocity, and 
(2) a final stage with monotonically decreasing velocities (down to $\sim$0) between 1\arcsec\ 
and 1\farcs5 from the nucleus. 
For these two stages (but excluding the innermost 0\farcs2 considerated as the nuclear outflows above)
we obtain
total outflow mass values of $\rm (9.3\pm0.9)\times 10^{6} [M_{\odot}]$ and $\rm (11.6\pm1.2)\times 10^{6} [M_{\odot}]$, 
respectively. 

\begin{figure}
\centering            
  \includegraphics[bb=35 120 375 455,width=0.22\textwidth,clip]{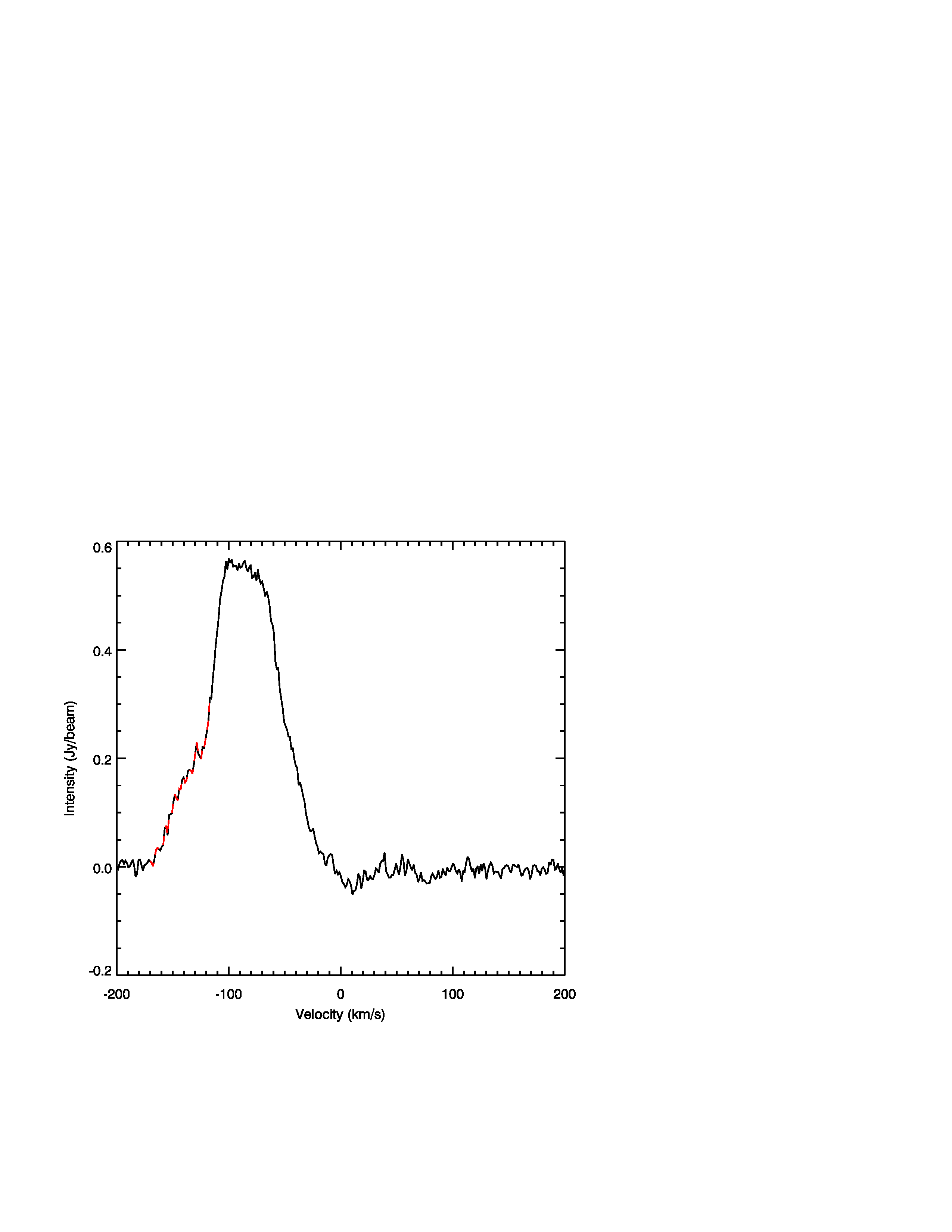}
  \includegraphics[bb=35 120 375 455,width=0.22\textwidth,clip]{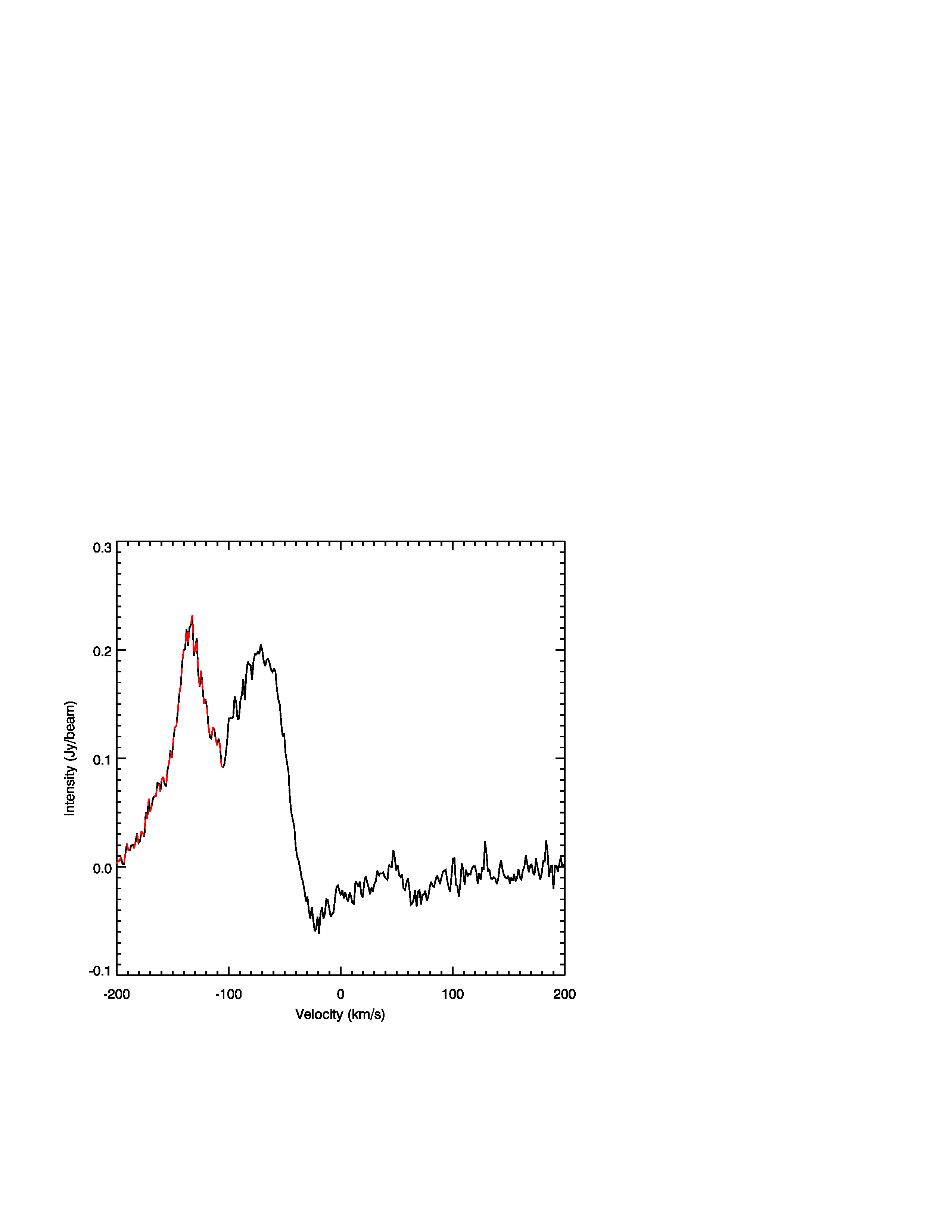} 
  \caption{CO J:2-1 spectra extracted from  apertures 1 and 2 of 
Fig.~\ref{aperturesfig} (the E inner spiral arm) are shown in black. 
The red overlay marks the velocities over which emission comes from gas potentially participating in streaming
inflows. 
  }
  \label{inflowspectfigs2}
\end{figure}

In the absence of clear signatures of inflowing gas we do not attempt to estimate 
an exhaustive streaming inflow rate. Instead we show two examples of spectra in
apertures for which we found the strongest signatures of non circular motions which
could be explained by our toy streaming model (see Sect.~\ref{streamsect})
and use these to roughly estimate the mass involved in the inflow. 
In the E inner spiral we use the spectra of apertures 1 and 2 of Fig.~\ref{aperturesfig}.
These spectra are shown in Fig.~\ref{inflowspectfigs2}, with the red overlay denoting
the velocities over which the emission is from gas potentially participating in 
the streaming inflows. For each of these apertures we find
masses of $\sim(1.1\pm0.4)\times10^{5}~[M_{\odot}]$ potentially participating in 
a streaming inflow. 
We performed the same exercise with four apertures in the outer extension of the E spiral arm 
(not shown) and find a median mass of $\sim(4\pm0.4)\times10^{4}~[M_{\odot}]$ potentially
participating in a streaming inflow. 

\subsection{Mass of the SMBH in \galaxy}
\label{massbhsec} 
To estimate the black hole mass in \galaxy\ , we use the empirical correlation between 
SMBH mass and stellar velocity dispersion $\sigma_{\star}$ 
\citep{ferrandmerr2000,gebhardt2000,tremaine2002,gultekin2009,korandho2013}.
Note that this  M-$\sigma_{\star}$ relationship can also be reliably applied to nearby AGNs
\citep{nelson2004,woo2010,graham2011}.
While there are several versions of the M-$\sigma_{\star}$ relation we use that
of \citet{gultekin2009} for their entire sample for several reasons; they demonstrate that previous studies are biased 
by considering culled samples according whether the SMBH sphere of influence is resolved (see Sect.~4 of their paper). Therefore we get:
\begin{equation}
 \label{tremaineeq}
\left( \frac{M_{BH}}{M_{\odot}}\right)=10^{(8.12\pm0.08)} \cdot \left( \frac{\sigma}{200}\right) ^{(4.24\pm0.41)}
\end{equation}
where $\sigma$, is the central velocity dispersion in \kms. To obtain the latter, 
we fit the integrated spectrum from our GMOS/IFU datacube with the 
program \textit{pPXF} described in \citet{capandems2004}.
The spectrum from 5700\AA\ 
to 6300\AA\ was fit in \textit{pPXF} using SSPs templates 
from \citet{bruandchar2003},  obtaining
a central stellar velocity dispersion of $\sigma_{\star}=116\pm9$ \kms. 
This value is consistent with that obtained using $\sigma_{\star}$ 
from \citet{bottema92} for the innermost region in \galaxy, and larger 
than the bulge stellar velocity dispersion used in other previous studies 
\citep{vankruandfree84,nelandwhi95,wooandurr2002}. 

Our measured value of $\sigma$ implies an estimated
super massive black hole mass of $M_{BH}=1.3\pm0.6\times10^{7}~[M_{\odot}]$.

\subsection{Bolometric luminosity and accretion, inflow, and outflow rates}
\label{lumandmassratessect}
Given the SMBH mass estimated above, for which the 
Eddington Luminosity is $\rm L_{Edd}=(4.0\pm1.8)\times10^{11}~[L_{\odot}]$,
we would like to ascertain the bolometric luminosity of the AGN $\rm L_{Bol}$ 
as well the 
Eddington ratio ($\rm l_{Edd}$ = $\rm \frac{L_{Bol}}{L_{Edd}})$. 

Several differing values for $\rm L_{Bol}, $ for \galaxy\, have been obtained in previous studies.
Here we use two different approaches to estimate $L_{\rm Bol}$: from the nuclear \oiii\ luminosity, 
and from the nuclear hard X-ray luminosity. 
The nuclear 2--10~keV luminosity as measured by \textit{XMM} 
and scaled to our adopted distance (\distangal\ Mpc) is $\rm L_{X}=(7\pm3)\times10^{33}~[W]$ 
\citep{levenson2009}.
Using the hard X-ray to bolometric luminosity conversion of \citet{ulandho2001} 
($\rm L_{Bol}=6.7\times L_{X(2-10keV)}$) we obtain $\rm L_{Bol} = (4.69\pm2)\times10^{34}~[W]$.
The nuclear \oiii\ flux \citep{moustakas2010} at our adopted distance
gives $\rm L_{\oiii}=(8.2\pm1.8)\times10^{31}~[W]$.
Using the scaling of \citet{heckman2004} modified as recommended  in \citet{dumas2007} ($\rm L_{Bol}=90\times L_{\oiii}$) we obtain
$\rm L_{Bol} = (7.42\pm1.62)\times10^{33}~[W]$.
These two methods thus give relatively consistent values for $\rm L_{Bol}$ and we thus adopt the mean 
value of  $\rm L_{Bol} = (2.7\pm1.3)\times10^{34}~[W]$ which implies 
$\rm l_{Edd}$ $\approx2.0\times10^{-4}$, i.e. a relatively low 
efficiency regime for the SMBH, considering the fact that in the most active galaxies, gas is 
accreted onto the SMBH in a efficient regime with ratios between 0.01-1 \citep{khorunzhev2012}. 
We can now estimate the mass accretion rates as follows:
\begin{equation}
\label{accrateeq}
 \dot{m}=\frac{L_{Bol}}{c^{2}\eta}
\end{equation}
where $\eta$ is the accretion efficiency which in nearby galaxies with 
geometrically thin, optically thick accretion disks, is typically taken
to be 0.1 \citep{soltan82,fabandiwa99,yuandtre2002,davandlao2011}. 
We thus get an accretion rate of $\dot{m}=(4.8\pm2.3)\times10^{-5}~[M_{\odot}yr^{-1}]$. 

We note that very discrepant values of 
$L_{Bol}$ were obtained by \citet{wooandurr2002} who integrated the flux in the 
spectral energy distribution (SED) of \galaxy\ using data from NED, after averaging
multiple datapoints in the same band and correcting for dust. 
The value of $\rm L_{Bol}$ they obtained, scaled to our adopted distance, is 
L$_{Bol}$ = $7.2\times10^{36}~[W]$. This implies $\rm l_{Edd}=0.05$  and an accretion rate of
$\rm \dot{m}=1.3\times10^{-2}~[M_{\odot}yr^{-1}]$, 3 orders of magnitude higher than the values 
obtained above. 
Two potential reasons for this large discrepancy are 
(a) flux variability in the hard X-ray, see e.g. \citet{landi2005};
 (b) the use of large (galaxy-wide), rather than nuclear, apertures for the data in NED.
In Sect.~\ref{molmasssect} we roughly estimated the mass of the molecular gas potentially participating in 
streaming inflows using six apertures along the SSE to E spiral arm. This can be used
to estimate a rough mass inflow rate. 
The radius of each aperture (0\farcs2) corresponds to a linear diameter of 
\FPeval\Fpresult{round(0.4*\scaleimg:0)}\Fpresult~pc. 
A streaming inflow velocity of $\sim50$\kms\ would imply an aperture
crossing time of   $\rm \sim3.8\times10^{5}~[yr]$, and thus a mass inflow rate of
 $\rm 0.1~[M_{\odot}yr^{-1}]$ along this spiral arm. 

A similar procedure can be used to estimate the outflow rates. 
We use the molecular masses deduced for the blue and red components of the outflows in the nuclear 
(0\farcs2 radius) aperture (see Sect.~\ref{molmasssect} and Fig.~\ref{nuclearspectfig}).
In Sect.~\ref{outflowsect} 
we have argued that the velocity of the outflows in the nucleus is $\sim180$\kms\ in the plane of the disk,
which gives a crossing time of around $5.42\times10^{4}~[yr]$ for the nuclear
0\farcs2 aperture used. 
The nuclear molecular mass outflow rates are thus 
$\rm 3.7~[M_{\odot}yr^{-1}]$ for the blueshifted component and $1.9~[M_{\odot}yr^{-1}]$ for the redshifted component. Note that these outflows do not appear to escape from the nucleus, but instead
`pile up' in the inner gas ring (see the CO residual map and pv diagrams).
We can also estimate the kinetic power associated with 
cold molecular outflow following, e.g., \citet{harrison2014} and \citet{lenaphd2015}: 
\begin{equation}
\label{kinpoweq}
\dot{E}_{out}=\frac{\dot{M}_{out}}{2}(v_{outflow}^{2}+3\sigma^{2}).
\end{equation}
 Assuming a nuclear velocity dispersion of 60\kms (see Fig.~\ref{velreswithfluxfigs})
 and, as mentioned above, $\rm v_{outflow}=180$\kms, we obtain a total outflow kinetic 
 power (over both blue and red outflow components) of 
 $\rm \dot{E}_{out}=7.62\times10^{33}~[W]$. The 
 ratio between the outflow kinetic power and the AGN bolometric luminosity  is thus
 $\frac{\dot{E}_{out}}{L_{Bol}}\approx0.28$. 
 This is significantly larger than the values obtained in previous studies of
 nearby active galaxies which found $\frac{\dot{E}_{out}}{L_{Bol}}$ ranging between 0.1-10\% 
 \citep[e.g.][]{storchi2010,mullersan2011,harrison2014,lenaphd2015,mullersan2016} 
 or even lower \citep[e.g.][less than 0.01\%]{barbosa2009}. However all of these studies have 
 measured outflow masses using ionized gas, which is expected to be a minor fraction of the 
 total gas content and also, using measurements on larger spatial scales than those considered
 here (0\farcs4 or $\sim$24 pc).


\section{Summary and Conclusions}
\label{concsect}
We have analyzed the kinematics in the inner kiloparsec of the nearby active galaxy 
\galaxy, using ALMA observations of \co2-1\ along with GMOS/IFU data of 
ionized gas emission lines and stellar absorption lines. 
Our results allow us to conclude that:
\begin{itemize}
 \item \galaxy\ presents a cold molecular dense disk in the inner 144 pc, along with a clear two-arm spiral structure in the inner 96 pc. These structures 
 are also seen in our ionized gas (\nii) images, as well as in previous studies 
 using \coo3-2\ and optical/IR emission lines. The inner spiral arms and dense disk 
 appear to have some continuity 
 with more extended \co2-1\ spiral arms which extend out of the
 inner disk to larger scales, and which coincide with dust lanes seen in HST images. 
\item Ionized gas and stars are detected over the full IFU FOV at high signal to noise.
  The \nii\ emission is peaked at the nucleus and is also strong at the known star-forming
  region to the SW, about 1\farcs5 from the nucleus.
\item The superior 
spectral resolution ($\sim$2.6\kms) and image fidelity in this new
\co2-1\ datacube allows  improved constraints on the nuclear kinematics. Further,
our use of pv diagrams (rather than only intensity weighted velocity maps) allows the full exploitation
of our velocity resolution.
The molecular gas kinematics of the inner disk is dominated by rotation 
with peak velocities of $\sim\pm$140\kms. The residual velocity field shows clear signs
of non rotational motion especially in the innermost $\sim$2\arcsec\ region. 
We argue that the strongest deviations are the result 
of nuclear outflows, though the strong two-arm inner spiral structure and large-scale
bar also play a role.  
The CO rotational curve over the inner $\sim$3\arcsec\ is 
asymmetric, which could be the consequence of a warped disk. 
PV diagrams at PAs between $-$15\degr\ to 15\degr\ show some 
discrepancies from our rotation + outflow model at radii $\sim$1\farcs5 from the nucleus.
These discrepancies could be explained by perturbations due to the barred potential and/or streaming velocities. 
\item We argue for the presence of nuclear (inner $\sim$2\arcsec) 
molecular gas outflows in the
plane of the galaxy disk, without ruling out the presence of bar- or spiral-related perturbations (see Sect.~\ref{barpersect} for a detailed analysis of the potential bar influence). The arguments for a nuclear outflow include:
(a) several previous authors have claimed kinematic and morphological evidence of outflows in the NLR, which likely intersect the galaxy disk given the observed geometries; 
(b) The nucleus shows a double-peaked profile with 
FWZI 200\kms, higher than that seen in the lower fidelity maps of 
C14, and which imply inclination corrected outflow velocities of 
up to 180\kms\ in the plane of the disk. A large angle between the outflow axis and the line of sight is unlikely since this would imply extremely high true outflow velocities. Attributing these velocity features to other  perturbations in the plane of the disk and along the minor axis requires true radial velocities around $\sim$80--130\kms in the nuclear region where rotation velocities are expected to be $\leq$40\kms. In Sect.~\ref{barpersect} we analysed the large scale bar perturbation and its implications, showing that it does not reliably produce both the morphology and the large perturbations seen in the observed velocity field; 
(c) the pv diagram along the minor axis (bottom right panel of Fig.~\ref{pv1fig}) 
connects high-velocity components to the zero velocity components seen at 
r $\approx$1\farcs8 on both sides of the nucleus. The regions with the largest velocity 
deceleration are correlated with the regions brightest in CO and thus richest in molecular
gas, indicating that the outflows decelerate due to mass loading which is clearly seen in the inner arcsec to the N and in the r$\sim$ 1--2\arcsec\ range to the S; 
(d) the pv diagrams (Fig.~\ref{pv1fig}) show velocity deviations which are consistent with radial outflows not just in the minor axis, but also in the plane of the disk over all PAs and over apertures at distances of several synthesized beams from the nucleus (see Sect.~\ref{pvsect});
and 
(e) the consistence with the evidence of a nuclear spherical (or bipolar) outflow in our ionized gas kinematics.
We also argue that the molecular outflow is primarily detected within the galaxy disk. The supporting arguments
for this include the low velocity dispersion of the molecular gas, the absence of evidence for radio jet-related outflows, 
the posited deceleration of the outflow, which would be consistent with the high density of gas in the galaxy disk. 
Other potential scenarios as warped disk or non coplanar disk cannot be constrained by us due to the limited resolution, the sparse velocity field and the lack of a reliable circular rotation model. 

\item While the stellar and ionized gas kinematics predominantly traces rotation in the galaxy
disk, we find evidence of two components in the ionized gas: a narrow ($\sigma \sim$60\kms)
component detected over almost the full field of view and which traces rotation in the disk,
and a broad component detected in the inner $\sim$3\arcsec radius. We postulate that the 
broad component of ionized gas is part of a nuclear spherical outflow. 
The broad component of the \nii\ emission line does not participate in the rotation of
the galaxy disk. Instead it is preferentially blueshifted, as expected
since nuclear dust obscures the receding side of the spherical outflow. Its velocity dispersion,
velocity, and the correlation with strong and weaker dust features in the nucleus all
strengthen the arguments for the spherical ionized gas outflow. 
Despite this, we are not ruling out the possibility of a bipolar outflow 
which currently is impeded by a poor resolution in the GMOS data. We are unable to test for
deceleration in this outflow. 

\item We have constrained and analyzed the circular vs. perturbed kinematics of the CO gas in the disk using \kinemetry, \textit{Diskfit}, and linear perturbation theory. Large radial velocity perturbations are clearly required. The \textit{Diskfit} model better fits the observed data as compared to the pure rotation model, but still does not attain the highest velocities seen in the inner arcsecs and fails to predict the large peak velocities seen along the minor axis in pv diagram (bottom middle panel in Fig.~\ref{pvbarmodfig}). The relatively good fits
obtained by \textit{Diskfit} do not necessarily imply that the perturbations are bar related. Our concerns here include the lack of 
definition of the physical parameters (e.g., pattern speed) used, and our findings that \textit{Diskfit} typically provides good
fits to perturbed kinematics of other similar datasets, even in the absence of a bar.
For our linear epicyclic perturbation (in the presence of a bar or m=2 mode) modelling, which provides a more physically-based model as 
compared to \textit{Diskfit}, we used two input rotation models, one based on the best-fit gas rotation model and one on the 
best-fit stellar rotation model. In both cases we varied the bar pattern speed and damping factor. 
While the perturbed velocity fields show the characteristic butterfly pattern  expected  from bar perturbations,
very high pattern speeds ($\gtrsim$ 300\kms\ kpc$^{-1}$) are required to cause a resonance close to r$\approx$1\arcsec\ and thus explain the observed high-velocity features along the minor axis. We note however that if the
true distance to  \galaxy\ is higher, the problem of high pattern speeds is mitigated. Even if the resonance radii
are matched, the velocity structures seen in the models are not aligned with or as sharp as those observed. 

\item We are unable to definitely prove the existence of streaming inflows based on the kinematics
of the molecular or ionized gas. This is in part due to the  line profiles
being complex with evidence of multiple velocity components, and also due to an asymmetry in the velocity 
profiles of the near and far sides of the disk, which could be
due to disk warping or the non-axisymmetric potential. We present and analyze specific apertures along
a spiral arm where the spectral profiles are similar to those
expected from our toy streaming inflow model and from these we estimate a potential
streaming inflow rate of $\rm 0.1~[M_{\odot}yr^{-1}]$.

\item We estimate the molecular mass of the unresolved nuclear outflow
(the innermost 0\farcs2 aperture)
as $(3.8\pm0.4)\times10^{5}~[M_{\odot}]$ and its momentum as 
$\rm (10.5\pm1.5)\times10^{6}~[M_{\odot}$ \kms] and 
$\rm (6.6\pm0.9)\times10^{6}~[M_{\odot}$ \kms] for the blue 
and red outflows, respectively. Summing all gas 
believed to be participating in the nuclear outflow 
(out to $\sim1\farcs5$ or $\sim$72 pc from the nucleus) we find a mass of 
$2.1\times10^{7}~[M_{\odot}]$. 
Given the nuclear velocities, the implied outflow rates 
from the nuclear 0\farcs2 region are $\rm 3.7~[M_{\odot}yr^{-1}]$ 
for the blueshifted component and $\rm 1.9~[M_{\odot}yr^{-1}]$ for the redshifted component.   
We emphasize that these outflows appear to decelerate within the inner 100~pc and thus
the gas is not lost to the galaxy nucleus.

\item We have used the results of three methods to estimate $\rm L_{Bol}$: from the nuclear \oiii\ 
luminosity, from the nuclear hard X-ray luminosity, and from a SED fit. The first two methods give
consistent results: a mean value of $\rm L_{Bol} = (2.7\pm1.3)\times10^{34}~[W]$, which implies
$\rm l_{Edd}$ of $\sim2.2\times10^{-4}$, indicating a relatively low efficiency regime 
for the SMBH, and an accretion rate of $\dot{m}=(4.8\pm2.3)\times10^{-5}~[M_{\odot}yr^{-1}]$, 
significantly smaller than the posited nuclear molecular outflow rate. 

\item A direct comparison between the molecular outflow kinetic power and the AGN bolometric luminosity gives a 
ratio of $\frac{\dot{E}_{out}}{L_{Bol}}\approx0.28$, a value  significantly larger than typical values
(between 0.01--10\%) found from previous studies ionized gas in nearby AGNs. This supports the idea that 
the ionized gas is a minor fraction of the total gas content.

\end{itemize}

\begin{acknowledgements}
We thank to the anonymous referee for the very constructive comments that made to improve this work.
 This paper makes use of the following ALMA data: ADS/JAO.ALMA\#2012.1.00474.S . ALMA is a partnership of ESO (representing its member states), NSF (USA) and NINS (Japan), together with NRC (Canada), NSC and ASIAA (Taiwan), and KASI (Republic of Korea), in cooperation with the Republic of Chile. The Joint ALMA Observatory is operated by ESO, AUI/NRAO and NAOJ. Based on observations obtained at the Gemini Observatory, which is operated by the Association of Universities for Research in Astronomy, Inc., under a cooperative agreement with the NSF on behalf of the Gemini partnership: the National Science Foundation (United States), the National Research Council (Canada), CONICYT (Chile), Ministerio de Ciencia, Tecnolog\'{i}a e Innovaci\'{o}n Productiva (Argentina), and Minist\'{e}rio da Ci\^{e}ncia, Tecnologia e Inova\c{c}\~{a}o (Brazil). This is research has made use of the services of the ESO Science Archive Facility. 
{\it N.N.} acknowledges support from BASAL-PFB/06 and extension, Fondecyt 1171506, and Anillo ACT172033.
{\it R.A.R.} acknowledges support from FAPERGS (project N0. 2366-2551/14-0) and CNPq (project N0. 470090/2013-8 and 302683/2013-5). We acknowledge the usage of the HyperLeda database (http://leda.univ-lyon1.fr). This research has made use of the NASA/IPAC Extragalactic Database (NED), which is operated by the Jet Propulsion Laboratory, California Institute of Technology, under contract with the National Aeronautics and Space Administration. IRAF is distributed by the National Optical Astronomy Observatory, which is operated by the Association of Universities for Research in Astronomy (AURA) under a cooperative agreement with the National Science Foundation. {\it G.O.} acknowledges the support provided by CONICYT(Chile) through FONDECYT postdoctoral research grant no 3170942.{\it R.S.} acknoledges support from CONICYT Beca/Nacional-Doctorado 21120516. 
\end{acknowledgements}
%

\begin{thebibliography}{dummy}
 \bibitem[Ag{\"u}ero et al.(2004)]{aguero2004} Ag{\"u}ero, E.~L., D{\'{\i}}az, R.~J., \& Bajaja, E.\ 2004, \aap, 414, 453 

 \bibitem[Alloin et al.(1985)]{alloin85} Alloin, D., Pelat, D., Phillips, M., \& Whittle, M.\ 1985, \apj, 288, 205 

 \bibitem[Bajaja et al.(1995)]{bajaja95} Bajaja, E., Wielebinski, R., Reuter, H.-P., Harnett, J.~I., \& Hummel, E.\ 1995, \aaps, 114, 147 

 \bibitem[Barbosa et al.(2009)]{barbosa2009} Barbosa, F.~K.~B., Storchi-Bergmann, T., Cid Fernandes, R., Winge, C., \& Schmitt, H.\ 2009, \mnras, 396, 2

 \bibitem[Barbosa et al.(2014)]{barbosa2014} Barbosa, F.~K.~B., Storchi-Bergmann, T., McGregor, P., Vale, T.~B., \& Rogemar Riffel, A.\ 2014, \mnras, 445, 2353 

\bibitem[Bertola et al.(1991)]{bertola91} Bertola, F., Bettoni, D., Danziger, J., et al.\ 1991, \apj, 373, 369 

 \bibitem[Bolatto et al.(2013)]{bolatto2013} Bolatto, A.~D., Wolfire, M., \& Leroy, A.~K.\ 2013, \araa, 51, 207 

 \bibitem[Bottema(1992)]{bottema92} Bottema, R.\ 1992, \aap, 257, 69 

 \bibitem[Bruzual \& Charlot(2003)]{bruandchar2003} Bruzual, G., \& Charlot, S.\ 2003, \mnras, 344, 1000 

 \bibitem[Cappellari \& Emsellem(2004)]{capandems2004} Cappellari, M., \& Emsellem, E.\ 2004, \pasp, 116, 138 

\bibitem[Cicone et al.(2014)]{cicone2014} Cicone, C., Maiolino, R., Sturm, E., et al.\ 2014, \aap, 562, A21 

 \bibitem[Combes et al.(2014)]{combes2014} Combes, F., Garc{\'{\i}}a-Burillo, S., Casasola, V., et al.\ 2014, \aap, 565, A97 

\bibitem[Comer{\'o}n et al.(2010)]{comeron2010} Comer{\'o}n, S., Knapen, J.~H., Beckman, J.~E., et al.\ 2010, \mnras, 402, 2462 

 \bibitem[Crenshaw et al.(2003)]{crenshaw2003} Crenshaw, D.~M., Kraemer, S.~B., \& Gabel, J.~R.\ 2003, \aj, 126, 1690 

\bibitem[da Silva et al.(2017)]{dasilva2017} da Silva, P., Steiner, J.~E., \& Menezes, R.~B.\ 2017, \mnras, 470, 3850 

\bibitem[Davies et al.(2016)]{davies2016} Davies, R.~L., Dopita, M.~A., Kewley, L., et al.\ 2016, \apj, 824, 50 

 \bibitem[Davis \& Laor(2011)]{davandlao2011} Davis, S.~W., \& Laor, A.\ 2011, \apj, 728, 98 

 \bibitem[de Vaucouleurs(1973)]{vauco73} de Vaucouleurs, G.\ 1973, \apj, 181, 31 

 \bibitem[Dicaire et al.(2008)]{dicaire2008} Dicaire, I., Carignan, C., Amram, P., et al.\ 2008, \mnras, 385, 553 

 \bibitem[Diniz et al.(2015)]{diniz2015} Diniz, M.~R., Riffel, R.~A., Storchi-Bergmann, T., \& Winge, C.\ 2015, \mnras, 453, 1727

 \bibitem[Dumas et al.(2007)]{dumas2007} Dumas, G., Mundell, C.~G., Emsellem, E., \& Nagar, N.~M.\ 2007, \mnras, 379, 1249 

 \bibitem[Ehle et al.(1996)]{ehle96} Ehle, M., Beck, R., Haynes, R.~F., et al.\ 1996, \aap, 306, 73 

 \bibitem[Elvis et al.(1989)]{elvis89} Elvis, M., Fassnacht, C., Wilson, A.~S., \& Briel, U.\ 1989, European Southern Observatory Conference and Workshop Proceedings, 32, 243 

 \bibitem[Erwin(2004)]{erwin2004} Erwin, P.\ 2004, \aap, 415, 941 

 \bibitem[Fabian \& Iwasawa(1999)]{fabandiwa99} Fabian, A.~C., \& Iwasawa, K.\ 1999, \mnras, 303, L34 

\bibitem[Fathi(2004)]{fathiphd2004} Fathi, K.\ 2004, Ph.D.~Thesis,  

\bibitem[Fathi et al.(2005)]{fathi2005} Fathi, K., van de Ven, G., Peletier, R.~F., et al.\ 2005, \mnras, 364, 773

 \bibitem[Ferrarese \& Ford(2005)]{ferrandford2005} Ferrarese, L., \& Ford, H.\ 2005, \ssr, 116, 523 

 \bibitem[Ferrarese \& Merritt(2000)]{ferrandmerr2000} Ferrarese, L., \& Merritt, D.\ 2000, \apjl, 539, L9 

 \bibitem[Finlez et al.(in prep.)]{finlezphd2017} Finlez, C., Nagar, N., and others\ 2018, PhD thesis work. 

\bibitem[Franx et al.(1994)]{franx94} Franx, M., van Gorkom, J.~H., \& de Zeeuw, T.\ 1994, \apj, 436, 642 

 \bibitem[Garc{\'{\i}}a-Burillo et al.(2014)]{gar-bur2014} Garc{\'{\i}}a-Burillo, S., Combes, F., Usero, A., et al.\ 2014, \aap, 567, A125 

 \bibitem[Gebhardt et al.(2000)]{gebhardt2000} Gebhardt, K., Bender, R., Bower, G., et al.\ 2000, \apjl, 539, L13 

 \bibitem[Graham et al.(2011)]{graham2011} Graham, A.~W., Onken, C.~A., Athanassoula, E., \& Combes, F.\ 2011, \mnras, 412, 2211 

 \bibitem[Greenhill et al.(2003)]{greenhill2003} Greenhill, L.~J., Booth, R.~S., Ellingsen, S.~P., et al.\ 2003, \apj, 590, 162 

\bibitem[Gruppioni et al.(2016)]{gruppioni2016} Gruppioni, C., Berta, S., Spinoglio, L., et al.\ 2016, \mnras, 458, 4297 

 \bibitem[G{\"u}ltekin et al.(2009)]{gultekin2009} G{\"u}ltekin, K., Richstone, D.~O., Gebhardt, K., et al.\ 2009, \apj, 698, 198 

\bibitem[Hackwell \& Schweizer(1983)]{hackandschwei83} Hackwell, J.~A., \& Schweizer, F.\ 1983, \apj, 265, 643 

 \bibitem[Harrison et al.(2014)]{harrison2014} Harrison, C.~M., Alexander, D.~M., Mullaney, J.~R., \& Swinbank, A.~M.\ 2014, \mnras, 441, 3306 

 \bibitem[Heckman \& Best(2014)]{heckandbest2014} Heckman, T.~M., \& Best, P.~N.\ 2014, \araa, 52, 589 

 \bibitem[Heckman et al.(2004)]{heckman2004} Heckman, T.~M., Kauffmann, G., Brinchmann, J., et al.\ 2004, \apj, 613, 109 

\bibitem[Hollyhead et al.(2016)]{hollyhead2016} Hollyhead, K., Adamo, A., Bastian, N., Gieles, M., \& Ryon, J.~E.\ 2016, \mnras, 460, 2087 

 \bibitem[Kawamuro et al.(2013)]{kawamuro2013} Kawamuro, T., Ueda, Y., Tazaki, F., \& Terashima, Y.\ 2013, \apj, 770, 157

 \bibitem[Kendall et al.(2011)]{kendall2011} Kendall, S., Kennicutt, R.~C., \& Clarke, C.\ 2011, \mnras, 414, 538 

 \bibitem[Khorunzhev et al.(2012)]{khorunzhev2012} Khorunzhev, G.~A., Sazonov, S.~Y., Burenin, R.~A., \& Tkachenko, A.~Y.\ 2012, Astronomy Letters, 38, 475 

 \bibitem[Kilborn et al.(2005)]{kilborn2005} Kilborn, V.~A., Koribalski, B.~S., Forbes, D.~A., Barnes, D.~G., \& Musgrave, R.~C.\ 2005, \mnras, 356, 77

 \bibitem[Korchagin et al.(2000)]{korchagin2000} Korchagin, V., Kikuchi, N., Miyama, S.~M., Orlova, N., \& Peterson, B.~A.\ 2000, \apj, 541, 565 

 \bibitem[Kormendy \& Ho(2013)]{korandho2013} Kormendy, J., \& Ho, L.~C.\ 2013, \araa, 51, 511 

 \bibitem[Krajnovi{\'c} et al.(2006)]{kraj2006} Krajnovi{\'c}, D., Cappellari, M., de Zeeuw, P.~T., \& Copin, Y.\ 2006, \mnras, 366, 787 

 \bibitem[Landi et al.(2005)]{landi2005} Landi, R., Malizia, A., \& Bassani, L.\ 2005, \aap, 441, 69 

 \bibitem[Lena et al.(2015)]{lena2015} Lena, D., Robinson, A., Storchi-Bergman, T., et al.\ 2015, \apj, 806, 84

 \bibitem[Lena(2015)]{lenaphd2015} Lena, D.\ 2015, Ph.D.~Thesis,  

\bibitem[Leroy et al.(2015)]{leroy2015} Leroy, A.~K., Walter, F., Martini, P., et al.\ 2015, \apj, 814, 83 

 \bibitem[Levenson et al.(2009)]{levenson2009} Levenson, N.~A., Radomski, J.~T., Packham, C., et al.\ 2009, \apj, 703, 390 

 \bibitem[Liu et al.(2013)]{liu2013} Liu, G., Zakamska, N.~L., Greene, J.~E., Nesvadba, N.~P.~H., \& Liu, X.\ 2013, \mnras, 436, 2576 

 \bibitem[Ma(2001)]{ma2001} Ma, J.\ 2001, \cjaa, 1,  

\bibitem[Makarov et al.(2014)]{LEDAIII} Makarov D., Prugniel P., Terekhova N., Courtois H., \& Vauglin I.\ 2014, \aap, 570, A13 

 \bibitem[Malkan et al.(1998)]{malkan98} Malkan, M.~A., Gorjian, V., \& Tam, R.\ 1998, \apjs, 117, 25 

\bibitem[McGaugh \& Schombert(2014)]{mcgaugh2014} McGaugh, S.~S., \& Schombert, J.~M.\ 2014, \aj, 148, 77 

 \bibitem[McMullin et al.(2007)]{casa} McMullin, J.~P., Waters, B., Schiebel, D., Young, W., \& Golap, K.\ 2007, Astronomical Data Analysis Software and Systems XVI, 376, 127  

 \bibitem[Mezcua et al.(2015)]{mezcua2015} Mezcua, M., Prieto, M.~A., Fern{\'a}ndez-Ontiveros, J.~A., et al.\ 2015, \mnras, 452, 4128 

 \bibitem[Morganti et al.(1999)]{morganti99} Morganti, R., Tsvetanov, Z.~I., Gallimore, J., \& Allen, M.~G.\ 1999, \aaps, 137, 457 

 \bibitem[Morganti et al.(2009)]{morganti2009} Morganti, R., Peck, A.~B., Oosterloo, T.~A., et al.\ 2009, \aap, 505, 559 

\bibitem[Morganti et al.(2013)]{morganti2013} Morganti, R., Frieswijk, W., Oonk, R.~J.~B., Oosterloo, T., \& Tadhunter, C.\ 2013, \aap, 552, L4 

 \bibitem[Moustakas et al.(2010)]{moustakas2010} Moustakas, J., Kennicutt, R.~C., Jr., Tremonti, C.~A., et al.\ 2010, \apjs, 190, 233-266 

\bibitem[Mulchaey et al.(1997)]{mulchaey97} Mulchaey, J.~S., Regan, M.~W., \& Kundu, A.\ 1997, \apjs, 110, 299 

 \bibitem[M{\"u}ller-S{\'a}nchez et al.(2011)]{mullersan2011} M{\"u}ller-S{\'a}nchez, F., Prieto, M.~A., Hicks, E.~K.~S., et al.\ 2011, \apj, 739, 69 

 \bibitem[M{\"u}ller-S{\'a}nchez et al.(2016)]{mullersan2016} M{\"u}ller-S{\'a}nchez, F., Comerford, J.~M., Stern, D., \& Harrison, F.~A.\ 2016, arXiv:1606.07446

 \bibitem[Nelson et al.(2004)]{nelson2004} Nelson, C.~H., Green, R.~F., Bower, G., Gebhardt, K., \& Weistrop, D.\ 2004, \apj, 615, 652

 \bibitem[Nelson \& Whittle(1995)]{nelandwhi95} Nelson, C.~H., \& Whittle, M.\ 1995, \apjs, 99, 67

 \bibitem[Nesvadba et al.(2010)]{nesvadba2010} Nesvadba, N.~P.~H., Boulanger, F., Salom{\'e}, P., et al.\ 2010, \aap, 521, A65

 \bibitem[Pence et al.(1990)]{pence90} Pence, W.~D., Taylor, K., \& Atherton, P.\ 1990, \apj, 357, 415 

 \bibitem[Ramakrishnan et al. (in prep.)]{venki2017} Ramakrishnan, V., Nagar, N., and others\ 2018. 

 \bibitem[Regan \& Teuben(2004)]{regandteub2004} Regan, M.~W., \& Teuben, P.~J.\ 2004, \apj, 600, 595 

 \bibitem[Riffel(2010)]{riffel2010} Riffel, R.~A.\ 2010, \apss, 327, 239 

 \bibitem[Riffel et al.(2013)]{riffel2013} Riffel, R.~A., Storchi-Bergmann, T., \& Winge, C.\ 2013, \mnras, 430, 2249

\bibitem[Roche et al.(2016)]{roche2016} Roche, N., Humphrey, A., Lagos, P., et al.\ 2016, \mnras, 459, 4259

 \bibitem[Roy et al.(1994)]{roy94} Roy, A.~L., Norris, R.~P., Kesteven, M.~J., Troup, E.~R., \& Reynolds, J.~E.\ 1994, \apj, 432, 496 

 \bibitem[Sakamoto et al.(1999)]{sakamoto99} Sakamoto, K., Okumura, S.~K., Ishizuki, S., \& Scoville, N.~Z.\ 1999, \apj, 525, 691 

\bibitem[Sakamoto et al.(2014)]{sakamoto2014} Sakamoto, K., Aalto, S., Combes, F., Evans, A., \& Peck, A.\ 2014, \apj, 797, 90 

\bibitem[Schinnerer et al.(2000)]{schin2000} Schinnerer, E., Eckart, A., Tacconi, L.~J., Genzel, R., \& Downes, D.\ 2000, \apj, 533, 850 

 \bibitem[Schmitt \& Kinney(1996)]{schm-kinn96} Schmitt, H.~R., \& Kinney, A.~L.\ 1996, \apj, 463, 498 

 \bibitem[Schnorr-M{\"u}ller et al.(2014b)]{schnorr2014b} Schnorr-M{\"u}ller, A., Storchi-Bergmann, T., Nagar, N.~M., et al.\ 2014, \mnras, 437, 1708 

 \bibitem[Schnorr-M{\"u}ller et al.(2014a)]{schnorr2014a} Schnorr-M{\"u}ller, A., Storchi-Bergmann, T., Nagar, N.~M., \& Ferrari, F.\ 2014, \mnras, 438, 3322 

\bibitem[Schnorr-M{\"u}ller et al.(2017b)]{schnorr2017b} Schnorr-M{\"u}ller, A., Storchi-Bergmann, T., Ferrari, F., \& Nagar, N.~M.\ 2017, \mnras, 466, 4370 

\bibitem[Schnorr-M{\"u}ller et al.(2017a)]{schnorr2017a} Schnorr-M{\"u}ller, A., Storchi-Bergmann, T., Nagar, N.~M., Robinson, A., \& Lena, D.\ 2017, \mnras, 471, 3888 

\bibitem[Schoenmakers et al.(1997)]{schoenmakers97} Schoenmakers, R.~H.~M., Franx, M., \& de Zeeuw, P.~T.\ 1997, \mnras, 292, 349 

 \bibitem[Sheth et al.(2005)]{sheth2005} Sheth, K., Vogel, S.~N., Regan, M.~W., Thornley, M.~D., \& Teuben, P.~J.\ 2005, \apj, 632, 217 

\bibitem[Sheth et al.(2010)]{sheth2010} Sheth, K., Regan, M., Hinz, J.~L., et al.\ 2010, \pasp, 122, 1397 

 \bibitem[Smaji{\'c} et al.(2015)]{smajic2015} Smaji{\'c}, S., Moser, L., Eckart, A., et al.\ 2015, arXiv:1508.02664 

 \bibitem[Sorce et al.(2014)]{sorce2014} Sorce, J.~G., Tully, R.~B., Courtois, H.~M., et al.\ 2014, \mnras, 444, 527

 \bibitem[Solomon \& Vanden Bout(2005)]{solbout2005} Solomon, P.~M., \& Vanden Bout, P.~A.\ 2005, \araa, 43, 677 

\bibitem[Soltan(1982)]{soltan82} Soltan, A.\ 1982, \mnras, 200, 115 

\bibitem[Spekkens \& Sellwood(2007)]{diskfit2007} Spekkens, K., \& Sellwood, J.~A.\ 2007, \apj, 664, 204  

 \bibitem[Storchi-Bergmann et al.(2007)]{storchi2007} Storchi-Bergmann, T., Dors, O.~L., Jr., Riffel, R.~A., et al.\ 2007, \apj, 670, 959

 \bibitem[Storchi-Bergmann et al.(2010)]{storchi2010} Storchi-Bergmann, T., Lopes, R.~D.~S., McGregor, P.~J., et al.\ 2010, \mnras, 402, 819 

 \bibitem[Tremaine et al.(2002)]{tremaine2002} Tremaine, S., Gebhardt, K., Bender, R., et al.\ 2002, \apj, 574, 740 

 \bibitem[Tully et al.(2013)]{tully2013} Tully, R.~B., Courtois, H.~M., Dolphin, A.~E., et al.\ 2013, \aj, 146, 86

 \bibitem[Ulvestad \& Ho(2001)]{ulandho2001} Ulvestad, J.~S., \& Ho, L.~C.\ 2001, \apj, 558, 561 

 \bibitem[van der Kruit \& Freeman(1984)]{vankruandfree84} van der Kruit, P.~C., \& Freeman, K.~C.\ 1984, \apj, 278, 81 

\bibitem[Veilleux \& Rupke(2002)]{veilandrub2002} Veilleux, S., \& Rupke, D.~S.\ 2002, \apjl, 565, L63 

\bibitem[Veilleux et al.(2005)]{veilleux2005} Veilleux, S., Cecil, G., \& Bland-Hawthorn, J.\ 2005, \araa, 43, 769 

 \bibitem[Westoby et al.(2012)]{westoby2012} Westoby, P.~B., Mundell, C.~G., Nagar, N.~M., et al.\ 2012, \apjs, 199, 1

\bibitem[Wada(1994)]{wada94} Wada, K.\ 1994, \pasj, 46, 165

 \bibitem[Woo et al.(2010)]{woo2010} Woo, J.-H., Treu, T., Barth, A.~J., et al.\ 2010, \apj, 716, 269 

 \bibitem[Woo \& Urry(2002)]{wooandurr2002} Woo, J.-H., \& Urry, C.~M.\ 2002, \apj, 579, 530 

\bibitem[Wong et al.(2004)]{wong2004} Wong, T., Blitz, L., \& Bosma, A.\ 2004, \apj, 605, 183 

 \bibitem[Yu \& Tremaine(2002)]{yuandtre2002} Yu, Q., \& Tremaine, S.\ 2002, \mnras, 335, 965

\end{thebibliography}
%

\end{document}